\documentclass[preprint,aps,prd,tightenlines,nofootinbib,eqsecnum,superscriptaddress]{revtex4-2}

\usepackage{graphicx}               
\usepackage{hyperref}               
\usepackage{xcolor}                 
\usepackage{amsmath}                
\usepackage{amssymb}
\usepackage{mathrsfs}
\usepackage{bm}
\usepackage{accents}
\usepackage[normalem]{ulem}         
\usepackage[bbgreekl]{mathbbol}

\DeclareSymbolFontAlphabet{\mathbbm}{bbold}
\DeclareSymbolFontAlphabet{\mathbb}{AMSb}

\newcommand{\beq}{\begin{equation}}
\newcommand{\eeq}{\end{equation}}
\newcommand{\ui}{\mathrm{i}}
\newcommand{\ud}{\mathrm{d}}
\newcommand{\Lie}{\mathcal{L}}
\newcommand{\calM}{\mathcal{M}}
\newcommand{\calH}{\mathcal{H}}
\newcommand{\calS}{\mathcal{S}}
\newcommand{\eqNEH}{\stackrel{\calH}{=}}
\newcommand{\eqEH}{\stackrel{\calH}{=}}
\newcommand{\Hajicek}{H\'aj\'{\i}\v{c}ek}
\newcommand{\pAdv}{\phi}

\makeatletter
\def\munderbar#1{\underline{\sbox\tw@{$#1$}\dp\tw@\z@\box\tw@}}
\makeatother

\definecolor{nicered}{rgb}{.647,.129,.149}
\definecolor{niceblue}{rgb}{.129,.149,.447}
\definecolor{nicegreen}{rgb}{.129,.647,.149}
\hypersetup{colorlinks=true,linkcolor=nicered,citecolor=niceblue,urlcolor=nicered}
\begin{document}

\title{Horizon Multipole Moments of a Kerr Black Hole}

\author{Eric Gourgoulhon}
\email{eric.gourgoulhon@obspm.fr}
\affiliation{LUX, Observatoire de Paris, Université PSL, CNRS, \\ Sorbonne Université, 92190 Meudon, France}
\affiliation{Laboratoire de Math\'ematiques de Bretagne Atlantique, CNRS UMR 6205, Universit\'e de Bretagne Occidentale, 6 avenue Victor Le Gorgeu, 29200 Brest, France}

\author{Alexandre Le Tiec}
\email{alexandre.letiec@obspm.fr}
\affiliation{LUX, Observatoire de Paris, Université PSL, CNRS, \\ Sorbonne Université, 92190 Meudon, France}
\affiliation{Instituto de F\'isica Te\'orica, UNESP -- Universidade Estadual Paulista, S\~ao Paulo 01140-070, SP, Brazil}

\author{Marc Casals}
\email{marc.casals@uni-leipzig.de}
\affiliation{Institut f\"ur Theoretische Physik, Universit\"at Leipzig,\\ Br\"uderstra{\ss}e 16, 04103 Leipzig, Germany}
\affiliation{School of Mathematics and Statistics, University College Dublin, Belfield D04 V1W8, Dublin 4,  Ireland}
\affiliation{Centro Brasileiro de Pesquisas F\'isicas (CBPF), Rio de Janeiro, CEP 22290-180, Brazil}

\date{25 March 2026}

\begin{abstract}
The horizon multipole moments of a Kerr black hole are computed from two distinct definitions that have been proposed in the literature. The first one [Ashtekar et al., Class. Quantum Grav. \textbf{21}, 2549 (2004)] regards axisymmetric isolated horizons, while the second one [Ashtekar et al., J. High Energ. Phys. \textbf{2022}, 28 (2022)] applies to generic (i.e., not necessarily axisymmetric) non-expanding horizons.
We review these definitions in a common frame and perform a detailed study of the resulting multipole moments for the Kerr event horizon.
The horizon multipoles are found to share several properties with the (Hansen) field multipoles, including parity constraints and the leading scaling behavior with respect to the Kerr spin parameter $a$ in the regime of small $a$. For the axisymmetry-based definition, we have obtained a closed-form expression of the
multipole moments in terms of $a$ and the spherical harmonic degree $\ell$. For the generic definition, we have established closed-form expressions for the conformal unit round metric, the `electric' and `magnetic' potentials related to the multipoles, and the values of the multipoles in the small $a$ limit. We show that the two definitions lead to different values of the Kerr horizon multipoles as soon as $\ell \geqslant 1$ (generic nonzero value of $a$) or $\ell\geqslant 2$ (small $a$ limit).
\end{abstract}

\maketitle

\newpage

\section{Introduction}

\subsection{Motivation}

The notion of multipole moment plays a central role in classical electrodynamics \cite{Jac} and Newtonian gravitational physics \cite{PoWi}. Indeed, whenever the details of the charge-current or mass distribution of the source can be ignored while addressing a specific problem, multipole expansions capture all of the physically relevant information about the structure of the electromagnetic or gravitational field. Two conceptually different notions of multipole moments need to be distinguished, however. On the one hand, the equations of motion for extended bodies can be formulated in terms of \textit{source} multipoles, defined as volume integrals over the charge-current or the mass density distribution; on the other hand, \textit{field} multipoles appear as coefficients in the asymptotic expansion of the (electromagnetic or gravitational) field itself, and are closely related to the fluxes of energy, angular momentum and linear momentum radiated by the source.

Because of such useful properties, there has been considerable interest in extending these notions to general relativity. In particular, for stationary, asymptotically flat spacetimes, the structure of the gravitational field in a neighborhood of spatial infinity has been shown to be fully characterized by two sets of field multipole moments $M_{\ell,m}$ and $S_{\ell,m}$, of mass-type and current-type, respectively \cite{Ge.70,Ge2.70,Ha.74,Th.80,BeSi.81,SiBe.83,Gu.83,Fo.al.89}. For instance, for a spinning (Kerr) black hole of mass $M > 0$ and angular momentum $S\in [0, M^2]$, the nonvanishing field multipoles are captured by the elegant formula (due to Hansen \cite{Ha.74})
\beq\label{Hansen}
    M_{\ell,0} + \ui S_{\ell,0} = M (\ui a)^\ell \, ,
\eeq
with $a \equiv S/ M$ the Kerr spin parameter and $\ell \in \mathbb{N}$ the spherical harmonic degree (thereafter we use the convention for which $0\in\mathbb{N}$). Regarding the source multipoles, Dixon \cite{Di.64,Di.70,Di.73,Di.74,Di.79} gave a definition for extended bodies in terms of the energy-momentum tensor $T_{ab}$, which was later generalized by Harte \cite{Ha.12,Ha.15} to account for self-interaction. The problem of defining the analog of source multipole moments for black holes, which correspond to \textit{vacuum} solutions ($T_{ab}=0$) of the Einstein equation, has only been tackled much more recently \cite{As.al.04,Sc.al.06,Ow.09,As.al.13,As.al.22}.

A first proposal (2004) by Ashtekar, Engle, Pawlowski and Van Den Broeck \cite{As.al.04} regards axisymmetric isolated horizons, while a more recent proposal (2022) by Ashtekar, Khera, Kolanowski and Lewandowski \cite{As.al.22} deals with non-expanding horizons and does not require any axisymmetry assumption. In both proposals, the Coulomb-type Weyl scalar $\Psi_2$ at the horizon plays the role of a complex-valued \textit{surface curvature density} and two sets of \emph{horizon multipole moments}, $I_{\ell,m}$ and $L_{\ell,m}$, are defined according to
\beq\label{def}
    I_{\ell,m} + \ui L_{\ell,m} \equiv - \oint_\calS \Psi_2 \, \mathring{Y}_{\ell,m} \, \ud S  \, .
\eeq
Here, the surface integral is taken over any (2-dimensional) cross-section $\calS$ of the horizon, $\ud S$ is the area element of $\calS$ induced by the spacetime metric $g_{ab}$, and
$\mathring{Y}_{\ell,m}$ are spherical harmonics with respect to a reference
\emph{unit round metric} $\mathring{q}_{ab}$ on $\calS$, i.e., a Riemannian metric
of constant scalar curvature $\mathring{\mathcal{R}} = 2$. The difference
between the two proposals lies in the choice of this unit round metric: in \cite{As.al.04}, $\mathring{q}_{ab}$ is uniquely defined from
the assumed axisymmetry of the horizon, while in \cite{As.al.22},
$\mathring{q}_{ab}$ is conformally related to the
physical metric $q_{ab}$ induced by $g_{ab}$ on $\calS$. In the main text, we shall denote the first metric by $\mathring{q}_{ab}^{\rm axi}$ to distinguish it from the second one. In both cases, it has been shown that the whole horizon geometry can be reconstructed from the knowledge of the multipole sequences $(I_{\ell,m}, L_{\ell,m})$ \cite{As.al.04,As.al.22}.

Interestingly, the axisymmetry-based definition \cite{As.al.04} has many applications in numerical relativity. Indeed, it has been used to assess the discrepancy between a black hole
numerical spacetime and the Kerr solution \cite{VaNoJa.09}. Furthermore, it has been extended to axisymmetric \textit{dynamical} horizons \cite{Sc.al.06}.\footnote{It has also been extended to non-axisymmetric dynamical horizons that tend to an axisymmetric isolated horizon in the asymptotic future \cite{As.al.13}.} The resulting horizon multipole moments have proven a powerful coordinate-invariant diagnostic of the strongly curved geometry in binary black hole mergers, either in the early inspiral phase (where each horizon is axisymmetric to a high degree of precision) \cite{Pr.21,Pr.al.22,Pr.24,Ri.al.24}, or in the post-merger phase (when the
common horizon becomes approximately axisymmetric) \cite{Re.al.10,JaMaMoRe.12,Pook.al.20,Pr.21,Mo.al.21,Gu.al.18,Ch.al.22,Pr.24,Ri.al.25}; see Ref.~\cite{AsKr.25} for a recent review.

In the context of Newtonian gravity (resp. classical electrodynamics), the linearity of the Poisson equation (resp. Maxwell equations) implies that field and source multipole moments coincide. In general relativity, however, this equality no longer holds: due to the nonlinear nature of the Einstein equation, the field multipoles are complicated functionals of the source multipoles \cite{Bl.24}. In particular, the relationship between the horizon multipole moments \eqref{def} of a non-expanding horizon and the corresponding field multipoles remain an open question. Even for the simplest case of an isolated, stationary rotating black hole, which is described by the Kerr metric in 4-dimensional general relativity by virtue of the no-hair theorem \cite{ChCoHe.12},
it turns out that no closed-form formula expressing the horizon multipoles in terms of the spin parameter $a$ and arbitrary spherical harmonic degree $\ell$ has been provided in the literature. It seems that the Kerr horizon multipoles have been computed only for the axisymmetry-based definition, and moreover only for $\ell \leqslant 3$ \cite{As.al.04} or  $\ell \leqslant 8$ \cite{Gu.al.18}.
Our aim here is to compute the Kerr horizon multipoles for all values of $\ell$ for both definitions, to compare them, and to study their relationship with the field multipoles \eqref{Hansen}. We achieve these goals as reviewed in the following subsection.

\subsection{Main results} \label{sec:main_results}

In this work, we review the definitions of horizon multipoles given in Refs.~\cite{As.al.04} and \cite{As.al.22}, and apply them to the event horizon of a Kerr black hole.
This event horizon is a Killing horizon, and thus an isolated horizon, and is moreover axisymmetric, so that both definitions are applicable.
For the axisymmetry-based definition \cite{As.al.04}, we derived formula~\eqref{truc} below,
which generalizes to all $\ell \in \mathbb{N}$ the expression of the multipole moments obtained in Ref.~\cite{Gu.al.18} for $2 \leqslant \ell \leqslant 8$. For the generic definition \cite{As.al.22}, we have computed the conformal factor $\psi$ between the unit round metric metric $\mathring{q}_{ab}$ and the physical metric $q_{ab}$ of the horizon cross-sections [Eq.~\eqref{psi_Kerr} below], thereby correcting a previous formula given in Ref.~\cite{As.al.22}. We have also obtained a rather simple expression of $\mathring{q}_{ab}$ in terms of the standard Kerr angular coordinates $(\theta,\varphi)$ [Eq.~\eqref{round_metric_Kerr}]. Moreover, we have derived a closed-form expression for both the `electric' and `magnetic' potentials involved in the construction of Ref.~\cite{As.al.22} [Eq.~\eqref{E_B_Kerr}]. We have expressed the resulting horizon multipole moments \eqref{def} in terms of a complicated integral [Eq.~\eqref{I-L_Kerr2}]; while we could not express this integral in terms of standard functions for a generic (non small) value of $a$, we devised a numerical code to compute it to an arbitrary precision and have made it publicly available (cf. App.~\ref{app:Sage}).

Using either definition, the obtained horizon multipoles obey the same parity constraints as the Hansen field multipoles \eqref{Hansen}: the axisymmetry of the Kerr metric implies that the only nonvanishing multipoles have $m=0$, and its discrete symmetry across the equatorial plane further implies that $I_{2n+1,0} = 0$ and $L_{2n,0} = 0$ for all $n \in \mathbb{N}$. In the regime $0 < a \ll M$ of small-spin values, we have shown that the nonvanishing horizon multipoles of a Kerr black hole behave as [Eqs.~\eqref{I-L_Kerr3-axi} and \eqref{I-L_Kerr3} below]
\beq\label{scaling}
    I_{\ell,0} + \ui L_{\ell,0} \sim \, \frac{\sqrt{(2\ell+1)\pi}}{M^\ell} \, (\ui a)^\ell \times
    \begin{cases}
        \frac{\ell!\, (\ell+2)!}{2(2\ell+1)!} & \text{(axisymmetric)} \\
        2^{-\ell}\alpha_\ell & \text{(generic)}
    \end{cases} \, ,
\eeq
where $(\alpha_\ell)_{\ell \in \mathbb{N}}$ is a sequence of rational numbers that increases `slowly' with $\ell$ (see Eq.~\eqref{alpha_ell} and Table~\ref{table:sequence}). The scaling behavior \eqref{scaling} is clearly reminiscent of the corresponding field multipoles \eqref{Hansen}. As the Kerr black hole spin increases, however, that scaling breaks down and the horizon multipole moments deviate significantly from Hansen's formula (see Figs.~\ref{fig:Kerr_I_L_axi}, \ref{fig:Kerr_M_ell}, \ref{fig:Kerr_S_ell}, \ref{fig:Kerr_I} and \ref{fig:Kerr_L}). Moreover, even for small spin values, the coefficient in front of $(\ui a)^\ell$ differs between field and horizon multipoles, except for $\ell=0$ and $\ell=1$ for the (properly rescaled) axisymmetry-based multipoles.
This coefficient also differs between the two families of horizon multipoles, except for $\ell=0$ and $\ell=1$, thereby showing that the generic definition \cite{As.al.22} yields values of the Kerr horizon multipoles distinct from those arising from the axisymmetry-based definition \cite{As.al.04}. For finite values of $a$, the difference occurs for any $\ell\geqslant 1$. In particular,
the ratio between a multipole of the generic family and the axisymmetry-based one of the same $\ell$ diverges
as $\ell\to +\infty$ [Eq.~\eqref{diverging_multipole_ratio} and Figs.~\ref{fig:Kerr_I_comp_axi}--\ref{fig:Kerr_L_comp_axi}].

The remainder of this paper is organized as follows. The various geometrical objects involved in the two definitions of horizon multipole moments are introduced first at the level of a generic null hypersurface (Sec.~\ref{sec:horizon}), and
then at the level of a non-expanding horizon (Sec.~\ref{sec:NEH}). The axisymmetry-based definition of horizon multipoles is reviewed in Sec.~\ref{sec:axi_multipoles}, while that
for generic non-expanding horizons is reviewed in Sec.~\ref{sec:generic_multipoles}. The application to the Kerr black hole
is performed in Sec.~\ref{sec:Kerr}, which presents the results summarized above.
Section~\ref{sec:concl} gives some concluding remarks and future prospects.
Some technical aspects are relegated to appendices: App.~\ref{app:integ_Kerr_axi} is devoted to the computation of the integral involved
in the axisymmetry-based multipoles of the Kerr horizon; App.~\ref{app:magnetic} provides
an alternative derivation of the electric-type and magnetic-type potentials of the Kerr horizon; App.~\ref{app:asymptotics} regards the small spin behavior of the Kerr horizon multipoles of the generic family; App.~\ref{app:Sage} provides the links to SageMath notebooks used for some symbolic or numerical computations.

Our conventions are those of Wald \cite{Wal}. In particular, the metric signature is $(-,+,+,+)$ and the Riemann curvature tensor $R_{abc}^{\phantom{acd}d}$ is defined by $2\nabla_{[a} \nabla_{b]} \omega_c \!=\! R_{abc}^{\phantom{acd}d} \omega_d$ for any 1-form $\omega_a$. The Latin letters $(a,b,c,\dots)$ over tensors denote abstract indices for tensor fields defined over a manifold, independently of the dimension of the latter, which can be 4 (spacetime), 3 (horizon) or 2 (horizon cross-sections). In the calculations regarding Kerr spacetime, we use advanced Kerr coordinates $(v,r,\theta,\phi)$, which are regular on the future event horizon. Throughout the paper, an overbar denotes complex conjugation and we set $G = c = 1$.

\section{Basic geometry of null hypersurfaces}\label{sec:horizon}

The multipole moments introduced in Refs.~\cite{As.al.04} and \cite{As.al.22} regard non-expanding horizons, which are a specific type of null hypersurfaces adapted
to the description of the event horizon of a black hole in equilibrium.
In this section, we therefore review basic properties of null hypersurfaces and introduce geometrical objects---such as the
rotation 1-form---which play a role in the definition of the horizon multipole moments.

\subsection{First and second fundamental forms, expansion and shear} \label{s:basic_geom_hor}

We consider a 4-dimensional time-oriented spacetime $(\calM,g_{ab})$ and a null hypersurface $\calH$ of $\calM$ with the topology
\beq \label{H_topology}
    \calH \sim \mathbb{R}\times\mathbb{S}^2 \, .
\eeq
For a stationary black hole with a connected event horizon, this is the only possible topology in 4-dimensional general relativity (assuming vacuum or the null energy condition) \cite{Ha.72,Ch.Wa.94}.

The \emph{first fundamental form} of $\calH$ is the ``metric'' $h_{ab}$ induced by
$g_{ab}$, i.e., the pullback of $g_{ab}$ on $\calH$ by the inclusion map
$\iota: \calH \hookrightarrow  \calM$: $\bm{h} \equiv \iota^* \bm{g}$. In other words, for any vectors
$u^a$ and $v^a$ tangent to $\calH$, $h_{ab} u^a v^b \equiv g_{ab} u^a v^b$.
By definition of a null hypersurface, at each point $p\in\calH$,
$h_{ab}$ is a degenerate symmetric bilinear form, of signature $(0,+,+)$.

\begin{figure}[t]
    \begin{center}
        \includegraphics[height=0.26\textheight]{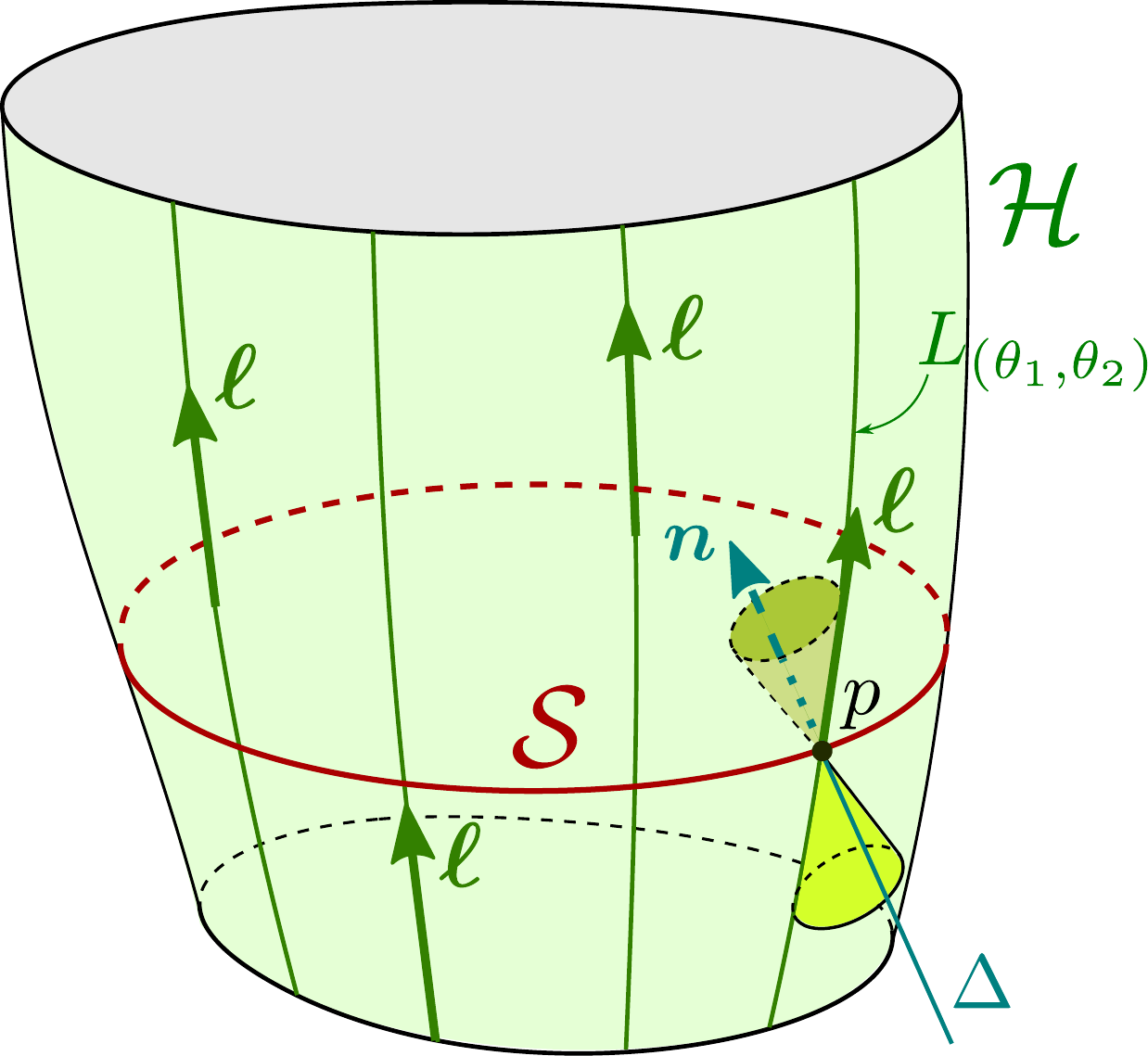}
        \caption{Null hypersurface $\calH$ of topology $\mathbb{R}\times\mathbb{S}^2$. Some of the null geodesic generators $L_{(\theta_1,\theta_2)}$ ruling $\calH$ are depicted as green curves with tangent vectors $\bm{\ell}$, the latter being null normals to $\calH$. $\calS$ is a cross-section of $\calH$; it is drawn as a curve (of topology $\mathbb{S}^1$), instead of a surface (of topology $\mathbb{S}^2$), due to the dimensional reduction of the graphic.
        The metric null cone is depicted at some point $p\in\calS$; this cone is tangent to $\calH$ along the null generator $L_{(\theta_1,\theta_2)}$ through $p$, with its past (resp. future) nappe lying outside (resp. inside) $\calH$. $\bm{n}$ is a future-directed null vector transverse to $\calH$ and normal to $\calS$ at $p$, while $\Delta$ is an ingoing null geodesic admitting $\bm{n}$ as a tangent vector.}
        \label{fig:null_hypersurface}
    \end{center}
\end{figure}

As a null hypersurface, $\calH$ is ruled by a 2-parameter family $L_{(\theta_1,\theta_2)}$ of null
geodesics, called the \emph{generators of} $\calH$ (cf. Fig.~\ref{fig:null_hypersurface}). Given the topology \eqref{H_topology}, the parameters
$(\theta_1,\theta_2)$ span the sphere $\mathbb{S}^2$. For instance they can be polar coordinates,
but any other coordinate system on $\mathbb{S}^2$ will do; what matters is that $(\theta_1,\theta_2)$ stays constant along a given generator. For example, for the event horizon of a Kerr black hole described in terms of advanced Kerr coordinates $(v,r,\theta,\phi)$ (cf. Sec.~\ref{sec:Kerr}),
one may choose $\theta_1 = \theta$ and $\theta_2 = \phi - \varOmega_{\calH} v$, where $\varOmega_{\calH}$ is the horizon's (constant) angular velocity.

Any vector field $\ell^a$ normal to $\calH$ is null, tangent to $\calH$ and
lies in the kernel of $h_{ab}$, so that $h_{ab} \ell^b = 0$.
In what follows, we consider only future-directed null normals $\ell^a$.
Since the kernel of $h_{ab}$ is 1-dimensional, given the signature $(0,+,+)$,
$\ell^a$ is necessarily tangent to the null geodesic generators
$L_{(\theta_1,\theta_2)}$ (cf. Fig.~\ref{fig:null_hypersurface}). It follows that $\ell^a$ is a \emph{pregeodesic} vector field:
there exists a scalar field $\kappa_{(\bm{\ell})}$ on $\calH$ such that
\beq \label{ell_pregeodesic}
    \ell^b \nabla_b \ell^a = \kappa_{(\bm{\ell})} \ell^a \, .
\eeq
Along a given generator $L_{(\theta_1,\theta_2)}$, $\ell^a$ can be viewed as the tangent
vector associated to
some parameter $v$ of $L_{(\theta_1,\theta_2)}$:
\beq \label{l_velocity_v}
    \ell^a = \left. \frac{\ud x^a}{\ud v} \right| _{L_{(\theta_1,\theta_2)}} .
\eeq
The quantity $\kappa_{(\bm{\ell})}$ measures the lack of affinity of $v$: $v$ is an affine parameter of the geodesic $L_{(\theta_1,\theta_2)}$ if, and only if,
$\kappa_{(\bm{\ell})} = 0$. If this happens, the right-hand side of \eqref{ell_pregeodesic} is
zero and $\ell^a$ is called a \emph{geodesic} vector.
By letting $v$ vary smoothly from one generator to the other, one obtains
a coordinate system $(v,\theta_1,\theta_2)$ on $\calH$ and, by construction, $\ell^a$
is the first vector of the associated coordinate basis:
$\ell^a = (\partial_v)^a$. It follows that $\ell^a \nabla_a v = 1$.

Since $h_{ab}$ is degenerate, there is no canonical volume 3-form on $\calH$ associated to it. However, there exists a unique (up to a sign) 2-form ${}^\calH\varepsilon_{ab}$ such that ${}^\calH\varepsilon_{ab}\, e_1^a e_2^b = \pm 1$ for any pair of unit spacelike vectors $(e_1^a,e_2^a)$ tangent to $\calH$ and orthogonal to each other.
It is defined as follows: for any pair of vectors $(u^a, v^a)$ in
$T_p\calH$,
\beq \label{def_area_form_H}
  {}^\calH\varepsilon_{ab} u^a v^b \equiv \varepsilon_{cdab} n^c \ell^d u^a v^b \, ,
\eeq
where $\varepsilon_{abcd}$ is the Levi-Civita tensor (volume 4-form) associated to the spacetime metric $g_{ab}$, $\ell^a$ is any null normal to $\calH$ and $n^a$ is any null vector such that $\ell_a n^a = -1$ (which implies that $n^a$ is transverse to $\calH$; cf. Fig.~\ref{fig:null_hypersurface}). The definition \eqref{def_area_form_H} is independent of the choice of $\ell^a$ and $n^a$. Indeed, any other pair $({\ell'}^a,{n'}^a)$ is necessarily related
to $(\ell^a,n^a)$ by ${\ell'}^a = f \ell^a$ and
${n'}^a = f^{-1} n^a + \alpha^0 \ell^a + \alpha^1 w_1^a + \alpha^2 w_2^a$, where $w_1^a$ and $w_2^a$ are two vectors tangent to $\calH$, such that $(\ell^a,w_1^a,w_2^a)$ is a basis
of $T_p\calH$. Accordingly,
\[
  \varepsilon_{cdab} {n'}^c {\ell'}^d u^a v^b
  = \varepsilon_{cdab} n^c \ell^d u^a v^b + f \alpha^1
     \underbrace{\varepsilon_{cdab}  w_1^c \ell^d u^a v^b}_{0}
   + f \alpha^2
     \underbrace{\varepsilon_{cdab} w_2^c \ell^d  u^a v^b}_{0}
     = {}^\calH\varepsilon_{ab} u^a v^b \, ,
\]
where the $0$'s occur because $\varepsilon_{abcd}$ is a 4-form and the vectors $w_{1/2}^a$, $\ell^a$, $u^a$ and $v^a$ are linearly dependent (four vectors in the 3-space $T_p\calH$).
The 2-form ${}^\calH\varepsilon_{ab}$ is called the \emph{area 2-form of $\calH$}. Note that, by construction,
\beq \label{epsilon_H_ell_zero}
  {}^\calH\varepsilon_{ab} \ell^b = 0 \, .
\eeq

The extrinsic geometry of $\calH$ is defined by the
variation of the normal $\ell^a$ along $\calH$ with respect to the ambient connection $\nabla_a$. It is measured by the \emph{Weingarten map} $\bm{\chi}$ \cite{Da.79,Da.82b},
which, at each point $p\in\calH$, is an endomorphism of the
tangent space $T_p\calH$ defined by
\beq \label{def_Weingarten}
    \chi^a_{\ \, b} u^b \equiv u^b \nabla_b \ell^a \, ,
\eeq
for any vector $u^a\in T_p\calH$.
One has $\ell_a \chi^a_{\ \, b} u^b = u^b \ell_a  \nabla_b \ell^a = 0$,
which shows that $\chi^a_{\ \, b} u^b \in T_p\calH$. Hence
$\bm{\chi}$ is well defined as an endomorphism of $T_p\calH$. Moreover $\bm{\chi}$
is self-adjoint with respect to $h_{ab}$:
$h_{ab} u^a \chi^b_{\ \, c} v^c = h_{ab} \chi^a_{\ \, c} u^c v^b$.
This follows from $\ell^a$ being normal to a hypersurface.
In view of Eq.~\eqref{ell_pregeodesic}, $\ell^a$ is an eigenvector of
the Weingarten map, of eigenvalue $\kappa_{(\bm{\ell})}$:
\beq \label{l_eigenvect_chi}
    \chi^a_{\ \, b} \ell^b =  \kappa_{(\bm{\ell})} \ell^a \, .
\eeq

The \emph{second fundamental form} of $\calH$ is the field of symmetric bilinear
forms $\Theta_{ab}$ on $\calH$ defined by
\beq \label{def_2nd_fund_form}
    \Theta_{ab} \equiv h_{ac} \chi^c_{\ \, b} \, .
\eeq
It is symmetric because $\bm{\chi}$ is self-adjoint.
Given Eq.~\eqref{def_Weingarten}, one has
$\Theta_{ab} u^a v^b = h_{ac} u^a v^b \nabla_b \ell^c = g_{ac}u^a v^b \nabla_b \ell^c
= (\nabla_b \ell_a) u^a v^b$, so that $\Theta_{ab}$ can be viewed as the pullback of the tensor field $(\nabla \ell^\flat)_{ab} = \nabla_b\ell_a$ (defined on $\calM$, given any
extension of $\ell^a$ in some neighborhood of $\calH$)
by the inclusion map $\iota: \calH \hookrightarrow \calM$:
\beq \label{Theta_pullback_nab_l}
    \bm{\Theta} = \iota^* \bm{\nabla\ell^\flat} \, .
\eeq
Just like $h_{ab}$, $\Theta_{ab}$ is a degenerate symmetric bilinear form with $\ell^a$ in its kernel: $\Theta_{ab} \ell^b = 0$.
The second fundamental form $\Theta_{ab}$ can be viewed as
the deformation tensor of $h_{ab}$ along the normal $\ell^a$, since
$\Theta_{ab}$ is (one half of) the Lie derivative of
$h_{ab}$ along $\ell^a$:
\beq \label{Theta_Lie_l_q}
    \Theta_{ab} = \tfrac{1}{2} \, \Lie_{\bm{\ell}} h_{ab} \, .
\eeq
This relationship easily follows from the definition of the Lie derivative.
Indeed, $\Lie_{\bm{\ell}} \bm{h} \equiv \lim_{\varepsilon\to 0} (\Phi_\varepsilon^* \bm{h} - \bm{h})/\varepsilon$, where $\Phi_\varepsilon^* \bm{h}$ is the pullback of $\bm{h}$ by the flow map $\Phi_\varepsilon$ of displacement $\varepsilon$ along
$\ell^a$. Now, since $\bm{h} = \iota^* \bm{g}$, we have $\Phi_\varepsilon^* \bm{h}= \iota^* \Phi_\varepsilon^* \bm{g}$, so that $\Lie_{\bm{\ell}} \bm{h} = \iota^* \Lie_{\bm{\ell}} \bm{g}$.
The identity $\Lie_{\bm{\ell}} g_{ab} = \nabla_a \ell_b + \nabla_b \ell_a$
along with Eq.~\eqref{Theta_pullback_nab_l} and the symmetry of $\Theta_{ab}$ leads to
Eq.~\eqref{Theta_Lie_l_q}.

The \emph{expansion} $\theta_{(\bm{\ell})}$ of the null normal $\ell^a$ is the scalar field defined on $\calH$ by
\beq \label{def_expansion}
  \theta_{(\bm{\ell})} \equiv \chi^a_{\ \, a} - \kappa_{(\bm{\ell})} \, .
\eeq
Notice that $\chi^a_{\ \, a}$ is the trace of the endomorphism $\bm{\chi}$, and is thus a well defined scalar field on $\calH$.
In terms of the coordinate system $(v,\theta_1,\theta_2)$ introduced above, it follows from Eq.~\eqref{l_eigenvect_chi} that $ \theta_{(\bm{\ell})} =  \chi^{\theta_1}_{\ \, \theta_1} + \chi^{\theta_2}_{\ \, \theta_2}$.
Moreover, it is easy to see that $\theta_{(\bm{\ell})}$ is related to the Lie derivative of the area 2-form ${}^\calH \varepsilon_{ab}$ along $\ell^a$ by
\beq \label{Lie_l_Heps}
  \Lie_{\bm{\ell}} {}^\calH \varepsilon_{ab} = \theta_{(\bm{\ell})} \, {}^\calH \varepsilon_{ab} \, .
\eeq

The \emph{shear tensor} $\sigma_{ab}$ of the null normal $\ell^a$ is the tensor field defined on $\calH$
by
\beq \label{def_shear_tensor}
  \sigma_{ab} \equiv \Theta_{ab} - \tfrac{1}{2} \, \theta_{(\bm{\ell})} \, h_{ab} \, .
\eeq
Like $h_{ab}$ and $\Theta_{ab}$, $\sigma_{ab}$ is a degenerate symmetric bilinear form, with $\ell^a$ in its kernel:
$\sigma_{ab} \ell^b = 0$.

The above definitions of the first and second fundamental forms, $\bm{h}$ and $\bm{\Theta}$, and of the Weingarten map $\bm{\chi}$ follow those for spacelike and timelike hypersurfaces. There is a major difference though: for a spacelike (resp. timelike) hypersurface, the normal vector $\ell^a$ can always be normalized by
$\ell_a \ell^a = -1$ (resp. $\ell_a \ell^a = 1$).
As a result, for non-null hypersurfaces, the tensor fields $\bm{\chi}$ and $\bm{\Theta}$ are unique (up to a sign). On the contrary, for the null hypersurface $\calH$, $\bm{\chi}$ and
$\bm{\Theta}$ depend on $\ell^a$. The latter can be arbitrarily rescaled as
\beq \label{rescale_ell}
  \ell^a \to {\ell'}^a = f \ell^a
\eeq
for any smooth positive scalar field $f$ on $\calH$, with $f>0$ to preserve the future orientation. It is easy to see that, under the rescaling \eqref{rescale_ell}, the geometric quantities introduced so far change as follows:\footnote{Notice a slight asymmetry in our notations: we have kept the dependency on $\ell^a$ explicit for the scalar quantities $\kappa_{(\bm{\ell})}$ and $\theta_{(\bm{\ell})}$, but have dropped it for the tensor quantities $\chi^a_{\ \, b}$, $\Theta_{ab}$ and $\sigma_{ab}$, to avoid cluttering with tensor indices.}
\begin{subequations}
\begin{align}
& \kappa_{(\bm{\ell})} \to \kappa_{(\bm{\ell}')} = f \kappa_{(\bm{\ell})} +  \ell^a (\ud f)_a \, ,   \label{rescaled_kappa} \\
& \chi^a_{\phantom{a}b} \to  \chi'^a_{\phantom{'a}b} = f \chi^a_{\phantom{a}b} + \ell^a (\ud f)_b \, , \label{rescaled_chi} \\
& \Theta_{ab} \to \Theta'_{ab} = f \Theta_{ab} \, , \\
& \theta_{(\bm{\ell})} \to \theta_{(\bm{\ell}')} = f  \theta_{(\bm{\ell})} \, , \label{theta_rescale_ell} \\
& \sigma_{ab} \to \sigma'_{ab} = f \sigma_{ab} \, .
\end{align}
\end{subequations}


\subsection{Cross-sections of a null hypersurface} \label{s:cross_section_slicings}

A \emph{cross-section} of $\calH$ is a 2-dimensional submanifold $\calS$ of $\calH$ such that each null geodesic generator $L_{(\theta_1,\theta_2)}$ intersects $\calS$ once, and only once,
without being tangent to $\calS$  (cf. Fig.~\ref{fig:null_hypersurface}). From \eqref{H_topology} it is clear that all cross-sections have
the topology of $\mathbb{S}^2$. Moreover, given the signature $(0,+,+)$ of
$\bm{h}$, any cross-section $\calS$ is necessarily a spacelike surface. Hence the pullback $\bm{q}$ of $\bm{h}$ by the inclusion
map $\jmath : \calS \hookrightarrow \calH$ is a Riemannian (i.e. positive definite) metric: for any pair of vectors $(u^a,v^a)$ tangent to $\calS$,
$q_{ab} u^a v^b \equiv h_{ab} u^a v^b = g_{ab} u^a v^b$
and $q_{ab} u^a u^b > 0$ as long as $u^a \neq 0$. One can easily see that the
Levi-Civita tensor (area 2-form) $\varepsilon_{ab}$ of $q_{ab}$ coincides with the pullback of the area 2-form of $\calH$ [Eq.~\eqref{def_area_form_H}] by the inclusion map:
\beq \label{Seps_pullback_Heps}
  \varepsilon_{ab} = \jmath^* \, {}^{\calH} \varepsilon_{ab} \, .
\eeq

\begin{figure}[t]
    \begin{center}
        \includegraphics[height=0.28\textheight]{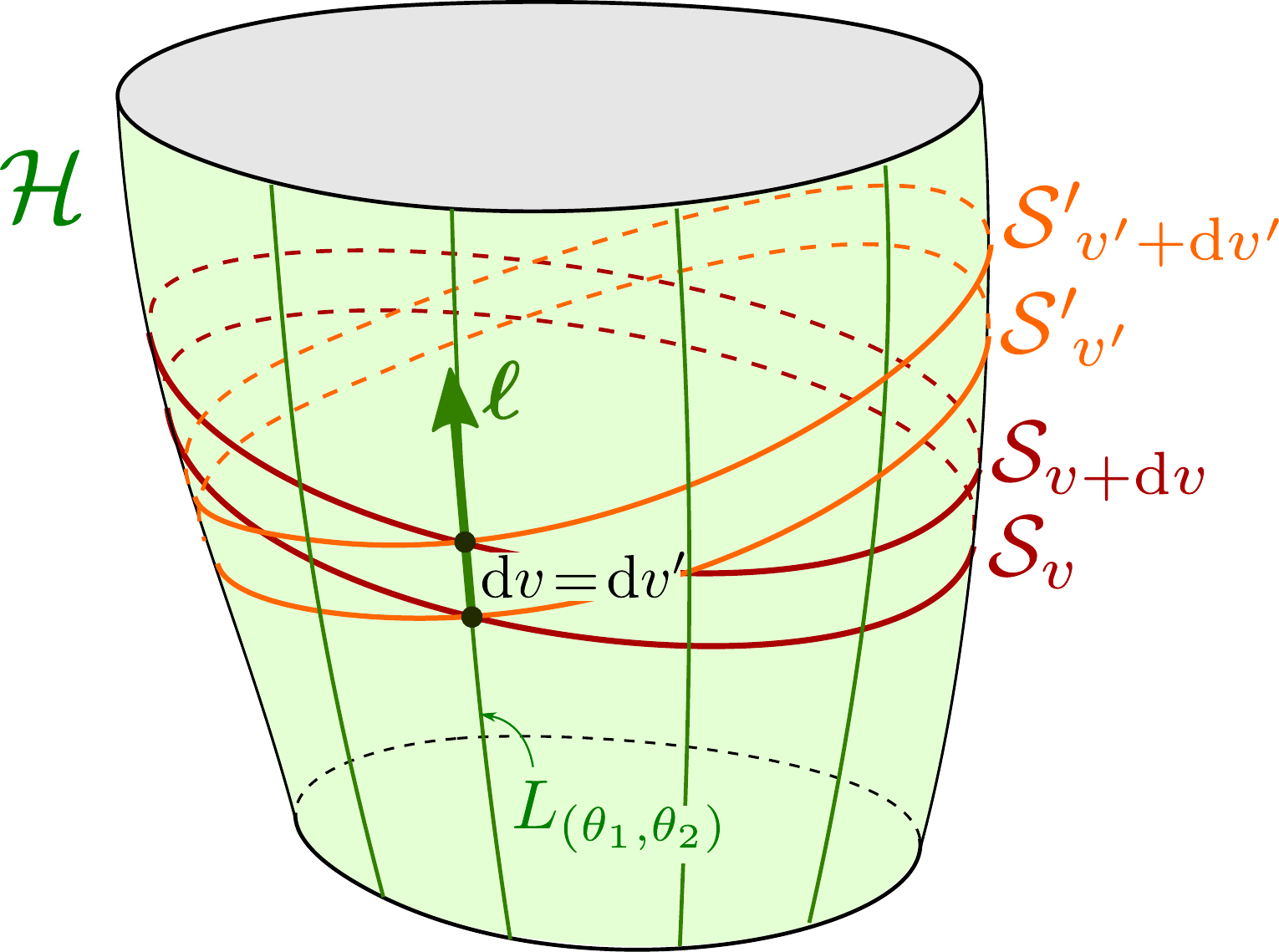}
        \caption{Two cross-section slicings $(\calS_v)_{v\in\mathbb{R}}$
and $(\calS'_{v'})_{v'\in\mathbb{R}}$ of a null hypersurface $\calH$, both defining the same null normal $\bm{\ell}$ to $\calH$:
$\ell^a = \ud x^a / \ud v = \ud x^a / \ud v'$ along the generators $L_{(\theta_1,\theta_2)}$ of $\calH$.}
        \label{fig:cross_section_slicings}
    \end{center}
\end{figure}

A \emph{cross-section slicing of} $\calH$ is a 1-parameter family
$(\calS_v)_{v\in\mathbb{R}}$ of non-intersecting cross-sections such
that $\calH = \bigcup_{v\in\mathbb{R}} \calS_v$.
Combining the slicing parameter $v$ with the parameters
$(\theta_1,\theta_2)$ labelling the generators of $\calH$ yields a coordinate system $(v,\theta_1,\theta_2)$ on
$\calH$.
The slicing parameter $v$ can also be considered as a regular parameter along each of the generators $L_{(\theta_1,\theta_2)}$.
The tangent vector associated to
this parametrization of $L_{(\theta_1,\theta_2)}$ by
\eqref{l_velocity_v} provides a normal vector field
$\ell^a$ to $\calH$, which is called the
\emph{null normal adapted to the slicing} $(\calS_v)_{v\in\mathbb{R}}$.
By definition, the point where $\calS_v$ intersects a given generator
$L_{(\theta_1,\theta_2)}$ is connected to the nearby point
on $\calS_{v+\ud v}$ along the same generator by the infinitesimal
vector $\ud v\, \ell^a$ (cf. Fig.~\ref{fig:cross_section_slicings}).

For a given cross-section slicing, there is a unique null normal $\ell^a$  to $\calH$ that is adapted to it. The converse is not true: given a normal $\ell^a$ to $\calH$, there are many distinct slicings for which $\ell^a$ is an adapted normal (cf. Fig.~\ref{fig:cross_section_slicings}).
Two such slicings, $(\calS_v)_{v\in\mathbb{R}}$
and $(\calS'_{v'})_{v'\in\mathbb{R}}$, say, are connected by
\beq \label{v_vp_H}
  v = v' + H(\theta_1,\theta_2) \, ,
\eeq
where $H$ is a (real-valued) smooth function. Indeed, Eq.~\eqref{v_vp_H} is a necessary and sufficient condition for having $\ud v' = \ud v$ along any geodesic generator $L_{(\theta_1,\theta_2)}$ of $\calH$, which ensures that  both $v$ and $v'$ parametrizations of $L_{(\theta_1,\theta_2)}$ lead to the same normal $\ell^a$ via \eqref{l_velocity_v}.

Not only does a slicing $(\calS_v)_{v\in\mathbb{R}}$ of $\calH$ define a unique null normal $\ell^a$ to $\calH$, but it defines as well a unique
vector field $n^a$ on $\calH$ that is null,  normal to $\calS_v$, and obeys
\beq \label{ln_minus_one}
  \ell_a n^a = -1 \, .
\eeq
Notice that \eqref{ln_minus_one} implies that $n^a$ is necessarily transverse to $\calH$  (cf. Fig.~\ref{fig:null_hypersurface}).
The uniqueness of $n^a$ follows from the spacelike character of the cross-sections. Indeed, the spacelike character of the tangent space $T_p\calS_v$ at any point $p\in\calS_v$
allows one to express $T_p\calM$ as the direct sum
$T_p\calM = T_p\calS_v \oplus T_p^\perp \calS_v$, with the
timelike 2-plane $T_p^\perp \calS_v$ intersecting the null cone of $g_{ab}$ at $p$ along two directions: one is necessarily along $\ell^a$---for $\ell^a$ is normal to $\calH$ and hence to $\calS_v$---and   $n^a$, as defined above, lies along the second direction. The normalization relation \eqref{ln_minus_one} then determines $n^a$ uniquely.
Moreover the pullback of the 1-form $n_a$ to $\calH$ by the inclusion map $\iota: \calH \hookrightarrow  \calM$ coincides with minus the differential
of $v$ considered as a scalar field on $\calH$:
\beq \label{pb_n_dv}
    \iota^* n_a = - (\ud v)_a \, .
\eeq
Indeed, $n_a \ell^a = -1 = - \ell^a \nabla_a v = - (\ud v)_a \ell^a$ by \eqref{ln_minus_one} and \eqref{l_velocity_v} and for any vector $u^a$ tangent to $\calS_v$, $n_a u^a = 0 = - u^a \nabla_a v = - (\ud v)_a u^a$ (since $n^a$ is normal to $\calS_v$), which proves \eqref{pb_n_dv}; see also Eq.~(4.34) in Ref.~\cite{GoJa.06}.


\subsection{Rotation 1-form and \Hajicek{} 1-form}
\label{s:rotation_form}

Let $(\calS_v)_{v\in\mathbb{R}}$ be a cross-section slicing of $\calH$ and $\ell^a$ the associated normal to $\calH$. The~\emph{rotation 1-form} $\omega_a$
associated to $(\calS_v)_{v\in\mathbb{R}}$ is the 1-form defined on $\calH$ by
\beq\label{omega_n_chi}
  \omega_a  \equiv (\ud v)_b \chi^b_{\ \, a} = - n_b \chi^b_{\ \, a} \, ,
\eeq
where $\chi^b_{\ \, a}$ is the Weingarten map associated to $\ell^a$ and $n^a$ be the unique null normal to $\calS_v$ such that
$\ell_a n^a = -1$  [Eq.~\eqref{ln_minus_one}].
The second equality in Eq.~\eqref{omega_n_chi} stems from Eq.~\eqref{pb_n_dv},
with the understanding that $n_a$ actually stands for
$\iota^* n_a$.
An immediate consequence of \eqref{l_eigenvect_chi} and \eqref{ln_minus_one} is
\beq \label{omega_ell_kappa}
  \omega_a \ell^a = \kappa_{(\bm{\ell})} \, .
\eeq

Under a rescaling $\ell^a \to f \ell^a$ of the null normal to $\calH$,
keeping the cross-sections $\calS_v$ fixed (albeit relabelling
them to $\calS_{v'}$ with $\ud v' \!=\! f^{-1} \ud v$), so that
$n^a$ is merely rescaled as $n^a \!\to\! f^{-1} n^a$,
one has $\chi^b_{\ \, a} \to f \chi^b_{\ \, a} + \ell^b (\ud f)_a$ [Eq.~\eqref{rescaled_chi}], so that the rotation 1-form changes only by the addition of the
differential of $\ln f$:
\beq \label{rescaled_omega}
     \omega_a \to \omega_a + (\ud \ln f)_a \, .
\eeq

It is worth stressing that the 1-form $\omega_a$ not only depends
on the scaling of $\ell^a$, as expressed by \eqref{rescaled_omega},
but it depends as well on the chosen cross-section slicing of $\calH$, contrary to $h_{ab}$,
$\chi^a_{\ \, b}$, $\Theta_{ab}$, $\theta_{(\bm{\ell})}$ and $\sigma_{ab}$.
This is clear from the definition \eqref{omega_n_chi}, which involves the $\calH$-transverse null normal $n^a$ to the slices
(cf. Sec.~\ref{s:cross_section_slicings}).

The \emph{\Hajicek{} 1-form} \cite{Ha3.73,Da.82b,GoJa.06} associated to the cross-section slicing  $(\calS_v)_{v\in\mathbb{R}}$ is the 1-form defined on $\calH$ by
\beq \label{def_Hajicek}
  \Omega_a \equiv \omega_b q^b_{\ \, a} \, ,
\eeq
where
\beq \label{def_projector_S_H}
  q^b_{\ \, a} \equiv \delta^b_{\ \, a} + \ell^b n_a
\eeq
is the orthogonal projector on $\calS_v$ within $\calH$. (The orthogonal projector on $\calS_v$ within $\mathcal{M}$ is $q^b_{\ \, a} \equiv \delta^b_{\ \, a} + \ell^b n_a + n^b \ell_a$. Given that $\iota^* \ell_a = 0$, the two operators coincide when applied to vectors tangent to $\calH$.) Since $q^b_{\ \, a} \ell^a = 0$, an immediate property is $\Omega_a \ell^a = 0$.
We deduce from \eqref{def_Hajicek}--\eqref{def_projector_S_H}
and $\omega_b \ell^b = \kappa_{(\bm{\ell})}$ [Eq.~\eqref{omega_ell_kappa}] that
\beq \label{Omega_omega_kappa_n}
  \Omega_a = \omega_a + \kappa_{(\bm{\ell})} \, n_a = \omega_a -\kappa_{(\bm{\ell})} \, (\ud v)_a \, ,
\eeq
where the second equality stems from \eqref{pb_n_dv}.
By combining Eqs.~\eqref{def_Hajicek}, \eqref{omega_n_chi} and
\eqref{def_Weingarten}, one gets an alternative expression of the
\Hajicek{} 1-form:
\beq \label{Hajicek_n_nabla_l_q}
  \Omega_a = - n_c \nabla_b \ell^c \, q^b_{\ \, a} \, .
\eeq

Under a rescaling $\ell^a \to f \ell^a$ of the null normal,
keeping the cross-sections $\calS_v$ fixed but relabelled
to $\calS_{v'}$ with $\ud v' = f^{-1} \ud v$, so that
$n_a \to f^{-1} n_a$, we see from  Eqs.~\eqref{rescaled_omega} and \eqref{rescaled_kappa} that
the \Hajicek{} 1-form changes as
\beq \label{rescale_Hajicek}
  \Omega_a \to \Omega_a + (\ud\ln f)_b \, q^b_{\ \, a} \, .
\eeq

\section{Non-expanding horizons}\label{sec:NEH}

The event horizon of an isolated, stationary black hole is a Killing horizon; see e.g. \cite{Wal}. Non-expanding horizons generalize the concept of Killing horizon for ``stationary'' black holes embedded in a non necessarily stationary gravitational environment \cite{Ha3.73,Ha3.74,As.al2.00,As.al.02,GoJa.06,AsKr.25}. In this section we recall the definition of a non-expanding horizon and review its main properties (Sec.~\ref{subsec:def_NEH}), including the induced geometry and the characterization of its shape an angular momentum structure (Sec.~\ref{subsec:structure}), out of one of the five Weyl curvature scalars (Sec.~\ref{subsec:Weyl_scalars}).

\subsection{Definition and main properties} \label{subsec:def_NEH}

A null hypersurface $\calH$ is a \emph{non-expanding horizon (NEH)} if, and only if, (i) $\calH$ has the topology \eqref{H_topology}: $\calH \sim \mathbb{R} \times \mathbb{S}^2$,
(ii) the expansion $\theta_{(\bm{\ell})}$ of any null normal $\ell^a$ to $\calH$, as defined by Eq.~\eqref{def_expansion}, vanishes
and (iii) the Ricci tensor of $g_{ab}$ admits any null normal to $\calH$ as an eigenvector: $R^a_{\ \, b} \ell^b \eqEH \alpha \ell^a$ for some (possibly zero) scalar field $\alpha$ on $\calH$
\cite{Haj.73,As.al2.00,As.al.02,AsKr.25}.  Given the scaling law \eqref{theta_rescale_ell}, if $\theta_{(\bm{\ell})}=0$ for some null normal, it remains zero for any other null normal.
From now on, $\calH$ will denote a NEH.
In general relativity (i.e. assuming the Einstein equation), condition (iii) is implied by the null dominant energy condition. In our work, it is automatically fulfilled since we
are considering vacuum spacetimes in the vicinity of $\calH$. Thanks to (ii) and (iii),
the Raychaudhuri equation implies the vanishing of the shear tensor $\sigma_{ab}$ of $\ell^a$ \cite{As.al2.00,As.al.02}. It follows then from Eq.~\eqref{def_shear_tensor} that the second fundamental form $\Theta_{ab}$ associated to any null normal
$\ell^a$ vanishes identically: $\Theta_{ab}=0$. Equation~\eqref{Theta_Lie_l_q} implies then the invariance of the first fundamental form by Lie transport along the null normal:
\beq \label{Lie_ell_q_zero_NEH}
  \Lie_{\bm{\ell}} h_{ab} = 0 \, .
\eeq
If $h_{ab}$ were a genuine (i.e. non-degenerate) metric, this would mean that $\ell^a$ is a Killing vector of $h_{ab}$. In addition,
it follows from $\theta_{(\bm{\ell})}=0$ and Eq.~\eqref{Lie_l_Heps}
that the area 2-form of a NEH is invariant as well by Lie transport along the null normal:
\beq \label{Lie_ell_Neps}
  \Lie_{\bm{\ell}} \, {}^\calH\!\varepsilon_{ab} = 0 \, .
\eeq

\begin{figure}[t]
    \begin{center}
        \includegraphics[height=0.28\textheight]{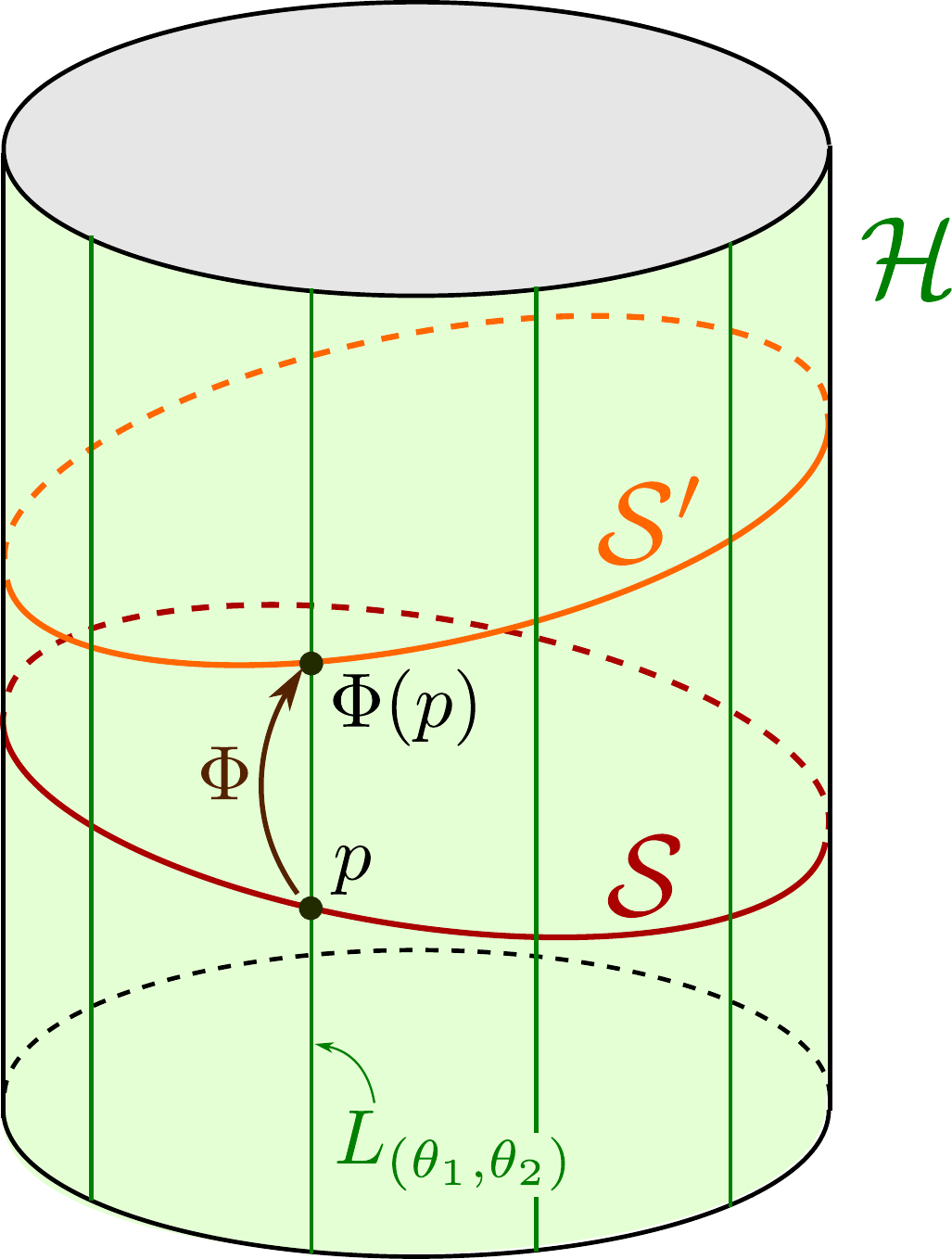}
        \caption{Isometry $\Phi$ between two arbitrary cross-sections $\calS$ and $\calS'$ of a NEH $\calH$, each being endowed with the metric induced by the spacetime metric $g_{ab}$.
        The green vertical lines depict some null geodesic generators $L_{(\theta_1,\theta_2)}$ of $\calH$.}
        \label{fig:neh_isometry}
    \end{center}
\end{figure}

A consequence of \eqref{Lie_ell_q_zero_NEH} is that all the cross-sections $\calS$ of $\calH$, equipped with the Riemannian metric $q_{ab}$ induced by $h_{ab}$ (or equivalently by $g_{ab}$) (cf. Sec.~\ref{s:cross_section_slicings}), are isometric.
The isometry between two arbitrary cross-sections $\calS$ and
$\calS'$ of $\calH$ is the diffeomorphism $\Phi: \calS \to \calS'$ defined by $\Phi(p)$ lying on the same null geodesic generator of $\calH$ as $p$ for every $p\in\calS$
(cf. Fig.~\ref{fig:neh_isometry}). Given that each generator of $\calH$ intersects a given cross-section exactly once, this clearly defines a bijective map. To show that $\Phi$ is an isometry, let us introduce a 1-parameter family
$(\calS_v)_{v\in[0,1]}$ of cross-sections of $\calH$ such that $\calS_0 = \calS$ and $\calS_1 = \calS'$. If $\calS\cap\calS'\neq\varnothing$ such a family does not constitute a slicing, as defined in Sec.~\ref{s:cross_section_slicings}, but this is not an issue here. Along the portion of each geodesic generator $L_{(\theta_1,\theta_2)}$ between $\calS_0$ and $\calS_1$, $v$ can be considered as a parameter of $L_{(\theta_1,\theta_2)}$. Let $\ell^a$ be the tangent vector
corresponding to this parametrization [Eq.~\eqref{ell_pregeodesic}], with
possibly $\ell^a = 0$ at points where $\calS$ and $\calS'$ intersect.
Let us consider the family of flow maps $\Phi_v: \calS_0 \to \calS_v$, such that
$\Phi_v(p)$ is the point of parameter $v$ along the generator  $L_{(\theta_1,\theta_2)}$ through $p$. By definition of a Lie derivative, we have
$\Lie_{\bm{\ell}} h_{ab} = \lim_{v\to 0} (\Phi^*_v h_{ab} - h_{ab})/v$, where
$\Phi^*_v h_{ab}$ stands for the pullback of $h_{ab}$ by $\Phi_v$. The property
\eqref{Lie_ell_q_zero_NEH} translates then to $\Phi^*_v h_{ab} = h_{ab}$.
In particular, for $v=1$, $\Phi_1 = \Phi$ and we get $\Phi^* h_{ab} = h_{ab}$. Given that the metric
$q_{ab}$ (resp. $q'_{ab}$) of $\calS$ (resp. $\calS'$) is that induced by $h_{ab}$, there comes
$\Phi^*  q'_{ab} = q_{ab}$, i.e., $\Phi$ is an isometry
from
$(\calS, q_{ab})$ to $(\calS', q'_{ab})$.

Since all the cross-sections of a NEH are isometric, they share the same area
\beq \label{def_area_NEH}
  A \equiv \oint_{\calS} \bm{\varepsilon} =  \oint_{\calS} \ud S
  =  \oint_{\calS}  \sqrt{q} \, \ud x^1 \ud x^2 \, ,
\eeq
where $\bm{\varepsilon}$ is the area 2-form of the metric $q_{ab}$ on the cross-section $\calS$ (recall Sec.~\ref{s:cross_section_slicings})
and $\ud S = \sqrt{q} \, \ud x^1 \ud x^2$ is the area element of $\calS$, expressed in terms of a
coordinate system $(x^1, x^2)$ covering\footnote{For instance, one may choose $(x^1,x^2)$ to coincide with the parameters $(\theta_1,\theta_2)$ labelling the null generators $L_{(\theta_1,\theta_2)}$ of $\calH$.} $\calS$ and the determinant $q$
of $q_{ab}$ with respect to  $(x^1, x^2)$.
Equivalently, $\bm{\varepsilon}  = \sqrt{q}\,  \bm{\ud} x^1 \wedge \bm{\ud} x^2$.
Being independent of $\calS$, the quantity $A$ is called the \emph{area of the NEH} $\calH$.
From $A$, one defines the \emph{areal radius of} $\calH$ by
\beq \label{def_areal_radius}
    R \equiv \sqrt{\frac{A}{4\pi}} \, .
\eeq

Another important consequence of $\Theta_{ab}=0$ is that the spacetime connection $\nabla_a$ induces a torsion-free affine connection $\mathcal{D}_a$ on $\calH$, by setting
$u^a \mathcal{D}_a v^b \equiv u^a \nabla_a v^b$
for any pair of vectors $(u^a,v^a)$ tangent to $\calH$. Indeed,
since $\ell_b v^b = 0$ one has
$\ell_b u^a \nabla_a v^b = - v_b u^a \nabla_a \ell^b = -v_b \chi^b_{\ \, a} u^a = - h_{ab} v^a \chi^b_{\ \, c} u^c = -\Theta_{ab} v^a u^b$,
where use has been made of Eq.~\eqref{def_2nd_fund_form}. Hence
$\Theta_{ab} = 0$ implies $\ell_b u^a \nabla_a v^b = 0$, which shows that $u^a \nabla_a v^b$ is tangent to $\calH$, i.e., that $\mathcal{D}_a$ is well defined.
Moreover this connection is compatible with $h_{ab}$ and ${}^\calH\!\varepsilon_{ab}$, as\footnote{Note, however, that $\mathcal{D}_a$ is not a Levi-Civita connection associated to the induced ``metric'' $h_{ab}$ on $\calH$, for the latter is degenerate. In particular, $\mathcal{D}_a$ cannot be reconstructed from the knowledge of $h_{ab}$ alone.}
\beq
  \mathcal{D}_c h_{ab} = 0 \quad \text{and} \quad \mathcal{D}_c {}^\calH\!\varepsilon_{ab} = 0 \, .
\eeq

For a NEH, the rotation 1-form $\omega_a$ introduced in Sec.~\ref{s:rotation_form} depends only on the null normal $\ell^a$
and not on the cross-section slicing $(\calS_v)_{v\in\mathbb{R}}$ compatible with $\ell^a$.
Indeed, a change of slicing $(\calS_v)_{v\in\mathbb{R}} \to (\calS'_{v'})_{v'\in\mathbb{R}}$ of the type \eqref{v_vp_H} leads to
a change of the transverse null vector of the type $n^a \to {n'}^a = n^a + w^a$, with $w^a$ tangent to $\calH$ and the coefficient
$+1$ in front of $n^a$ ensuring $\ell_a {n'}^a = -1$. Then, according to the definition~\eqref{omega_n_chi} of the rotation 1-form,
for any vector $u^a\in T_p\calH$, one has
$\omega'_a u^a = - n'_b \chi^b_{\ \, a} u^a = \omega_a u^a - w_b \chi^b_{\ \, a} u^a$,
with, thanks to Eqs.~\eqref{def_Weingarten} and \eqref{Theta_pullback_nab_l},
$w_b \chi^b_{\ \, a} u^a = w_b u^a \nabla_a \ell^b = \nabla_a \ell_b u^a w^b = \Theta_{ab} u^a w^b = 0$ since $\Theta_{ab}=0$.
Hence $\omega_a$ depends only on the null normal $\ell^a$,
through the scaling law \eqref{rescaled_omega}. The independence from the cross-section slicing is also made clear by the
following expression of the Weingarten map of $\calH$ relative to $\ell^a$:
\beq \label{Dell_omega_ell}
    \chi^a_{\ \, b} = \mathcal{D}_b \ell^a = \omega_b \ell^a \, .
\eeq
The first equality is a direct consequence of the definition \eqref{def_Weingarten} of $\chi^a_{\ \, b}$, while the second one follows from
setting $\Theta_{ab} = 0$ in Eq.~\eqref{def_2nd_fund_form}: one gets $h_{ac} \mathcal{D}_b \ell^c = 0$, which means that for any
vector $v^a$ tangent to $\calH$, the vector $v^b \mathcal{D}_b \ell^a$ lies along the degenerate direction of $h_{ab}$, i.e.
is collinear to $\ell^a$. There exists then a 1-form $\alpha_a$ on $\calH$ such that $v^b \mathcal{D}_b \ell^a = \alpha_b v^b \ell^a$,
hence $\mathcal{D}_b \ell^a = \alpha_b \ell^a$. Equation~\eqref{omega_n_chi} leads then to
$\omega_a = (\ud v)_b \alpha_a \ell^b = \alpha_a$ since $\ell^b (\ud v)_b = 1$ [Eq.~\eqref{l_velocity_v}], which completes the proof
of \eqref{Dell_omega_ell}.

By means of the Cartan identity and Eq.~\eqref{omega_ell_kappa}, we get $\Lie_{\bm{\ell}} \, \omega_a = \ell^b (\ud\omega)_{ba}
+ (\ud \kappa_{(\bm{\ell})})_a$. But thanks to \eqref{ImPsi2_NEH} below and the property ${}^\calH\varepsilon_{ab} \ell^b = 0$ [Eq.~\eqref{epsilon_H_ell_zero}], one readily obtains \cite{As.al2.00,As.al.02}
\beq\label{tutu}
    \Lie_{\bm{\ell}} \, \omega_a = (\ud \kappa_{(\bm{\ell})})_a \, .
\eeq
Hence the rotation 1-form $\omega_a$ is Lie-transported along the null normal $\ell^a$ to a NEH $\calH$ if, and only if, $\kappa_{(\bm{\ell})}$ is constant.
Applied to a Killing horizon, which is a special case of NEH, this property yields the zeroth law of black hole mechanics. Indeed, it suffices to choose $\ell^a$
to be the Killing vector $\xi^a$ generating $\calH$ and normalized by $\xi_a\xi^a = -1$ at the asymptotically flat end of $(\calM,g_{ab})$; $\kappa_{(\bm{\ell})}$ is then the so-called \emph{surface gravity} and one has necessarily $\Lie_{\bm{\ell}} \, \omega_a = 0$, so that
Eq.~\eqref{tutu} implies $\kappa_{(\bm{\ell})} = \mathrm{const}$ \cite{Da.82b}.

\subsection{Weyl curvature scalars on a NEH}\label{subsec:Weyl_scalars}

Let us consider a Newman-Penrose null tetrad $(\ell^a,n^a,m^a,\bar m^a)$ such that $\ell^a$ is normal to $\calH$. By definition, $n^a$ is a null vector fulfilling $\ell_a n^a = -1$
and $m^a$ is a complex null vector orthogonal to $\ell^a$ and $n^a$ fulfilling $m_a m^a = 0$ and $m_a \bar{m}^a = 1$, where $\bar{m}^a$ is the complex conjugate of $m^a$.
Note that the two vectors $\ell^a$ and $n^a$ discussed in Sec.~\ref{s:cross_section_slicings} can be considered as the first two legs of such a null tetrad; cf. Eq.~\eqref{ln_minus_one} and Fig.~\ref{fig:null_hypersurface}.
The ten independent components of the Weyl tensor $C_{abcd}$ are neatly encoded into the five complex-valued curvature scalars
\begin{subequations}\label{eq:Weyl sc}
    \begin{align}
        \Psi_0 &\equiv C_{abcd} \ell^a m^b \ell^c m^d \, , \\
        \Psi_1 &\equiv C_{abcd} \ell^a m^b \ell^c n^d \, , \\
        \Psi_2 &\equiv C_{abcd} \ell^a m^b \bar m^c n^d \, , \label{Psi2} \\
        \Psi_3 &\equiv C_{abcd} \ell^a n^b \bar m^c n^d \, , \\
        \Psi_4 &\equiv C_{abcd} n^a \bar m^b n^c \bar m^d \, .
    \end{align}
\end{subequations}
Generally, $\Psi_0$ and $\Psi_4$ are associated with ingoing and outgoing transverse gravitational radiation, $\Psi_1$ and $\Psi_3$ with ingoing and outgoing longitudinal radiation, and $\Psi_2$ \mbox{with a Coulomb} field; see for instance Ref.~\cite{Sz.65}.

It can be shown that the Weyl scalars $\Psi_0$ and $\Psi_1$ vanish identically on a NEH $\calH$ \cite{As.al2.00}. It then follows from the transformation laws of the Weyl scalars under a change of null tetrad while keeping the direction of $\ell^a$ fixed (the so-called rotations of class I and III \cite{Cha}) that $\Psi_2$ is
\textit{tetrad-invariant} on $\calH$.

\subsection{Shape and angular momentum structure}\label{subsec:structure}

The frame-invariance on $\calH$ of the Weyl scalar $\Psi_2$ implies that it ``encodes" physically meaningful (Coulomb-type) curvature information about the NEH. More precisely,
the geometry of $\calH$ is encoded into the real and imaginary parts of $\Psi_2$, according to \cite{Haj.73,As.al.02,As.al.04,AsBa.19}
\begin{subequations}\label{Re-Im}
    \begin{align}
        & \text{Re}\,\Psi_2 \eqNEH - \tfrac{1}{4} \, \mathcal{R} \, , \label{RePsi2_NEH} \\
        & (\text{Im\,}\Psi_2) \, {}^\calH\!\varepsilon_{ab} \eqNEH \tfrac{1}{2} \, (\ud \omega)_{ab} \, ,  \label{ImPsi2_NEH}
    \end{align}
\end{subequations}
where $\mathcal{R}$ is the scalar curvature of an arbitrary 2-sphere cross-section $\calS$ of $\calH$ (i.e., $\mathcal{R}$ is the Ricci scalar of the Riemannian metric $q_{ab}$ induced by $g_{ab}$ on $\calS$) and ${}^\calH\!\varepsilon_{ab}$ is the area 2-form of $\calH$ [cf. Eq.~\eqref{def_area_form_H}]. Since all cross-sections are isometric (cf. Sec.~\ref{subsec:def_NEH}), the value of $\mathcal{R}$ at a given point $p\in\calH$ does not depend on the specific cross-section $\calS$ through $p$.
Accordingly, $\mathcal{R}$ can be considered as a scalar field on $\calH$, so that Eq.~\eqref{RePsi2_NEH} makes sense. As for the second equation, we note
that the exterior derivative $(\mathrm{d}\omega)_{ab}$ of the rotation 1-form $\omega_a$ is independent of the choice of the
null normal $\ell^a$, thanks to the behavior
\eqref{rescaled_omega} under a generic rescaling $\ell^a \to f \ell^a$.\footnote{This is analogous to the invariance of the Maxwell tensor $F_{ab} = (\ud A)_{ab}$ under a gauge transformation $A_a \to A_a + (\ud\Phi)_a$ of the electromagnetic potential $A_a$ generated by a scalar potential $\Phi$.} This is in agreement with $\Psi_2$ and
${}^\calH\!\varepsilon_{ab}$ in \eqref{ImPsi2_NEH} being independent
of $\ell^a$ (recall that $\Psi_2$ on a NEH is tetrad independent). Moreover, for quasilocal horizons, among which NEHs, it is known that $\omega_a$
is involved in the definition of angular momentum (hence the name \emph{rotation 1-form}); cf. Refs.~\cite{Da.82b,As.al.01,Go.05,Ha.06,JaGo.11} and Eq.~\eqref{def_angu_mom} below.
In view of Eqs.~\eqref{Re-Im}, we conclude that the real and imaginary parts of the Weyl curvature scalar $\Psi_2$ neatly encode the shape and angular momentum structure of $\calH$, respectively.

Equation~\eqref{ImPsi2_NEH} is an identity between two 2-forms on the 3-dimensional manifold $\calH$.
We can recast it as an identity between two 2-forms on a 2-dimensional manifold by pulling it back
on a generic cross-section $\calS$ of $\calH$ via the inclusion map $\jmath : \calS \hookrightarrow \calH$.
Thanks to Eq.~\eqref{Seps_pullback_Heps}, one has $\jmath^* \, {}^\calH\!\varepsilon_{ab} = \varepsilon_{ab}$.
For evaluating $\jmath^* (\ud \omega)_{ab}$, let us consider that $\calS$ is the element $v=0$ of a cross-section
slicing $(\calS_v)_{v\in\mathbb{R}}$ of $\calH$ and let $\Omega_a$ be the associated \Hajicek{} 1-form
(cf. Sec.~\ref{s:rotation_form}). Thanks to relation~\eqref{Omega_omega_kappa_n} between $\omega_a$ and
$\Omega_a$, one has $\bm{\ud}\bm{\omega} = \bm{\ud}\bm{\Omega} + \bm{\ud} \kappa_{(\bm{\ell})} \wedge \bm{\ud} v$.
Since $\jmath^* \bm{\ud} v = 0$ (for $v$ is constant over $\calS$), there comes
$\jmath^* \bm{\ud}\bm{\omega} = \jmath^* \bm{\ud}\bm{\Omega} = \bm{\ud} \jmath^* \bm{\Omega}$,
where use has been made of the commuting property of pullbacks and exterior derivatives
to write the second equality. Accordingly, the pullback of Eq.~\eqref{ImPsi2_NEH} onto $\calS$ yields
\beq \label{ImPsi2_eps_dOmega}
    (\text{Im\,}\Psi_2) \, \varepsilon_{ab}  \stackrel{\calS}{=}
  \tfrac{1}{2} \, (\ud \Omega)_{ab} \, ,
\eeq
where we use $\Omega_a$ to also denote the 1-form $\jmath^* \Omega_a$ on $\calS$. This slight abuse of notation is permissible since the \Hajicek{} 1-form $\Omega_a$
``essentially lives'' in the cross-sections, as shown by Eq.~\eqref{def_Hajicek}.

Let $\ell^a$ be any null normal to $\calH$.
Because all cross-sections of $\calH$ are isometric by transport along the null generators of $\calH$ (Sec.~\ref{subsec:def_NEH}), we have
$\Lie_{\bm{\ell}} \mathcal{R} = 0$ and Eq.~\eqref{RePsi2_NEH} yields $\text{Re}\,\Lie_{\bm{\ell}}\Psi_2 = 0$. On the other hand, \eqref{ImPsi2_NEH} and \eqref{Lie_ell_Neps} lead to
$2\Lie_{\bm{\ell}}(\text{Im}\,\Psi_2)\,  {}^\calH\!\bm{\varepsilon} = \Lie_{\bm{\ell}} \bm{\ud} \bm{\omega} =
\bm{\ud} \Lie_{\bm{\ell}} \bm{\omega} =
\bm{\ud} \bm{\ud} \kappa_{(\bm{\ell})} = 0$, where we have used successively the commutation of Lie and exterior derivatives, Eq.~\eqref{tutu} and
the nilpotence of the
exterior derivative. Hence, we get $\text{Im}\,\Lie_{\bm{\ell}}\Psi_2 = 0$. Combining with $\text{Re}\,\Lie_{\bm{\ell}}\Psi_2 = 0$, we conclude that
\beq\label{LiePsi2}
    \Lie_{\bm{\ell}} \Psi_2 \eqNEH 0 \, .
\eeq
Hence, on a NEH, not only the Weyl scalar $\Psi_2$ is frame-invariant, but it is also constant along the NEH's generators.

\section{Multipole moments of an axisymmetric NEH}\label{sec:axi_multipoles}

In Ref.~\cite{As.al.04}, Ashtekar, Engle, Pawlowski and Van Den Broeck have defined the geometrical multipole moments $I^\text{axi}_\ell$ and $L^\text{axi}_\ell$ of an axisymmetric isolated horizon.  An
\emph{isolated horizon} is a pair $(\calH, [\ell^a])$, where $\calH$ is a NEH and $[\ell^a]$ is an equivalence class of null normals to $\calH$, defined by ${\ell'}^a \sim \ell^a$ if, and only if, ${\ell'}^a = c \ell^a$ with $c$ constant over $\calH$, such that
the affine connection $\mathcal{D}_a$ (cf. Sec.~\ref{subsec:def_NEH}) is invariant by transport along the null normals belonging to $[\ell^a]$, i.e. $\mathcal{D}_a$ commutes with the Lie derivative along any
representative $\ell^a$ of the equivalence class:
$[\Lie_{\bm{\ell}}, \mathcal{D}_a] = 0$ \cite{As.al2.00,As.al.02}.
This condition supplements the NEH property $\Lie_{\bm{\ell}} h_{ab} = 0$ [Eq.~\eqref{Lie_ell_q_zero_NEH}] to make the full geometry $(h_{ab},\mathcal{D}_a)$ of $\calH$ be ``time independent''
(see Ref.~\cite{AsKr.25} for a review).
However it plays no role in the definition of the multipole moments $I^\text{axi}_\ell$ and $L^\text{axi}_\ell$ by formula \eqref{def}, since the latter relies only on the     frame-independent quantity $\Psi_2$, the metric $q_{ab}$ of an arbitrary cross-section of $\calH$ and the axisymmetric character of $q_{ab}$. The isolated horizon property is required only to reconstruct the whole horizon geometry
from the knowledge of $I^\text{axi}_\ell$ and $L^\text{axi}_\ell$ \cite{As.al.04}, as well as to define the \emph{source} multipole moments $M_\ell$ and $S_\ell$ to be discussed in Sec.~\ref{subsec:mass-current}.
We shall therefore provide the definition of the geometric multipole moments
following the prescription of Ref.~\cite{As.al.04}, but without requiring the isolated horizon structure atop of the NEH one. This has the advantage to put the definition on the same footing as that of multipole moments for a generic (non-axisymmetric) NEH to be presented in Sec.~\ref{sec:generic_multipoles}.
We start by constructing a privileged unit round metric on cross-sections of an axisymmetric NEH (Sec.~\ref{subsec:axi}); then we provide the definition of horizon multipole moments and discuss some of their properties (Sec.~\ref{subsec:multipole_axi}), and the related notions of mass and current source multipoles (Sec.~\ref{subsec:mass-current}).

\subsection{Axisymmetric NEHs and their cross-section geometry}\label{subsec:axi}

Let us define a NEH $\calH$ to be \emph{axisymmetric} if, and only if, there exists a SO(2) group action on $\calH$, of generating vector field $\eta^a$, such that (i) $\eta^a$ vanishes on exactly two null geodesic generators of $\calH$, named the \emph{North} and \emph{South pole generators}, (ii) apart from these two generators, the orbits of $\eta^a$ are closed spacelike curves and their parameter length with respect to $\eta^a$ is $2\pi$, (iii) there exists a null normal $\ell^a$ to $\calH$ that commutes with $\eta^a$: $[\bm{\ell},  \bm{\eta}]^a = 0$, and (iv) $\calH$'s intrinsic ``metric'' $h_{ab}$ and extrinsic curvature, represented by the Weingarten map $\chi^a_{\ \, b}$ with respect to $\ell^a$, are preserved by the group action: $\Lie_{\bm{\eta}} h_{ab} = 0$ and $\Lie_{\bm{\eta}} \chi^a_{\ \, b} = 0$.

The null normal $\ell^a$ obeying (iii) and (iv) (recall that $\chi^a_{\ \, b}$ depends on $\ell^a$; cf. Sec.~\ref{s:basic_geom_hor})
is far from unique: (iii) and (iv) are fulfilled by any other normal ${\ell'}^a = f \ell^a$, with $f$ a positive scalar field on $\calH$
respecting the axisymmetry, i.e. obeying $\eta^a (\ud f)_a = 0$. Indeed, one has $[\bm{\ell}', \bm{\eta}]^a = f [\bm{\ell},  \bm{\eta}]^a - \eta^b (\ud f)_b \ell^a = 0 - 0 = 0$
and the Weingarten map ${\chi'}^a_{\ \, b}$ associated to ${\ell'}^a$ obeys $\Lie_{\bm{\eta}} {\chi'}^a_{\ \, b} = 0$
thanks to the transformation law \eqref{rescaled_chi} and  $\eta^a (\ud f)_a = 0$.
We shall denote by $\{\ell^a\}_{\rm axi}$ the set of all such normals.
Note that the second requirement in (iv) is equivalent to $\Lie_{\bm{\eta}} \omega_a = 0$, where $\omega_a$ is the
rotation 1-form associated to $\ell^a$. Indeed, from the NEH identity $\chi^a_{\ \, b} = \ell^a \omega_b$
[Eq.~\eqref{Dell_omega_ell}], we have
$\Lie_{\bm{\eta}} \chi^a_{\ \, b} = \Lie_{\bm{\eta}} \ell^a \, \omega_b + \ell^a \Lie_{\bm{\eta}} \omega_b = \ell^a \Lie_{\bm{\eta}} \omega_b$ since $\Lie_{\bm{\eta}} \ell^a = [\bm{\eta}, \bm{\ell}]^a = 0$ by (iii).
Accordingly, (iv) can be restated as
\beq \label{Lie_eta_h_omega}
  \Lie_{\bm{\eta}} h_{ab} = 0 \quad\mbox{and}\quad \Lie_{\bm{\eta}} \omega_a = 0 \, .
\eeq

Let $\calS$ be a cross-section of an axisymmetric NEH $\calH$, such that $\eta^a$ is tangent to $\calS$.
Since $\Lie_{\bm{\eta}} h_{ab} = 0$, the induced metric $q_{ab} = \jmath^* h_{ab}$ on $\calS$ obeys $\Lie_{\bm{\eta}} q_{ab} = 0$, i.e. $\eta^a$ is a Killing vector of $q_{ab}$,
or, in other words, $(\calS,q_{ab})$ is an axisymmetric Riemannian manifold.\footnote{If $\calS'$ is a cross-section of $\calH$ to which $\eta^a$ is not tangent everywhere, it is not globally invariant
by the SO(2) action on $\calH$. It is nevertheless
axisymmetric, given that all cross-sections of a NEH are isometric (cf. Sec.~\ref{subsec:def_NEH}). To show it explicitly, let us consider the SO(2) group action on $\calS'$
defined by $\Psi'_{\phi} \equiv \Phi\circ \Psi_\phi \circ \Phi^{-1}$, where $\Psi_\phi$
is the rotation of angle $\phi$ acting on the axisymmetric cross-section $\calS$ considered above
and $\Phi$ is the isometry $\calS \to \calS'$ along the null generators of $\calH$ considered in Sec.~\ref{subsec:def_NEH}.
Since $\Phi$ and $\Psi_\phi$ are isometries, $\Psi'_\phi$ is clearly an isometry of $(\calS', q'_{ab})$. The corresponding Killing vector $\eta'_a$ is then nothing but the pushforward of $\eta^a$ (considered as
a vector field tangent to $\calS$) by $\Phi$.}
Based on the axisymmetric structure, one may construct a fiducial unit round metric $\mathring{q}^{\rm axi}_{ab}$ on $\calS$ as follows \cite{As.al.04}. First, let us consider the Hodge dual $\star\eta_a$ of the 1-form $\eta_a \equiv q_{ab} \eta^b$ on $\calS$, i.e. the 1-form defined by
$\star\eta_a \equiv \eta^b \varepsilon_{ba}$, where $\varepsilon_{ab}$ is the area 2-form of $q_{ab}$ (cf. Sec.~\ref{s:cross_section_slicings}).
One has necessarily $\Lie_{\bm{\eta}} \varepsilon_{ab} = 0$, so that
the Cartan identity $\Lie_{\bm{\eta}} \bm{\varepsilon} = \bm{\eta}\cdot \bm{\ud} \bm{\varepsilon} + \bm{\ud} \star\!\bm{\eta}$ and $\bm{\ud}\bm{\varepsilon}=0$ (vanishing of any 3-form
on a 2-manifold)
imply $\bm{\ud}\star\!\bm{\eta} = 0$, i.e. $\star\bm{\eta}$ is closed.
Now, on a simply connected manifold, such as $\calS\sim \mathbb{S}^2$, any closed 1-form is exact: $\star\bm{\eta} = \bm{\ud} \mu$ for some  scalar field $\mu$ on $\calS$.
A priori $\mu$ is determined up to a constant; one can uniquely fix the latter by demanding that the integral of $\mu$ over $\calS$ vanishes. Rescaling $\mu$ by $\calH$'s areal radius
$R$ [Eq.~\eqref{def_areal_radius}], which is constant, we conclude that there exists a unique  scalar field $\zeta$ on $\calS$ such that
\beq \label{theta_prime_from_axi}
   (\ud \zeta)_a = \frac{1}{R^2} \, \eta^b \varepsilon_{ba}  \quad \mbox{and} \quad
   \oint_{\calS} \zeta \bm{\varepsilon} = 0 \, .
\eeq
Moreover, $\eta^a  (\ud \zeta)_a = R^{-2} \varepsilon_{ba} \eta^b \eta^a = 0$, which implies that $\zeta$ is constant along the orbits of the Killing vector $\eta^a$.
One can then introduce a parameter $\phi$ along these orbits such that $x^{A'} = (\zeta,\phi)$ is a coordinate system on $\calS$ with $\zeta\in [-1,1]$, $\zeta=-1$ (resp. $+1$) at the South (resp. North)
pole, $\phi\in[0,2\pi)$, $\eta^a = \partial_\phi^a$ and the metric $q_{ab}$ takes the form (see Sec.~2.2 of Ref.~\cite{As.al.04} for details)
\beq \label{q_in_axisym_coord}
   q_{A'B'} \, \ud x^{A'} \ud x^{B'} = R^2 \left( f^{-1} \, \ud \zeta^2 + f \, \ud\phi^2 \right), \quad \mbox{where} \quad
   f = f(\zeta) \equiv q_{ab} \eta^a \eta^b / R^2 .
\eeq
The coordinates $x^{A'} = (\zeta,\phi)$ defined above are unique, up to a constant shift in $\phi$. Let us introduce
the coordinates $x^A=(\theta,\phi)$ with $\theta\in [0,\pi]$
defined via $ \cos\theta \equiv \zeta$
and define from them the metric $\mathring{q}^{\rm axi}_{ab}$ on $\calS$ by
\beq \label{round_metric_axi}
    \mathring{q}^{\rm axi}_{AB} \, \ud x^A \ud x^B \equiv \ud \theta^2 + \sin^2\theta \, \ud \phi^2 .
\eeq
Clearly, $\mathring{q}^{\rm axi}_{ab}$ is a Riemannian metric of constant scalar curvature equal to 2, i.e. a unit round metric. Moreover, by construction, this metric is unique
(since the coordinates $(\theta,\phi)$ are uniquely defined---up to a shift in $\phi$) and follows from the axisymmetry of $\calS$, via Eq.~\eqref{theta_prime_from_axi}, which defines $\zeta$ in terms of
the rotational Killing vector $\eta^a$. The metric $\mathring{q}^{\rm axi}_{ab}$ admits the same Killing vector $\eta^a=\partial_\phi^a$ as the physical metric $q_{ab}$.
Expressed in terms of the coordinates $(\zeta,\phi)$ instead of $(\theta,\phi)$, it reads
$\mathring{q}^{\rm axi}_{A'B'} \, \ud x^{A'} \ud x^{B'} = (1 - \zeta^2)^{-1} \, \ud \zeta^2 +  (1 - \zeta^2) \, \ud\phi^2$,
which takes the same form as in Eq.~\eqref{q_in_axisym_coord} with $R$ replaced by $1$ and $f(\zeta)$ replaced by $1 - \zeta^2$.
Moreover, we note from Eq.~\eqref{q_in_axisym_coord} that $\det(q_{A'B'}) = R^4$, so that $\mathring{q}^{\rm axi}_{ab}$ shares the same area 2-form
as the physical metric $q_{ab}$, up to a constant factor $R^2$:
\beq \label{eps_R2_eps_axi}
    \varepsilon_{ab} = R^2 \, \mathring{\varepsilon}^{\rm axi}_{ab} \, .
\eeq

\subsection{Horizon multipole moments}\label{subsec:multipole_axi}

Given the unit round metric $\mathring{q}^{\rm axi}_{ab}$ introduced above on the cross-section $\calS$,
we follow Ref.~\cite{As.al.04} to define the \textit{shape} and \textit{current} multipole moments $I_{\ell}^{\rm axi}$ and $L_{\ell}^{\rm axi}$
of the axisymmetric NEH $\calH$ from the Weyl scalar $\Psi_2$ (cf. Sec.~\ref{subsec:structure}) by
\beq\label{I-L_axi}
    I_{\ell}^{\rm axi} + \ui L_{\ell}^{\rm axi} \equiv - \oint_\calS \Psi_2 \, \mathring{Y}_{\ell,0}^{\rm axi} \, \ud S =
    - R^2 \oint_\calS \Psi_2 \, \mathring{Y}_{\ell,0}^{\rm axi} \, \ud \mathring{S}^{\rm axi} \, .
\eeq
Here, $\ud S$ (resp. $\ud \mathring{S}^{\rm axi}$) is the area element of $\calS$ associated to the physical metric $q_{ab}$
(resp. the unit round metric  $\mathring{q}^{\rm axi}_{ab}$)
[recall Eq.~\eqref{def_area_NEH}], while $\mathring{Y}_{\ell,0}^{\rm axi}$ stands for the $m=0$ spherical harmonic
of degree $\ell$ with respect to the unit round metric $\mathring{q}^{\rm axi}_{ab}$; hence $\mathring{Y}_{\ell,0}^{\rm axi}$ is the $m=0$ eigenfunction of the Laplace operator of $\mathring{q}^{\rm axi}_{ab}$ corresponding to the eigenvalue $-\ell(\ell+1)$.
In terms of the coordinates $(\theta,\phi)$ in which $\mathring{q}^{\rm axi}_{ab}$ takes the canonical form \eqref{round_metric_axi}, we have
\beq \label{def_mathringYlm_axi}
    \mathring{Y}_{\ell,0}^{\rm axi} = Y_{\ell,0}(\theta,\phi) \, ,
\eeq
where $Y_{\ell,0}$ are the ordinary spherical harmonic functions.
The second equality in Eq.~\eqref{I-L_axi} holds because $\ud S = R^2 \, \ud \mathring{S}^{\rm axi}$, as a consequence of \eqref{eps_R2_eps_axi}.
Notice that the multipole moments $I_{\ell}^{\rm axi}$ and $L_{\ell}^{\rm axi}$ are dimensionless, given that $\Psi_2$ has the dimension of an inverse squared length and $\ud S$ that of a squared length.

Since all cross-sections of $\calH$ are isometric (Sec.~\ref{subsec:def_NEH}) and $\Psi_2$ is invariant by the isometry map along the null generators, thanks to Eq.~\eqref{LiePsi2}, the definition
\eqref{I-L_axi} of the multipole moments is clearly independent of the choice of the cross-section $\calS$ of $\calH$.

Thanks to expressions \eqref{RePsi2_NEH} and \eqref{ImPsi2_eps_dOmega} for, respectively, $\text{Re\,}\Psi_2$ and $\text{Im\,}\Psi_2$, the definition \eqref{I-L_axi} is equivalent
to
\beq \label{I-L_axi_R_Omega}
        I_{\ell}^{\rm axi} \equiv \frac{1}{4}  \oint_\calS \mathcal{R} \, \mathring{Y}_{\ell,0}^{\rm axi} \, \ud S
        \qquad\mbox{and}\qquad
        L_{\ell}^{\rm axi} \equiv -\frac{1}{2}  \oint_\calS   \mathring{Y}_{\ell,0}^{\rm axi} \, \bm{\ud}\bm{\Omega} \, .
\eeq
For $\ell=0$, $\mathring{Y}_{\ell,0}^{\rm axi} = Y_{0,0}(\theta,\phi) = 1/(2\sqrt{\pi})$ and Eq.~\eqref{I-L_axi_R_Omega} yields
\beq \label{monopoles_GB_Stokes}
   I_0^{\rm axi}  = \frac{1}{8\sqrt{\pi}} \underbrace{\oint_\calS \mathcal{R} \, \ud S}_{8\pi} = \sqrt{\pi}
        \qquad\mbox{and}\qquad
   L_0^{\rm axi} =
   - \frac{1}{4\sqrt{\pi}}
    \underbrace{\oint_\calS \bm{\ud} \bm{\Omega}}_{0}  = 0 \, ,
\eeq
where use has been made of the Gauss-Bonnet theorem for the first integral and of Stokes' theorem on a manifold without boundary for the second one. Hence we conclude that any NEH has a shape monopole $I_0^{\rm axi}$ equal to $\sqrt{\pi}$ and a vanishing current monopole $L_0^{\rm axi}$.
Moreover, thanks to the identity $\mathcal{R} = - f''(\zeta)/R^2$, one can show that the shape dipole $I_1^{\rm axi}$ always vanishes on an axisymmetric NEH
\cite{As.al.04}. We thus have the identities
\beq \label{universal_values_axi}
    I_0^{\rm axi} = \sqrt{\pi},\quad
    L_0^{\rm axi} = 0 \quad\mbox{and}\quad
    I_1^{\rm axi} = 0 \, .
\eeq

\subsection{Source multipole moments}\label{subsec:mass-current}

The physical interpretation of the dimensionless, \textit{geometrical} multipole moments \eqref{I-L_axi} of a given axisymmetric NEH $\calH$ is not straightforward. However, by introducing appropriate powers of the NEH mass $M_\calH$ and areal radius $R$, one can define \textit{physical source} multipole moments $M_{\ell}$ and $S_{\ell}$ of mass-type and current-type  that have the appropriate physical dimensions
\cite{As.al.04}.
The radius $R$ is defined unambigously from the NEH geometry by Eq.~\eqref{def_areal_radius}. On the other hand, defining the mass $M_\calH$
of an axisymmetric NEH requires some Hamiltonian framework. The latter has been developed in Ref.~\cite{As.al.01} by adding some extra structure on $\calH$,
namely a privileged constant-rescaling equivalence class $[\ell^a]$
such that $[\Lie_{\bm{\ell}}, \mathcal{D}_a] \ell^b= 0$ for any $\ell^a \in [\ell^a]$. The pair $(\calH,[\ell^a])$ is called a \emph{weakly isolated horizon (WIH)} \cite{As.al.01,AsKr.25}. Indeed the defining requirement is
weaker than that of isolated horizons, which is $[\Lie_{\bm{\ell}}, \mathcal{D}_a] v^b= 0$ for
any $v^b$ tangent to $\calH$ (cf. the beginning of Sec.~\ref{sec:axi_multipoles}). Thanks to the NEH identity
$\mathcal{D}_a \ell^b = \omega_a \ell^b$ [Eq.~\eqref{Dell_omega_ell}], the WIH condition is
equivalent to
\beq \label{def_WIH}
\Lie_{\bm{\ell}}\,\omega_a = 0
\eeq
for any $\ell^a \in [\ell^a]$.
Note that the rotation 1-form $\omega_a$ is unique for a given WIH structure: the scaling law \eqref{rescaled_omega} with $f = c = \mathrm{const}$ shows that $\omega_a$ does not depend on the representative $\ell^a$ in $[\ell^a]$. In view of Eq.~\eqref{tutu}, property~\eqref{def_WIH} implies that $\kappa_{(\bm{\ell})}$ is uniform over $\calH$, so that WIHs can
be seen as objects for which a kind of ``zeroth law'' holds, keeping in mind that the value of $\kappa_{(\bm{\ell})}$ depends on the choice of the representative $\ell^a$
in the class $[\ell^a]$, since ${\ell'}^a = c \ell^a$ implies $\kappa_{(\bm{\ell}')} = c \kappa_{(\bm{\ell})}$ (scaling law \eqref{rescaled_kappa} with $f=c=\mathrm{const}$).
For any NEH $\calH$, there exists infinitely many constant-rescaling classes $[\ell^a]$ such that $(\calH, [\ell^a])$ is a WIH \cite{As.al.02}.\footnote{On the contrary, it is not always possible to endow a NEH with an isolated horizon structure  \cite{As.al.02,AsKr.25}.}

Given an axisymmetric NEH $\calH$, let us select a WIH structure $(\calH, [\ell^a])$ such that $[\ell^a] \subset \{\ell^a\}_{\rm axi}$, i.e. any $\ell^a\in [\ell^a]$ commutes with the axisymmetry generator
$\eta^a$ (cf. Sec.~\ref{subsec:axi}). One may then build a Hamiltonian framework such that $\eta^a$ is the value on $\calH$
of a vector field $\varphi^a$ on $\calM$ that vanishes outside a compact neighborhood of $\calH$ and that generates a Hamiltonian vector field on the phase space \cite{As.al.01}.
The value of the corresponding Hamiltonian reduces to a surface integral on $\calH$, which one defines as the \emph{angular momentum of $\calH$} \cite{As.al.01,AsKr.25}:
\beq \label{def_angu_mom}
    J_{\calH} \equiv - \frac{1}{8\pi} \oint_{\calS} \omega_a \eta^a \, \ud S = - \frac{1}{8\pi} \oint_{\calS} \Omega_a \eta^a \, \ud S \, ,
\eeq
where $\calS$ is any cross-section of $\calH$ with $\eta^a$ as a tangent vector. The second integral involves the \Hajicek{} 1-form  $\Omega_a$ of $\ell^a$ with respect to $\calS$ and is equivalent to the first one
thanks to Eq.~\eqref{def_Hajicek} and $\eta^a$ being tangent to $\calS$.
The value of $J_{\calH}$ does not depend on the choice of the cross-section $\calS$. This follows from the Hamiltonian framework,
but it is easy to show it directly by considering another cross-section $\calS'$ to which $\eta^a$ is tangent.
$\calS'$ is necessarily connected to $\calS$ by an isometry flow along the null generators of $\calH$
(cf. Sec.~\ref{subsec:def_NEH})
generated by a vector field ${\ell'}^a \in \{\ell^a\}_{\rm axi}$. Writing ${\ell'}^a = f\ell^a$ with $\ell^a \in [\ell^a]$, we have then $\Lie_{\ell'} (\omega_a \eta^a) = f \Lie_{\bm{\ell}}(\omega_a \eta^a) = f (\eta^a \Lie_{\bm{\ell}}\omega_a +  \omega_a \Lie_{\bm{\ell}} \eta^a) = 0$ thanks to Eq.~\eqref{def_WIH} and $[\bm{\ell},  \bm{\eta}]^a = 0$. Since the area element $\ud S$ is preserved as well be the isometry flow, it follows
that the integral \eqref{def_angu_mom} taken on $\calS'$ equals that taken on $\calS$.
Furthermore, it can be shown \cite{As.al.01} that if the whole spacetime $(\calM,g_{ab})$ is axisymmetric, with the corresponding Killing vector
coinciding with $\eta^a$ on $\calH$, then $J_{\calH}$ is nothing but the standard Komar angular momentum.

The integral in the expression \eqref{def_angu_mom} of $J_{\calH}$ can be written as the integral of the 2-form $(\bm{\Omega}\cdot\bm{\eta})\bm{\varepsilon}$ on the 2-surface $\calS$.
It is easy to check that $(\bm{\Omega}\cdot\bm{\eta})\bm{\varepsilon} = \bm{\Omega}\wedge \star\bm{\eta}$, where $\star\eta_a \equiv \eta^b \varepsilon_{ba}$ (cf. Sec.~\ref{subsec:axi}); in particular, as a 2-form, $\bm{\Omega}\wedge \star\bm{\eta}$ has to be proportional to $\bm{\varepsilon}$. Now, Eq.~\eqref{theta_prime_from_axi} yields $\star\bm{\eta} = R^2 \bm{\ud}\zeta$. Hence, we may write
\beq
    (\bm{\Omega}\cdot\bm{\eta}) \, \bm{\varepsilon} = - R^2 \bm{\ud}\zeta \wedge \bm{\Omega} = -R^2 \left[ \bm{\ud}(\zeta\bm{\Omega}) - \zeta \bm{\ud}\bm{\Omega} \right] .
\eeq
The integral over $\calS$ of $\bm{\ud}(\zeta\bm{\Omega})$ being zero by Stokes' theorem, we are left with
\beq
   J_{\calH} = - \frac{R^2}{8\pi} \oint_{\calS} \zeta \, \bm{\ud}\bm{\Omega} \, .
\eeq
By comparing with Eq.~\eqref{I-L_axi_R_Omega} for $\ell=1$, taking into account that $Y_{1,0}(\theta,\phi) = \sqrt{3/(4\pi)} \cos\theta = \sqrt{3/(4\pi)} \, \zeta$, we see that
the angular momentum is proportional to the current dipole $L_1^{\rm axi}$:
\beq \label{J_H_L_1_axi}
   J_{\calH} = \frac{R^2}{2\sqrt{3\pi}} \, L_1^{\rm axi} \, .
\eeq

The definition of the horizon mass $M_{\calH}$ is more tricky than that of the horizon angular momentum $J_{\calH}$ because the WIH structure does not allow one to single out a unique vector field $\xi^a$ playing the role of a time translation generating a Hamiltonian on the phase space, which would define the
energy, and hence the mass. On the contrary, the
rotational vector $\eta^a$ leading to $J_{\calH}$ could be defined uniquely from the axisymmetry hypothesis.
To be a symmetry generator of the WIH $(\calH,[\ell^a])$, any candidate $\xi^a$ must be such that its restriction to $\calH$ takes the form
$\xi^a \eqNEH c \ell^a - \varOmega \eta^a$, where $\ell^a\in[\ell^a]$ and $c$ and $\varOmega$ are two constants:
thanks to Eqs.~\eqref{Lie_ell_q_zero_NEH}, \eqref{Lie_eta_h_omega} and \eqref{def_WIH}, this guarantees $\Lie_{\bm{\xi}} h_{ab} = 0$ and $\Lie_{\bm{\xi}}\omega_a = 0$.
In Ref.~\cite{As.al.01}  (see also the reviews \cite{AsKr.04,AsKr.25}), it is shown that $\xi^a$ generates a Hamiltonian on the phase space if, and only if,
$\kappa \equiv c\kappa_{(\bm{\ell})}$ and $\varOmega$ are functions of $J_{\calH}$ and the horizon area $A$ only, and moreover obey
$\partial \kappa/\partial J_{\calH} = 8\pi \partial \varOmega / \partial A$.
The (on-shell) value of the Hamiltonian is then the difference $H_{(\bm{\xi})} = E^{(\bm{\xi})}_\infty - E^{(\bm{\xi})}_{\calH}$ between two surface integrals, one at spatial infinity,
defining the ADM energy $E^{(\bm{\xi})}_\infty$,
and the other one on $\calH$, defining the horizon energy $E^{(\bm{\xi})}_{\calH}$. The variations of $E^{(\bm{\xi})}_{\calH}$, $A$ and $J_{\calH}$ in the phase space
then obey $\delta E^{(\bm{\xi})}_{\calH} = \kappa/(8\pi) \delta A + \varOmega \delta J_{\calH}$, which is known as the \emph{first law of WIH mechanics}.
The \emph{horizon mass} $M_{\calH}$ is defined as the value of $E^{(\bm{\xi})}_{\calH}$ for a configuration where $\calH$ is ``at rest'' with respect to $\xi^a$,
i.e. when $\xi^a$ is a global Killing vector of spacetime.
It is argued in  Ref.~\cite{As.al.01} that, in order to recover the Kerr solution, with $\xi^a$ being the standard stationary Killing vector,
this fixes the functions $\kappa(A, J_{\calH})$ and $\varOmega(A, J_{\calH})$ to their Kerr values, thereby leading to the following
expression of the horizon mass:
\beq
    M_{\calH} = \frac{R}{2} \sqrt{1+ 4J_\calH^2/R^4} \, .
\eeq

In view of Eq.~\eqref{I-L_axi_R_Omega} and by analogy with the electrostatics and magnetostatics of a charged, conducting sphere, one can assign to $\calH$ a `surface mass density'
$M_\calH \mathcal{R} / 8\pi$, a `surface momentum density' $\Omega_a / 8\pi$, as well as the associated  mass and current multipole moments \cite{As.al.04,AsKr.25}
\beq\label{source}
    M_{\ell} \equiv \frac{M_\calH R^\ell}{\sqrt{(2\ell+1)\pi}}\, I_{\ell}^{\rm axi} \quad \text{and} \quad S_{\ell} \equiv \frac{R^{\ell+1}}{2\sqrt{(2\ell+1)\pi}} \, L_{\ell}^{\rm axi}\, .
\eeq
Incidentally, we note that interpreting $\Omega_a / 8\pi$ as a momentum density was first put forward in Refs.~\cite{Da.79,Da.82b}, while developing a ``fluid bubble'' analogy for the horizon dynamics.
The scaling \eqref{source} ensures that $M_0 = M_{\calH}$ [since $I_0^{\rm axi} = \sqrt{\pi}$, cf. Eq.~\eqref{universal_values_axi}] and
$S_1 = J_{\calH}$ [thanks to Eq.~\eqref{J_H_L_1_axi}]. Moreover, we shall see explicitly in Sec.~\ref{subsec:axi_Kerr}
that for a Kerr black hole, $M_0$ and $S_1$ reproduce the Komar mass and angular momentum, respectively.

A natural question is how the \textit{source} multipole moments \eqref{source} of a given axisymmetric NEH compare to the \textit{field} multipole moments associated with that same NEH, defined for instance according to Hansen \cite{Ha.74} [cf. Eq.~\eqref{Hansen} for a Kerr black hole] or Thorne \cite{Th.80}, which are equivalent for stationary spacetimes \cite{Gu.83}. In Newtonian gravity or electrodynamics, the linearity of the Poisson equation or the Maxwell equations imply that source and field multipoles necessarily coincide. In General Relativity, however, the nonlinear nature of the Einstein equation means that source and field multipoles generically differ \cite{As.al.04,Bl.24}. As we shall see in section \ref{subsec:axi_Kerr}, in the case of Kerr black holes, the source multipoles \eqref{source} share several properties with the field multipoles, including their behavior for small spin values, but they do differ in their magnitude, except for $\ell=0$ and $\ell=1$.


\section{Multipole moments of a generic NEH}\label{sec:generic_multipoles}

In this section we review the definition of multipole moments of a generic NEH recently proposed by Ashtekar, Khera, Kolanowski and Lewandowski \cite{As.al.22}.
Contrary to that discussed in Sec.~\ref{sec:axi_multipoles}, this definition does not require the NEH to be axisymmetric. Rather, it is based
on a conformal decomposition of the cross-section metric, devised by Korzy\'nski \cite{Ko.07} to define the horizon angular momentum (the $\ell=1$ source multipole $S_\ell$ in the axisymmetric case).
We successively discuss the conformal round metrics of horizon cross-sections (Secs.~\ref{subsec:round} and \ref{subsec:extension}), the definition of the NEH multipole moments (Sec.~\ref{subsec:multipoles}), the uniqueness of this definition by imposing the condition of a vanishing area dipole moment (Sec.~\ref{subsec:dipole}), and the expressions of the horizon multipoles in terms of an electric-type scalar potential and a magnetic-type pseudo-scalar potential (Sec.~\ref{subsec:scalars}).

\subsection{Conformal unit round metrics on a cross-section}\label{subsec:round}

Let us consider a cross-section $\calS$ of a NEH $\calH$.
As discussed in Sec.~\ref{s:cross_section_slicings}, $\calS$ is endowed with the Riemannian metric $q_{ab}$ induced by the spacetime metric $g_{ab}$. Let us denote by $D_a$ the associated Levi-Civita connection and by $\mathcal{R}$ the associated scalar curvature, as in \eqref{RePsi2_NEH}. Since $\calS$ is topologically a 2-sphere, the uniformization theorem implies that
there exists some unit round metrics $\mathring{q}_{ab}$ on $\calS$ (with Levi-Civita connection $\mathring{D}_a$ and constant scalar curvature $\mathring{\mathcal{R}} = 2$) that are conformally related to $q_{ab}$ (see e.g. Refs.~\cite{Ch.91,Gi.al.13,Ko.07}):
\beq\label{conformal_q}
    \mathring{q}_{ab} = \psi^2 q_{ab} \, ,
\eeq
where the conformal factor $\psi$ is a scalar field on $\calS$ with the dimension of an inverse length, since $\mathring{q}_{ab}$ is a round metric of a sphere of radius 1.\footnote{One could make $\psi$ dimensionless by introducing
$\breve{\psi} \equiv R \psi$, where $R$ is the areal radius \eqref{def_areal_radius}.}
Being a unit round metric means that there exists some coordinates $x^A = (\vartheta,\varphi)$ on $\calS$ such that
$\vartheta\in[0,\pi]$, $\varphi\in[0, 2\pi)$ and
\beq\label{canonical}
   \mathring{q}_{AB} \, \ud x^A \ud x^B = \ud \vartheta ^2 + \sin^2 \vartheta \, \ud \varphi ^2 \, .
\eeq
We will refer to $(\vartheta,\varphi)$ as \emph{polar coordinates adapted to $\mathring{q}_{ab}$}.
Note that such coordinates are by no means unique: because of the $\mathrm{SO}(3)$ symmetry of $\mathring{q}_{ab}$, there exists a 3-parameter  family of coordinates $(\vartheta,\varphi)$ such that
\eqref{canonical} holds, 2 parameters defining the axis $\vartheta\in\{0,\pi\}$ and 1 parameter setting the origin of $\varphi$.

According to the analysis performed in App.~D of Ref.~\cite{Wal},\footnote{See in particular Eq.~(D9) in Ref.~\cite{Wal}, with $n=2$, $\Omega=\psi$, $\tilde{R} = \mathring{\mathcal{R}} = 2$ and $R = \mathcal{R}$.} the conformal factor $\psi$ is a solution of the nonlinear, second-order, elliptic partial differential equation
\beq\label{plout}
    D^2 \ln\psi + \psi^2 = \tfrac{1}{2} \,  \mathcal{R} \, ,
\eeq
where $D^2 \equiv q^{ab} D_a D_b$ is the Laplace operator associated with $q_{ab}$. Alternatively, using the conformal transformation law $D^2 \Phi = \psi^2 \mathring{D}^2 \Phi$ valid for any scalar field $\Phi$ in dimension 2 \cite{Wal}, we may write the partial differential equation \eqref{plout} as
\beq\label{D0ln_psi}
   \mathring{D}^2 \ln\psi + 1  =  \tfrac{1}{2} \, \psi^{-2} \mathcal{R} \, ,
\eeq
where $\mathring{D}^2 \equiv \mathring{q}^{ab} \mathring{D}_a \mathring{D}_b$ is the Laplace operator associated with $\mathring{q}_{ab}$.

For a given $q_{ab}$, the round metric $\mathring{q}_{ab}$ is by no means unique: any other unit round metric $\mathring{q}'_{ab}$ on $\calS$
that obeys $\mathring{q}'_{ab} = \psi'^2 q_{ab}$ for some conformal factor $\psi'$ is necessarily conformally related to $\mathring{q}_{ab}$
via $\alpha \equiv \psi' / \psi$:
\beq\label{ploutou}
    \mathring{q}'_{ab} = \alpha^2 \mathring{q}_{ab} \, , \quad \text{where}~\alpha~\text{satisfies} \quad \mathring{D}^2 \ln\alpha + \alpha^2 = 1 \, .
\eeq
The above partial differential equation\footnote{Note that there is a typo in the corresponding equation in Ref.~\cite{As.al.22}, i.e. Eq.~(2.6) there, which should be written as Eq.~\eqref{ploutou} above.} \mbox{for $\alpha$ follows directly from \eqref{plout} with $q_{ab}$ (within $D^2$)} substituted by $\mathring{q}_{ab}$ and $\psi$ substituted by $\alpha$. The most general solution to \eqref{ploutou} is given via a (suitably normalized) linear combination of the first four spherical harmonics of $\mathring{q}_{ab}$ \cite{AsBa.19}, namely
\beq\label{alpha}
    \alpha(\vartheta,\varphi)^{-1} = \alpha_0 + \alpha_1 \sin\vartheta \cos\varphi + \alpha_2 \sin\vartheta \sin\varphi + \alpha_3 \cos\vartheta \, ,
\eeq
where $\alpha_0$ and $\vec{\alpha} \equiv (\alpha_1,\alpha_2,\alpha_3)$ are real constants satisfying the constraint $\alpha_0^2 - |\vec{\alpha}|^2 = 1$.
We conclude that the set of unit round metrics $\mathring{q}_{ab}$ conformal to the physical metric $q_{ab}$ on $\calS$ forms a 3-parameter family.
The set of conformal factors $\psi$ does the same, since Eq.~\eqref{ploutou} is equivalent to $\psi' = \alpha\psi$.

\subsection{Extension of the conformal decomposition}\label{subsec:extension}

A priori, the conformal factor $\psi$ introduced in Eq.~\eqref{conformal_q} is defined only on
the considered cross-section $\calS$. We now show that, on a NEH, where all cross-sections are isometric, one can actually extend $\psi$ to a scalar field defined on the whole of $\calH$.
To do so, let us consider another cross-section $\calS'$ of $\calH$, with metric $q'_{ab}$ induced by $g_{ab}$, and the isometry $\Phi: \calS \to \calS'$ introduced in Sec.~\ref{subsec:def_NEH}.
By pulling back \eqref{conformal_q} to $\calS'$ via $\Phi^{-1}$, we
get $(\Phi^{-1})^*\, \mathring{q}_{ab} = (\psi\circ \Phi^{-1})^2\,  (\Phi^{-1})^* \, q_{ab}$.
But $(\Phi^{-1})^* \, q_{ab} = q'_{ab}$ since $\Phi$ is an isometry for the induced metrics.
Hence there comes
\beq  \label{conformal_q_Sp}
  \mathring{q}'_{ab} = {\psi'}^2 q'_{ab} \, ,
\eeq
with
\beq \label{round_metric_drag}
\mathring{q}'_{ab} \equiv (\Phi^{-1})^*\, \mathring{q}_{ab} \quad \mbox{and} \quad
\psi' \equiv \psi\circ \Phi^{-1} .
\eeq
As it is defined, the metric $\mathring{q}'_{ab}$ is isometric to $\mathring{q}_{ab}$ ---the isometry being $\Phi$. This implies
that $\mathring{q}'_{ab}$ is a unit round metric on $\calS'$, so that \eqref{conformal_q_Sp} is
exactly similar to \eqref{conformal_q}.
Let us then define $\psi$ on $\calH$ by $\psi(p) = \psi(p_0)$, where $p_0$ is the unique point on $\calS$ lying on the same null
geodesic generator $L_{(\theta_1,\theta_2)}$ as $p$.
On $\calS'$, the function $\psi$ hence defined coincides with $\psi'$.
Accordingly, \eqref{conformal_q} can be extended to all cross-sections of $\calH$, thereby
defining a unit round metric $\mathring{q}_{ab}$ on each of them.
By construction, the scalar field $\psi$ is invariant along
the generators of $\calH$. It follows that, for any null normal
$\ell^a$ to $\calH$,
\beq\label{Lie_ell_psi_zero}
    \Lie_{\bm{\ell}} \psi = 0 \, .
\eeq

\subsection{Horizon multipole moments}\label{subsec:multipoles}

We are ready to define the multipole moments of a generic NEH $\calH$, as introduced in \cite{As.al.22}. For a given cross-section $\calS$ of $\calH$ and a choice of the conformal unit round metric $\mathring{q}_{ab}$ fulfilling Eq.~\eqref{conformal_q}, the \textit{shape} and \textit{current} multipole moments $I_{\ell,m}$ and $L_{\ell,m}$ are defined according to
\beq\label{I-L}
    I_{\ell,m} \equiv - \oint_\calS (\text{Re}\,\Psi_2) \, \mathring{Y}_{\ell,m} \, \ud S \qquad \text{and} \qquad L_{\ell,m} \equiv - \oint_\calS (\text{Im}\,\Psi_2) \, \mathring{Y}_{\ell,m} \, \ud S \, .
\eeq
Here, $\ud S$ is the area element of $\calS$ corresponding to the physical metric $q_{ab}$ [Eq.~\eqref{def_area_NEH}], while $\mathring{Y}_{\ell,m}$
stands for the spherical harmonic with respect to the unit round metric $\mathring{q}_{ab}$, i.e., the eigenfunction of order $m$ corresponding to the eigenvalue $-\ell(\ell+1)$
of the Laplace operator $\mathring{D}_a \mathring{D}^a$ of $\mathring{q}_{ab}$. In terms of polar coordinates $(\vartheta,\varphi)$ adapted to $\mathring{q}_{ab}$
[cf. Eq.~\eqref{canonical}], we have
\beq \label{def_mathringYlm}
    \mathring{Y}_{\ell,m} = Y_{\ell,m}(\vartheta,\varphi) \, ,
\eeq
where $Y_{\ell,m}$ are the ordinary spherical harmonic functions. For $m=0$, the above relationship is similar to Eq.~\eqref{def_mathringYlm_axi}, but one should keep in mind that, on an axisymmetric NEH,  $(\vartheta,\varphi)$ and $(\theta,\phi)$ are a priori distinct coordinates on $\calS$, as we shall see explicitly in the Kerr case in Sec.~\ref{sec:Kerr}.

The multipole moments $I_{\ell,m}$ and $L_{\ell,m}$ are dimensionless, given that $\Psi_2$ has the dimension of an inverse squared length and $\ud S$ that of a squared length. Besides, for all $|m| \leqslant \ell$, the multipole moments \eqref{I-L} obey the symmetry $I_{\ell,-m} = (-)^m \bar{I}_{\ell,m}$ and $L_{\ell,-m} = (-)^m \bar{L}_{\ell,m}$, inherited from standard properties of spherical harmonics $Y_{\ell,m}$. Therefore, it is possible to determine \textit{separately} $I_{\ell,m}$ and $L_{\ell,m}$ from the knowledge of the convenient, complex-valued linear combination
\beq\label{I+iL}
    K_{\ell,m} \equiv I_{\ell,m} + \ui L_{\ell,m} = - \oint_\calS \Psi_2 \, \mathring{Y}_{\ell,m} \, \ud S = - \oint_\calS \psi^{-2} \, \Psi_2 \, \mathring{Y}_{\ell,m} \, \ud \mathring{S}  \, ,
\eeq
where the second integral is expressed in terms of the area
element $\ud\mathring{S}$ of $\mathring{q}_{ab}$, taking into
account the relation $\ud S = \psi^{-2} \ud\mathring{S}$, which
follows from Eq.~\eqref{conformal_q}.

Recall that $\Psi_2$ is frame-independent on a NEH (cf. Sec.~\ref{subsec:Weyl_scalars}). Moreover, the multipoles $I_{\ell,m}$ and $L_{\ell,m}$ do not depend on the choice of the cross-section $\calS$ over which the integrals \eqref{I-L} are carried out, once a unit round metric has been chosen within the 3-parameter family given by \eqref{ploutou}--\eqref{alpha}
on a fiducial cross-section and transported on $\calH$ via \eqref{round_metric_drag}. Indeed, let us consider a cross-section $\calS'$ of $\calH$, distinct from the cross-section $\calS$ on which the unit round metric $\mathring{q}_{ab}$ has been chosen. Let $K'_{\ell,m}$ be the integral \eqref{I+iL} taken on $\calS'$. Since $\Psi_2$ is invariant along the null generators of $\calH$ [cf. Eq.~\eqref{LiePsi2}], in the integral for $K'_{\ell,m}$, $\Psi_2 = \left. \Psi_2 \right|_{\calS'}$ is nothing but the pullback of $\left. \Psi_2 \right|_{\calS}$ by
the inverse of the isometry $\Phi: \calS \to \calS'$ introduced in Sec.~\ref{subsec:def_NEH}. Moreover, the unit round metric $\mathring{q}'_{ab}$ on $\calS'$ is also the pullback by $\Phi^{-1}$ of the unit round metric $\mathring{q}_{ab}$ on $\calS$; cf. Eq.~\eqref{round_metric_drag}.
Therefore, the spherical harmonic $\mathring{Y}_{\ell,m}$ in the integral for $K'_{\ell,m}$ is the pullback of that on $\calS$ by $\Phi^{-1}$. Finally, the area element $\ud S'$ on $\calS'$ is the pullback by $\Phi^{-1}$ of the area element of $\calS$ since $\Phi$ is an isometry $(\calS, q_{ab}) \to (\calS', q'_{ab})$. Hence all the integrand of $K'_{\ell,m}$ is the pullback of the integrand of $K_{\ell,m}$ by $\Phi^{-1}$. We conclude that $K'_{\ell,m} = K_{\ell,m}$. One may thus refer to \eqref{I-L} as \textit{the} multipole moments of the NEH $\calH$, irrespective of the choice of 2-sphere cross-section $\calS$.

It should be noticed that the monopole moments $I_{0,0}$ and $L_{0,0}$ take the same values for all NEHs:
\beq \label{monopole_universal}
    I_{0,0} = \sqrt{\pi} \quad\mbox{and}\quad L_{0,0} = 0 \, .
\eeq
The proof is the same as for the axisymmetry-based monopoles [Eq.\eqref{universal_values_axi}], given that $\mathring{Y}_{0,0}$ is the constant $1/(2\sqrt{\pi})$, independently of the unit round metric
$\mathring{q}_{ab}$, so that Eq.~\eqref{monopoles_GB_Stokes} does not involve the choice of any $\mathring{q}_{ab}$ and is valid outside axisymmetry.
On the contrary the proof of $I_1^{\rm axi} = 0$ in Sec.~\ref{subsec:multipole_axi} is not applicable to $I_{1,0}$ here, even if $q_{ab}$ is assumed to be axisymmetric.

\subsection{Canonical round metric from vanishing area dipole moment}\label{subsec:dipole}

The definition \eqref{I-L} does not provide a unique set of horizon multipole moments, because the involved spherical harmonics $\mathring{Y}_{\ell,m}$ are relative to a unit round metric $\mathring{q}_{ab}$ on $\calS$ that is arbitrary among the 3-parameter family described by Eqs.~\eqref{ploutou}--\eqref{alpha}, all members of this family being conformal to the physical metric $q_{ab}$. To obtain a uniquely-defined, `canonical' set of horizon multipoles, one must pick a `canonical' unit round metric $\undertilde{\mathring{q}}_{ab}$, and accordingly a `canonical' conformal factor $\undertilde{\psi}$. One possible choice is to require the vanishing of the \textit{area dipole moment} \cite{As.al.22}
\beq\label{d}
    d^i \equiv \oint_\calS n^i \, \ud S = \oint_{\calS} n^i \, \psi^{-2} \, \ud\mathring{S} \, ,
\eeq
where
$n^i = (\sin\vartheta \cos\varphi, \sin\vartheta \sin\varphi, \cos\vartheta)$ can be viewed as a unit direction normal to $\mathbb{S}^2$ in terms of some polar coordinates $(\vartheta,\varphi)$
adapted to $\mathring{q}_{ab}$, i.e., obeying Eq.~\eqref{canonical}. Hence $d^i = 0$ for $\mathring{q}_{ab} = \undertilde{\mathring{q}}_{ab}$ and $\psi = \undertilde{\psi}$. (This is analogous to the choice of mass-centered coordinates in other contexts). As shown in App.~A of Ref.~\cite{As.al.22}, the metric $\undertilde{\mathring{q}}_{ab}$ always exists and is \textit{unique}.

Equivalently, using the bijection between the angular bases of symmetric trace-free unit tensors $n^{\langle i_1} \cdots n^{i_\ell \rangle}$ and the spherical harmonics $Y_{\ell,m}$ \cite{BlDa.86}, the condition $d^i = 0$ of a vanishing area dipole moment can be recast in terms of the $\ell = 1$ spherical harmonic modes of $\psi^{-2}$ in adapted polar coordinates $(\vartheta,\varphi)$, namely
\beq\label{d modes}
    d_{1,m} \equiv \oint_\calS \mathring{Y}_{1,m} \, \ud S = \oint_{\calS} \mathring{Y}_{1,m} \, \psi^{-2} \, \ud \mathring{S} = \int_0^{2\pi} \ud\varphi \int_{-1}^1 \ud(\cos{\vartheta}) \, Y_{1,m}(\vartheta,\varphi) \, \psi(\vartheta,\varphi)^{-2} \, .
\eeq
Setting $d_{1,m} = 0$ for $m \in \{-1,0,1\}$ provides three conditions obeyed by the four parameters $\alpha_0$ and $\vec{\alpha}$ in Eq.~\eqref{alpha}, in addition to the constraint $-\alpha_0^2 + |\vec{\alpha}|^2 = - 1$, and thus specifies \textit{uniquely} the unit round metric
$\mathring{q}_{ab}$.

Alternatively, one could require the vanishing of the `mass dipole moment' $\propto \oint_\calS \mathcal{R} \, n^i \, \ud S$, instead of the area dipole moment \eqref{d}. However, as discussed in App.~A of Ref.~\cite{As.al.22}, while there exist unit round metrics for which the mass dipole moment vanishes, in general that metric is not unique and additional conditions have to be imposed to guarantee uniqueness.

\subsection{Electric and magnetic scalar potentials}\label{subsec:scalars}

Expression \eqref{ImPsi2_eps_dOmega} of $\text{Im\,}\Psi_2$ involves the \Hajicek{} 1-form $\Omega_a$ relative to any null normal $\ell^a$ with respect to the cross-section $\calS$. As noticed in Ref.~\cite{As.al.22}, one can select null normals such that $\Omega_a$ is a divergence-free 1-form on $\calS$; we shall then denote it by
$\hat{\Omega}_a$:
\beq \label{Omega_divfree}
  q^{ab} D_a \hat{\Omega}_b = 0 \, .
\eeq
By demanding further that the null normal are geodesic ($\kappa_{(\bm{\ell})} = 0$), condition \eqref{Omega_divfree} reduces
the choice of null normals to $\calH$ to a single constant-rescaling equivalence class $[\ell^a]$ \cite{As.al.22}.

Equation~\eqref{Omega_divfree}
is equivalent to $\hat{\Omega}_a$ being co-exact: there
exists a (pseudo-)scalar field $B$ on $\calS$, which is referred to as the \textit{magnetic potential} \cite{As.al.22}, such that
\beq \label{bar_Omega_coexact}
    \hat{\Omega}_a = \varepsilon_a^{\phantom{a}b} D_b B =  \mathring{\varepsilon}_a^{\phantom{a}b} \mathring{D}_b B \, .
\eeq
The second equality involves the area 2-form $\mathring{\varepsilon}_{ab}$ of the unit round metric $\mathring{q}_{ab}$
and results from the conformal invariance of the Hodge dual of 1-forms on a 2-dimensional manifold.\footnote{This is easy to
establish: Eq.~\eqref{conformal_q} implies $\mathring{\varepsilon}_{ab} = \psi^2 \varepsilon_{ab}$, so that
$ \mathring{\varepsilon}_a^{\phantom{a}b} = \mathring{\varepsilon}_{ac} \mathring{q}^{cb}  = \psi^2 \varepsilon_{ac} \psi^{-2} q^{cb} = \varepsilon_a^{\phantom{a}b}$.}
It follows from Eq.~\eqref{bar_Omega_coexact} that
$(\ud \hat{\Omega})_{ab} = - (\mathring{D}^2 B) \, \mathring{\varepsilon}_{ab}$.
Given that
$\varepsilon_{ab} = \psi^{-2} \mathring{\varepsilon}_{ab}$,
Eq.~\eqref{ImPsi2_eps_dOmega} is then equivalent to
\beq \label{ImPsi2_Lap_B}
  \psi^{-2} (\text{Im\,}\Psi_2) \stackrel{\calS}{=}  - \tfrac{1}{2} \, \mathring{D}^2 B \, .
\eeq
On the other hand, by combining \eqref{RePsi2_NEH} and
\eqref{D0ln_psi}, we get
$\psi^{-2} (\text{Re\,}\Psi_2)  \!=\! - \tfrac{1}{2} (\mathring{D}^2 \ln \psi + 1)$.
In view of Eq.~\eqref{ImPsi2_Lap_B}, this yields
\beq \label{psi2_epsilon}
    \psi^{-2} \, \Psi_2 \stackrel{\calS}{=}  - \tfrac{1}{2} \, \bigl[ (1 + \mathring{D}^2 E) + \ui \mathring{D}^2 B \bigr]  \, ,
\eeq
where
\beq \label{def_E_electric}
  E \equiv \ln{(R\,\psi)}
\eeq
is a scalar field on $\calH$, which we shall refer to  as the \textit{electric potential}.\footnote{The equivalent quantity introduced in Ref.~\cite{As.al.22} is $E = \ln\psi$; it differs from Eq.~\eqref{def_E_electric} by the additive constant $\ln R$, which plays no role in what follows. We simply have introduced the factor $R$ in Eq.~\eqref{def_E_electric} to make the argument of the logarithm dimensionless.}
By virtue of \eqref{Lie_ell_psi_zero} $E$ is Lie-dragged along the null geodesic generators of $\calH$. Notice the constant term in the electric sector in Eq.~\eqref{psi2_epsilon}, absent from the magnetic sector, which originates from a topological constraint: indeed, by integrating over $\calS$ the scalar curvature $\mathcal{R}$ (a topological invariant), we readily find
\beq\label{Gauss-Bonnet}
    \tfrac{1}{2} \oint_\calS \mathcal{R} \, \ud S = - 2 \oint_\calS (\text{Re}\,\Psi_2) \, \ud S = - 2 \oint_\calS \psi^{-2} \, (\text{Re}\,\Psi_2) \, \ud \mathring{S} =  \oint_\calS \ud \mathring{S} = 4\pi \, ,
\eeq
where we successively used Eq.~\eqref{RePsi2_NEH}, $\ud S = \psi^{-2} \ud\mathring{S}$, the real part of Eq.~\eqref{psi2_epsilon}, as well as $\oint_\calS (\mathring{D}^2 f) \, \ud \mathring{S} = \int_{\partial\calS} \mathring{\varepsilon}_{ab} \mathring{D}^b f = 0$ for any scalar field $f$ defined on $\calS \sim \mathbb{S}^2$, as a consequence of Stokes' theorem and $\partial \calS = \varnothing$. The formula \eqref{Gauss-Bonnet} is in agreement with the Gauss-Bonnet theorem applied to a closed orientable 2-surface with \textit{genus} $g = 0$.

Now, by substituting expression \eqref{psi2_epsilon} for $\psi^{-2} \Psi_2$ into \eqref{I+iL}, we obtain, for any $\ell \geqslant 1$,
\beq\label{I+iL_bis}
    K_{\ell,m} = \tfrac{1}{2} \oint_{\calS} \mathring{D}^2 (E + \ui B) \, \mathring{Y}_{\ell,m} \, \ud \mathring{S} \, .
\eeq
Integrating by parts twice while using Stokes' theorem and the fact that $\mathring{Y}_{\ell,m}$ is an eigenfunction of the Laplace operator $\mathring{D}^2$, with eigenvalue $-\ell(\ell+1)$,  Eq.~\eqref{I+iL_bis} can be expressed in the simpler form
\beq\label{I+iL_ter}
    K_{\ell,m} = - \tfrac{1}{2} \, \ell(\ell+1) \oint_{\calS} (E + \ui B) \, \mathring{Y}_{\ell,m} \, \ud \mathring{S} \, , \qquad \ell \geqslant 1 \, .
\eeq
Consequently, the geometry of the NEH $\calH$ is encoded into the electric potential $E$ and the magnetic pseudo-potential $B$, which can then be reconstructed from the multipole moments $I_{\ell,m}$ and $L_{\ell,m}$ according to\footnote{We believe that there is a sign error in Eq.~(2.14) of Ref.~\cite{As.al.22}, which originates from a typo in the second line of Eq.~(2.13) there. Also the mean values $\langle E \rangle$ and $\langle B \rangle$ are missing in Eq.~(2.14) of Ref.~\cite{As.al.22}.} \cite{As.al.22}
\beq\label{E+iB}
    E + \ui B = \langle E \rangle + \ui \langle B \rangle - \sum_{\ell=1}^{+\infty} \sum_{m=-\ell}^{\ell} \; \frac{2K_{\ell,m}}{\ell(\ell+1)} \, \mathring{\overline{Y}}_{\!\ell,m} \, ,
\eeq
where $\langle E \rangle$ and $\langle B \rangle$ stand for the mean values of $E$ and $B$ over $\calS$, with respect to the measure $\mathring{\varepsilon}_{ab}$, respectively, and $\mathring{\overline{Y}}_{\!\ell,m}$ is the complex conjugate of $\mathring{Y}_{\ell,m}$.

\section{Application to Kerr black holes}\label{sec:Kerr}

In this section we apply the two distinct definitions of NEH multipole moments reviewed in Secs.~\ref{sec:axi_multipoles} and \ref{sec:generic_multipoles} to the event horizon of a Kerr black hole. We first discuss the main properties of Kerr horizon cross-sections in Sec.~\ref{subsec:cross}, before computing the multipole moments
$I_\ell^{\rm axi}$ and $L_\ell^{\rm axi}$ resulting from the axisymmetry-based definition in Sec.~\ref{subsec:axi_Kerr}. Then, we move to the multipole family defined for generic NEHs, by determining
first the conformal round metric with vanishing area dipole moment in Sec.~\ref{subsec:conformal}. We then evaluate in closed-form the magnetic-type potential on the Kerr horizon in Sec.~\ref{subsec:magnetic}. Finally, we explore the resulting horizon multipoles in Sec.~\ref{subsec:moments}, and compare them to
the axisymmetry-based multipoles in Sec.~\ref{subsec:comparison_Kerr}.

\subsection{Geometry of the horizon cross-sections}\label{subsec:cross}

Consider a Kerr black hole of mass $M$ and spin angular momentum $S$. In advanced Kerr coordinates $(x^\alpha) = (v,r,\theta,\pAdv)$,\footnote{In the literature these coordinates are also referred to as \emph{ingoing Kerr coordinates} \cite{TePr.74}, or more simply as \emph{Kerr coordinates} \cite{MTW}. They are related to the Boyer-Lindquist coordinates~\cite{BoLi.67} $(t,r,\theta,\phi_\text{BL})$ by $\ud v = \ud t + (r^2+a^2) \, \ud r / \Delta \equiv \ud t + \ud r_*$ and $\ud\phi = \ud\phi_\text{BL} + a \, \ud r / \Delta$, where $\Delta \equiv r^2 - 2 M r + a^2$.} which are adapted to the stationarity and axisymmetry of the Kerr metric, and are regular on the horizon (contrary to Boyer-Lindquist coordinates), the metric components $g_{\alpha\beta}$ read
\begin{align}\label{eq:metric}
    g_{\alpha\beta} \, \ud x^\alpha \ud x^\beta = &- \left( 1 - \frac{2Mr}{\Sigma} \right) \ud v^2 + 2 \ud v \ud r - \frac{4 M r}{\Sigma} \, a \sin^2{\theta} \, \ud v \ud \pAdv - 2a \sin^2{\theta} \, \ud r \ud \pAdv \nonumber \\ &+ \Sigma \, \ud \theta^2 + \left( r^2 + a^2 + \frac{2Mr}{\Sigma} \, a^2 \sin^2{\theta} \right) \sin^2{\theta} \, \ud \pAdv^2 \, ,
\end{align}
where $a \equiv S/M$ is the Kerr spin parameter and $\Sigma \equiv r^2 + a^2 \cos^2{\theta}$. The black hole event horizon $\calH$ is located at
$r = r_+ \equiv M + \sqrt{M^2 - a^2}$; it is
a \emph{Killing horizon}, namely a null hypersurface admitting a Killing vector field as normal. This constitutes a special case of NEH.
Therefore, according to the discussion in Sec.~\ref{subsec:Weyl_scalars}, the Weyl curvature scalar $\Psi_2$ is frame-invariant on $\calH$, provided that $\ell^a$ coincides with a null normal to $\calH$. In particular, it may be evaluated using either the Hartle-Hawking \cite{HaHa.72} or Hartle \cite{Ha2.74} tetrad. Using advanced Kerr coordinates, we explicitly have\footnote{See e.g. Eqs.~(2.30) and (2.33) in \cite{Po2.04}, which uses the opposite sign convention for the Weyl scalar $\Psi_2$.}
\beq\label{Psi2_Kerr}
    \Psi_2 = M \varrho^3 \, , \quad \text{where} \quad \varrho \equiv - \frac{1}{r-\ui a \cos\theta} \, .
\eeq
Moreover, by using a null frame adapted to the two repeated principal null directions of the Kerr metric, such as the Hartle-Hawking \cite{HaHa.72} or Kinnersley \cite{Ki.69} tetrad, the only nonvanishing Weyl scalar in Eqs.~\eqref{eq:Weyl sc} is precisely \eqref{Psi2_Kerr} \cite{Cha}. This is not the case for the Hartle \cite{Ha2.74} tetrad, for which $\Psi_3$ and $\Psi_4$ do not vanish \cite{FaLo.05}.

Let $\calS$ be a cross-section of $\calH$ defined by $v = \text{const}$.\footnote{Note that Hartle's tetrad \cite{Ha2.74} is adapted to such cross-sections,
since the vector $n^a$ of that tetrad is normal to the slices $v = \text{const}$ of $\calH$.
On the contrary, the Hartle-Hawking tetrad \cite{HaHa.72} is not, since its $n^a$ vector is not normal to any slicing of $\calH$ (the 2-planes $\mathrm{Span}(\ell^a,n^a)^\perp$ are not integrable into 2-surfaces).}
Since $\calH$ is located at a constant value of $r$, namely $r = r_+$, $\calS$ is spanned by the
Kerr angular coordinates $x^A = (\theta,\phi)$, and the induced metric $q_{ab}$ on $\calS$ is obtained by setting $\ud v = 0$, $\ud r = 0$ and $r=r_+$
in Eq.~\eqref{eq:metric}; one obtains
\beq\label{q_AB}
    q_{AB} \, \ud x^A \ud x^B = R^2 \left[ (1-\beta^2 \sin^2{\theta}) \, \ud \theta^2 + \frac{\sin^2{\theta}}{1-\beta^2 \sin^2{\theta}} \, \ud\phi^2 \right] ,
\eeq
where $R = \sqrt{r_+^2 + a^2} = \sqrt{2Mr_+}$ is the areal radius \eqref{def_areal_radius} and
\beq\label{def_beta}
    \beta \equiv \frac{a}{R} = \frac{a}{\sqrt{r_+^2 + a^2}} = \frac{a}{\sqrt{2Mr_+}} = \frac{a}{\sqrt{2M(M + \sqrt{M^2 - a^2})}}
\eeq
is the dimensionless distortion parameter introduced by Smarr \cite{Sm2.73}, who pointed out that all cross-sections of the Kerr horizon are isometric, in agreement with the general NEH result derived in Sec.~\ref{subsec:def_NEH}. Notice that $\beta$ ranges from 0 to $1/\sqrt{2}$ when $a$ ranges from $0$ to $M$ and that $\beta^2 = a\varOmega_{\calH}$, with $\varOmega_{\calH}=a/(r_+^2 + a^2)$ the (constant) angular velocity of the horizon.

We read immediately from expression~\eqref{q_AB} that $\det (q_{AB}) = R^4 \sin^2\theta$, so that the area 2-form $\varepsilon_{ab}$ of $(\mathcal{S}, q_{ab})$ is simply
\beq\label{epsilon_S_Kerr}
    \bm{\varepsilon} = R^2 \sin\theta \, \bm{\ud} \theta \wedge \bm{\ud} \phi \, .
\eeq
One also deduces from Eq.~\eqref{q_AB} that the scalar curvature $\mathcal{R}$ of $(\calS, q_{ab})$ is (cf. Ref.~\cite{Sm2.73} or notebook~1 in App.~\ref{app:Sage})
\beq \label{eq:S_curvature_Kerr}
    \mathcal{R} = \frac{2 \left[ 1 - \beta^2 (1 + 3 \cos^2\theta) \right]}{R^2(1 - \beta^2\sin^2\theta)^3} \, .
\eeq
It is easy to check that
$-\mathcal{R}/4$ coincides with the real part of $\Psi_2$ given by \eqref{Psi2_Kerr}, in compliance with Eq.~\eqref{RePsi2_NEH}.

\subsection{Axisymmetry-based horizon multipoles}\label{subsec:axi_Kerr}

By introducing the coordinates $x^{A'} = (\zeta,\phi)$, with $\zeta \equiv \cos\theta$, we may recast the metric \eqref{q_AB} and area 2-form \eqref{epsilon_S_Kerr}  as, respectively,\footnote{As indicated by the minus sign in Eq.~\eqref{epsilon_S_Kerr_zeta}, which
stems from $\bm{\ud}\zeta = - \sin\theta \, \bm{\ud}\theta$, $(\partial_\zeta,\partial_\phi)$ is a left-handed vector frame of $\calS$ for the orientation
set by Eqs.~\eqref{Seps_pullback_Heps} and \eqref{def_area_form_H}, given that $(\partial_v,\partial_r,\partial_\theta,\partial_\phi)$ is a right-handed vector frame of $\calM$.}
\beq \label{q_Kerr_zeta}
 q_{A'B'} \, \ud x^{A'} \ud x^{B'} = R^2 \left[ \left( \frac{1}{1-\zeta^2} - \beta^2 \right) \ud\zeta^2
    +  \left( \frac{1}{1-\zeta^2} - \beta^2 \right)^{-1} \! \ud \phi^2 \right] ,
\eeq
\beq\label{epsilon_S_Kerr_zeta}
    \bm{\varepsilon} = - R^2 \, \bm{\ud} \zeta \wedge \bm{\ud} \phi \, .
\eeq

We note that the metric \eqref{q_Kerr_zeta} is of the canonical form \eqref{q_in_axisym_coord} for an axisymmetric NEH, with $f = \left[ (1-\zeta^2)^{-1} - \beta^2 \right]^{-1}$. It follows that the coordinates $(\zeta,\phi)$ coincide with those of the axisymmetric construction of Sec.~\ref{subsec:axi}, and that the fiducial axisymmetry-based unit round metric \eqref{round_metric_axi} is given in terms of Kerr coordinates $(\theta,\phi)$ by simply
\beq \label{q_axi}
    \mathring{q}^{\rm axi}_{AB} \, \ud x^A \ud x^B = \ud {\theta}^2 + \sin^2\theta\,  \ud {\phi}^2 \, .
\eeq
Substituting Eqs.~\eqref{Psi2_Kerr} and \eqref{def_mathringYlm_axi} into the definition \eqref{I-L_axi} of the axisymmetry-based multipole moments, with
$\ud S =- R^2 \, \ud\zeta \, \ud\phi$ from Eq.~\eqref{epsilon_S_Kerr_zeta}, we readily find
\beq\label{I-L_Kerr_axi}
    I_\ell^{\rm axi} + \ui L_\ell^{\rm axi} = - M  \oint_\calS  \varrho_+^3 \, \mathring{Y}_{\ell,0}^{\rm axi} \, \ud S =
    - M R^2 \int_0^{2\pi} \ud\phi \int_{-1}^{1} \ud \zeta \, \varrho_+^3(\zeta) \, Y_{\ell,0}(\theta(\zeta),\phi) \, ,
\eeq
where $\varrho_+(\zeta)$ is the value of the coefficient $\varrho(r,\zeta)$ defined by \eqref{Psi2_Kerr} at $r=r_+$. Given expression \eqref{Psi2_Kerr} for $\varrho$ and the identity $Y_{\ell,0}(\theta(\zeta),\phi) = \sqrt{(2\ell+1)/(4\pi)} \, P_\ell(\zeta)$,
where $P_\ell$ is the Legendre polynomial of order $\ell$, it follows that
\beq\label{multipoles_axi}
    I_\ell^{\rm axi} + \ui L_\ell^{\rm axi} = \tfrac{1}{2} \, {(1+\hat{a}^2)}^2 \sqrt{(2\ell+1)\pi}  \int_{-1}^1 \ud\zeta \, \frac{P_\ell(\zeta)}{(1 - \ui \hat{a} \zeta)^3} \, ,
\eeq
where we have introduced the dimensionless parameter
\beq \label{def_hat_a}
    \hat{a} \equiv \frac{a}{r_+} = \frac{a}{M + \sqrt{M^2 - a^2}} \, ,
\eeq
and have made use of the identity $MR^2/r_+^3 = \tfrac{1}{2} \, (1 + \hat{a}^2)^2$.
The relation between $\hat{a}$ and the distortion parameter $\beta$ [Eq.~\eqref{def_beta}] is
\beq \label{hat_a_beta}
    \hat{a} = \frac{\beta}{\sqrt{1 - \beta^2}}  \quad \iff \quad  \beta = \frac{\hat{a}}{\sqrt{1 + \hat{a}^2}} \, .
\eeq
Note that $\hat{a}$ ranges from $0$ to $1$ when $a$ ranges from $0$ to $M$ and that,
for small spin values, $\beta \sim \hat{a} \sim \chi/2$, with $\chi  \equiv a/M = S/M^2$.

\begin{figure}[t]
    \begin{center}
        \includegraphics[width=0.6\textwidth]{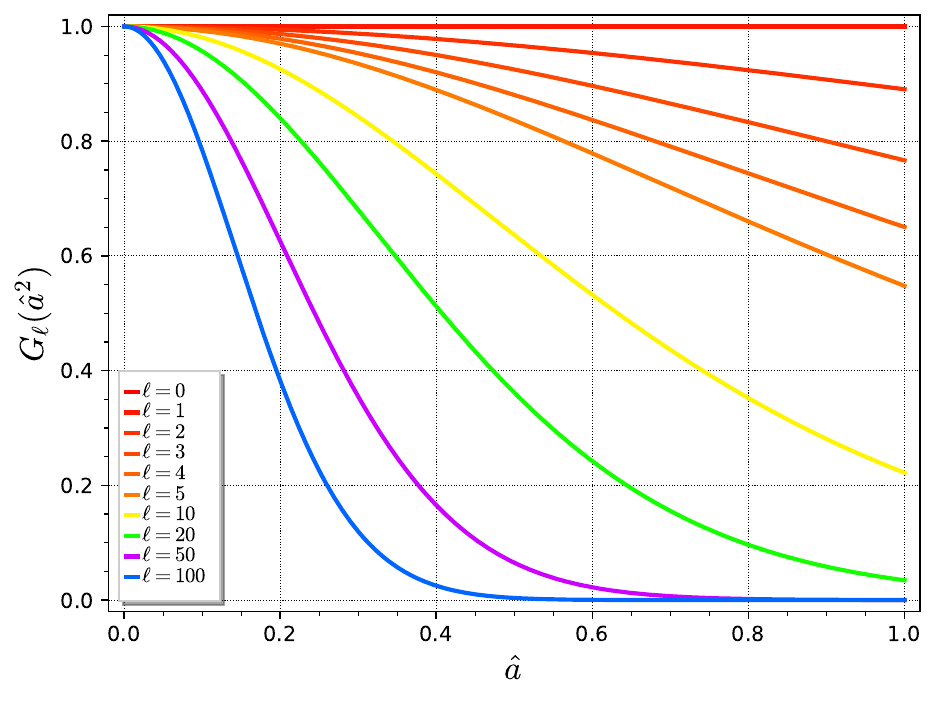}
        \caption{Function $G_\ell(\hat{a}^2)$ defined by Eq.~\eqref{bidule}, for selected values of $\ell$.}
        \label{fig:G_ell}
    \end{center}
\end{figure}

The integral in the right-hand side of Eq.~\eqref{multipoles_axi} is evaluated in App.~\ref{app:integ_Kerr_axi}, yielding
\beq \label{truc}
   I_\ell^\text{axi} + \ui L_\ell^\text{axi} = 2^{\ell-1} \sqrt{(2\ell+1)\pi} \; \frac{\ell!\, (\ell+2)!}{(2\ell+1)!} \, (\ui \hat{a})^\ell \, G_\ell(\hat{a}^2) \, ,
\eeq
where the real-valued function $G_\ell$ is defined in terms of the hypergeometric function ${}_2 F_1$ by
\beq \label{bidule}
    G_\ell(\hat{a}^2) \equiv {}_2F_1\left(\frac{\ell}{2}, \frac{\ell-1}{2},\ell+\frac{3}{2};-\hat{a}^2\right) ,
\eeq
and is plotted in Fig.~\ref{fig:G_ell}.
It is shown in App.~\ref{app:integ_Kerr_axi} that $G_\ell$ is expressible in terms of polynomials and the arctangent function as
\begin{subequations}
\label{G_ell_general_form}
\begin{align}
     & G_0(\hat{a}^2) = G_1(\hat{a}^2) = 1 \, , \label{G0_G1_1} \\
     & G_\ell(\hat{a}^2) = \frac{1}{\hat{a}^{2\ell}} \left[  \mathscr{P}_{\lceil \ell/2 \rceil}(\hat{a}^2)
        + (1 + \hat{a}^2)^2 \mathscr{Q}_{\lfloor \ell/2 \rfloor - 1}(\hat{a}^2) \, \frac{\arctan \hat{a}}{\hat{a}} \right] \qquad \ell \geqslant 2 \, ,  \label{G_l>2}
\end{align}
\end{subequations}
where $\mathscr{P}_{\lceil \ell/2 \rceil}(\hat{a}^2)$ stands for a polynomial of degree $\lceil \ell/2 \rceil$ in $\hat{a}^2$ and $\mathscr{Q}_{\lfloor \ell/2 \rfloor -1}(\hat{a}^2)$
for a polynomial of degree $\lfloor \ell/2 \rfloor -1$ in $\hat{a}^2$. Explicit expressions for these polynomials are given in \eqref{G_l_low_l} for $\ell \leqslant 5$ and
in the notebook~3 of App.~\ref{app:Sage} for $\ell \leqslant 12$ (see also notebook~4). For any value of $\ell$, such expressions
can be computed from formulas \eqref{P_Q_l_even} and \eqref{P_Q_l_odd}.
As apparent in Fig.~\ref{fig:G_ell}, despite the singular factor of $\hat{a}^{-2\ell}$ in expression \eqref{G_l>2}, $G_\ell(\hat{a}^2)$ is bounded between $0$ and $1$ for all $\hat{a}\in [0,1]$ and all $\ell\in\mathbb{N}$. Moreover, $G_\ell(0) = 1$ for all $\ell\in\mathbb{N}$, with the following small $\hat{a}$ behavior:
\beq \label{G_ell_small_ha}
    G_\ell(\hat{a}^2) = 1 - \frac{\ell(\ell-1)}{2(2\ell+3)} \, \hat{a}^2 + O(\hat{a}^4) \, .
\eeq
Inserting \eqref{G_ell_small_ha} into \eqref{truc}, we obtain the small spin behavior:
\beq\label{I-L_Kerr3-axi}
    I_\ell^\text{axi} + \ui L_\ell^\text{axi} = \sqrt{(2\ell+1)\pi} \; (\ui \hat{a})^\ell \left[ \alpha^\text{axi}_\ell + O(\hat{a}^2) \right] ,
    \quad
    \alpha^\text{axi}_\ell \equiv 2^{\ell-1}\frac{\ell!\, (\ell+2)!}{(2\ell+1)!}.
\eeq
For any spin, the factor $\ui^\ell$ in formula \eqref{truc} implies the following parity properties:
\beq\label{parity_axi}
   \forall n \in \mathbb{N},\quad  I_{2n+1}^\text{axi} = 0 \quad \text{and} \quad L_{2n}^\text{axi} = 0 \, .
\eeq
Taking into account the expressions \eqref{G_l_low_l} of $G_\ell(\hat{a}^2)$ for $0\leqslant \ell \leqslant 3$,
the values of the first non-vanishing multipole moments deduced from formula~\eqref{truc} are
\begin{subequations}
\begin{align}
    & I_0^{\rm axi} = \sqrt{\pi} \quad\mbox{and}\quad I_2^{\rm axi} = - \frac{\sqrt{5\pi}}{2\hat{a}^3} \left[3 (1 + \hat{a}^2)^2 \arctan \hat{a}  -3 \hat{a} - 5 \hat{a}^3 \right] , \\
    & L_1^{\rm axi} = \sqrt{3\pi} \, \hat{a} \quad\mbox{and}\quad L_3^{\rm axi} =  \frac{\sqrt{7\pi}}{2\hat{a}^4} \left[ 15 (1 + \hat{a}^2)^2 \arctan \hat{a} - 15 \hat{a} - 25 \hat{a}^3 - 8 \hat{a}^5 \right] .
    \label{L1_L3_axi_Kerr}
\end{align}
\end{subequations}
More values, up to $\ell=12$, can be found in the notebook~3 of App.~\ref{app:Sage}.
In particular, the monopole and dipole moments obey the general NEH relations \eqref{universal_values_axi}. Moreover, injecting the value \eqref{L1_L3_axi_Kerr} of $L_1^{\rm axi}$ in the expression \eqref{J_H_L_1_axi} for the horizon angular momentum $J_{\calH}$ and using the identities $\hat{a} = a / r_+$ and $R^2 = 2 M r_+$ leads to $J_{\calH} = a M$, i.e. one recovers the Komar angular momentum of Kerr spacetime.

\begin{figure}[t]
    \begin{center}
        \includegraphics[width=0.48\textwidth]{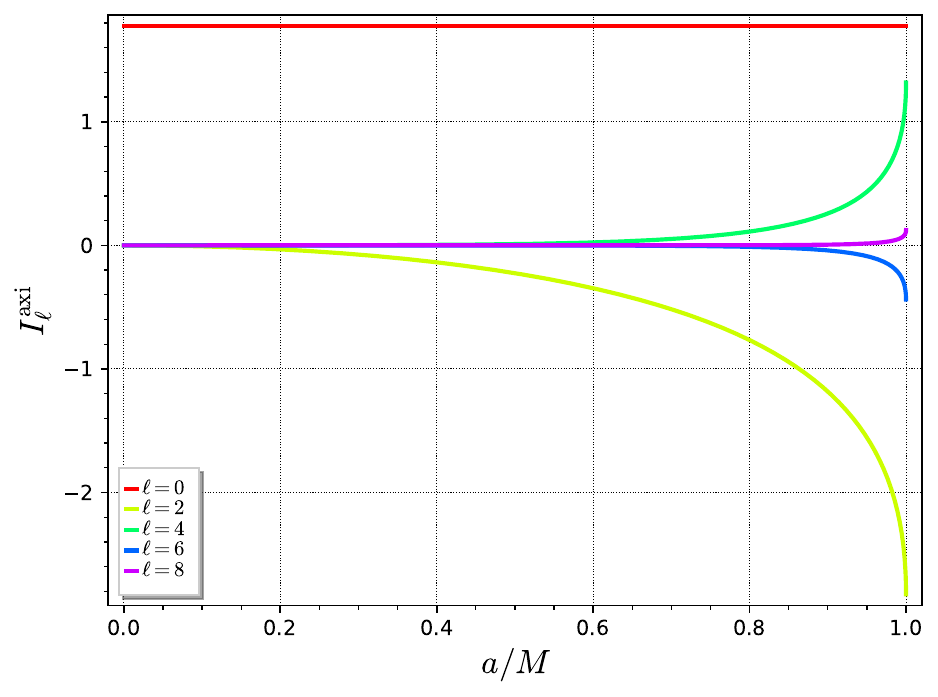}\quad
        \includegraphics[width=0.48\textwidth]{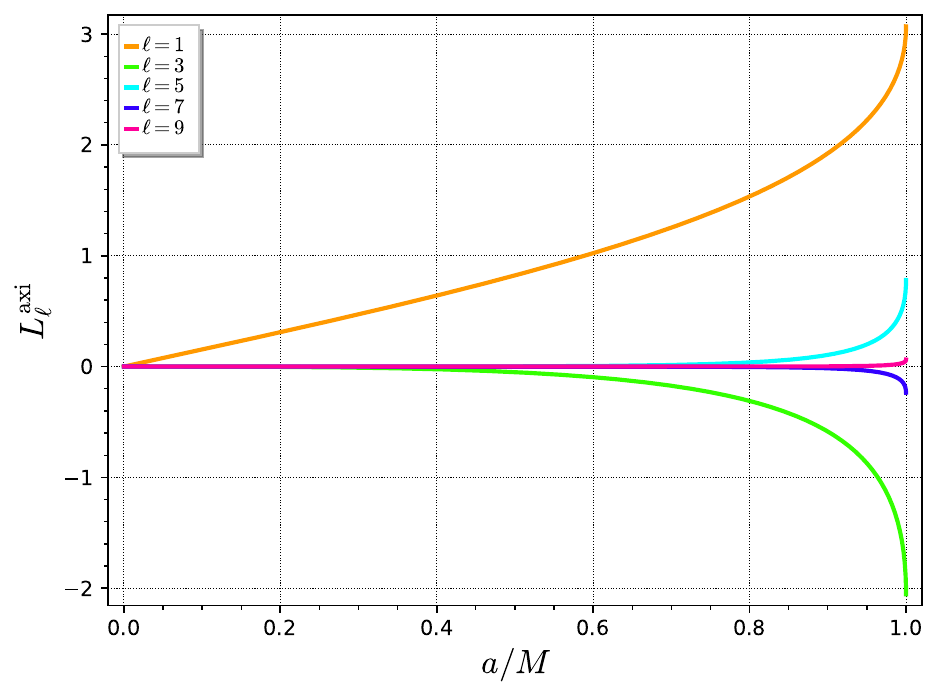}
        \caption{Shape and current horizon multipole moments $I_\ell^{\rm axi}$ and $L_\ell^{\rm axi}$ of the Kerr black hole, resulting from the axisymmetry-based definition, as functions of the Kerr spin parameter $a$.}
        \label{fig:Kerr_I_L_axi}
    \end{center}
\end{figure}

The multipoles $I_\ell^{\rm axi}$ and $L_\ell^{\rm axi}$ are plotted as functions of $a/M$ in Fig.~\ref{fig:Kerr_I_L_axi}, for $0\leqslant \ell\leqslant 9$.
For large values of $\ell$, Stirling's formula yields  $\ell!(\ell+2)!/(2\ell + 1)! \sim \sqrt{\pi} \, \ell^{3/2} / 2^{2\ell+1}$ for $\ell\to +\infty$.
Taking into account the asymptotic behavior \eqref{G_l_large_l} of the hypergeometric function $G_\ell(\hat{a}^2)$, we arrive at
\beq
    I_\ell^\text{axi} + \ui L_\ell^\text{axi} \sim \frac{\pi}{2} \frac{(1 + \hat{a}^2)^{3/4}}{(1 +  \sqrt{1 + \hat{a}^2})^{1/2}} \, \ell^2
    \left( \frac{\ui \hat{a}}{1 + \sqrt{1 + \hat{a}^2}} \right) ^\ell \quad\mbox{for}\quad \ell \to +\infty \, .
\eeq
Since $0 \leqslant \hat{a}/(1 + \sqrt{1 + \hat{a}^2}) < 1$, we deduce from this formula that, for a fixed value of $\hat{a}$, the multipole moments decay to zero when $\ell$ increases:
\beq
    \lim_{\ell\to+\infty}  I_\ell^\text{axi} + \ui L_\ell^\text{axi} = 0 \, ,
\eeq
which is consistent with the behavior observed in Fig.~\ref{fig:Kerr_I_L_axi}.

The source multipoles (cf. Sec.~\ref{subsec:mass-current}) associated with the geometric multipoles \eqref{truc} are obtained by setting $M_{\calH} = M$ in Eq.~\eqref{source}, yielding
\beq
    M_{\ell} + \ui S_{\ell} = \frac{M R^\ell}{\sqrt{(2\ell+1)\pi}} \left( I^\text{axi}_\ell + \ui \, \frac{R L^\text{axi}_\ell}{2M} \right) .
\eeq
Substituting $I^\text{axi}_\ell$ and $L^\text{axi}_\ell$ by their expressions from Eq.~\eqref{truc} and using the identities $R\hat{a} = a (1+\hat{a}^2)^{1/2}$ and
$R/(2M) = (1+\hat{a}^2)^{-1/2}$, there comes
\beq \label{source_multipoles_Kerr}
     M_{\ell} + \ui S_{\ell} =  2^{\ell-1}  \frac{\ell!\, (\ell+2)!}{(2\ell+1)!} \, (1 + \hat{a}^2)^{\lfloor\ell/2\rfloor} \, G_\ell(\hat{a}^2) \; M (\ui a)^\ell \ .
\eeq
Since $G_0(\hat{a}^2) = G_1(\hat{a}^2) = 1$ [Eq.~\eqref{G0_G1_1}], we get $M_0 = M$ and $S_1 = a M$.
In the small spin limit, we have, thanks to $G_\ell(\hat{a}^2) = 1 + O(\hat{a}^2)$ [recall Eq.~\eqref{G_ell_small_ha}],
\beq
    M_{\ell} + \ui S_{\ell} =  M (\ui a)^\ell \left[ 2^{\ell-1}  \frac{\ell!\, (\ell+2)!}{(2\ell+1)!} + O(a^2) \right] .
\eeq
Hence, to leading order in spin, the source multipole moments of a Kerr black hole share the same sign and scaling behavior as the corresponding field multipole moments, as given by the Hansen formula \eqref{Hansen}.

\begin{figure}[t]
    \begin{center}
        \includegraphics[width=0.48\textwidth]{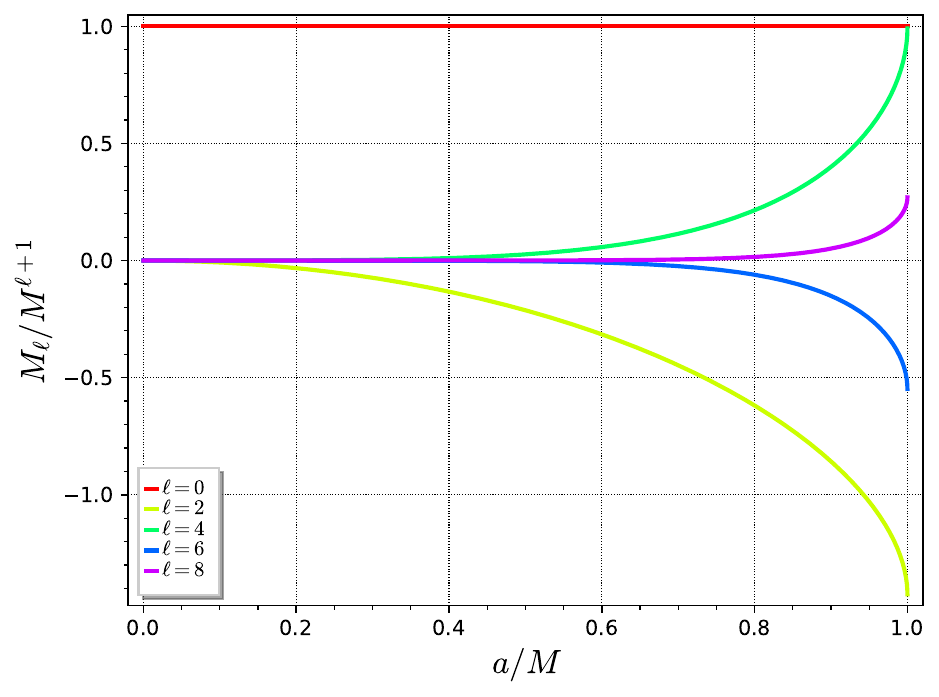}\quad
        \includegraphics[width=0.48\textwidth]{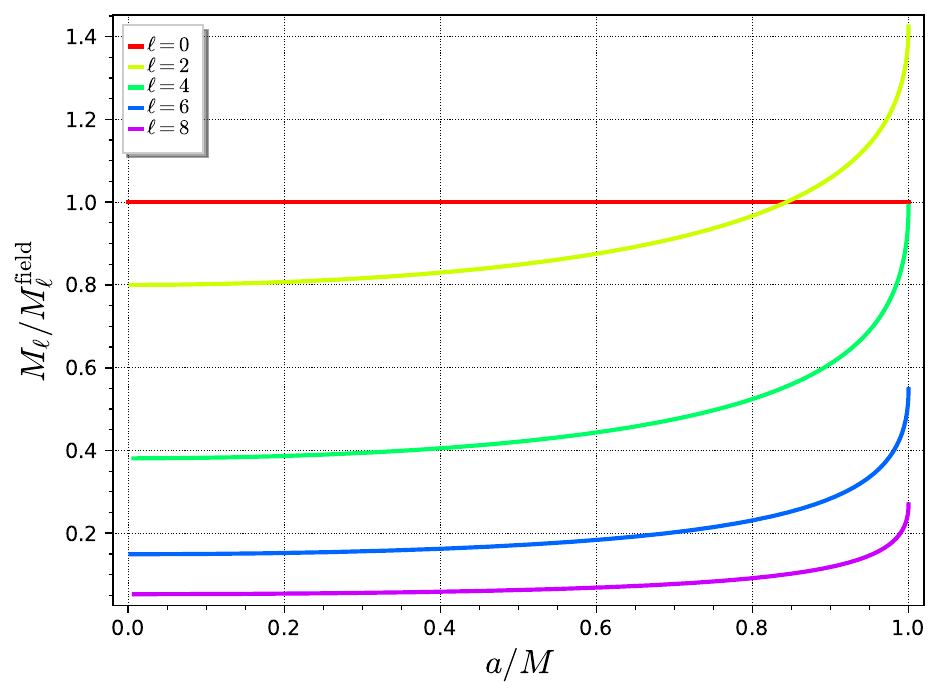}
        \caption{Mass-type source multipole moments $M_\ell$ of the Kerr horizon as functions of the Kerr spin parameter $a$. The right panel shows $M_\ell$ rescaled by the mass-type field multipole moments given by Hansen's formula \eqref{Hansen}: $M_\ell^{\rm field} = (-)^{\ell/2} M a^\ell$ for $\ell$ even.}
        \label{fig:Kerr_M_ell}
    \end{center}
\end{figure}
\begin{figure}[t]
    \begin{center}
        \includegraphics[width=0.48\textwidth]{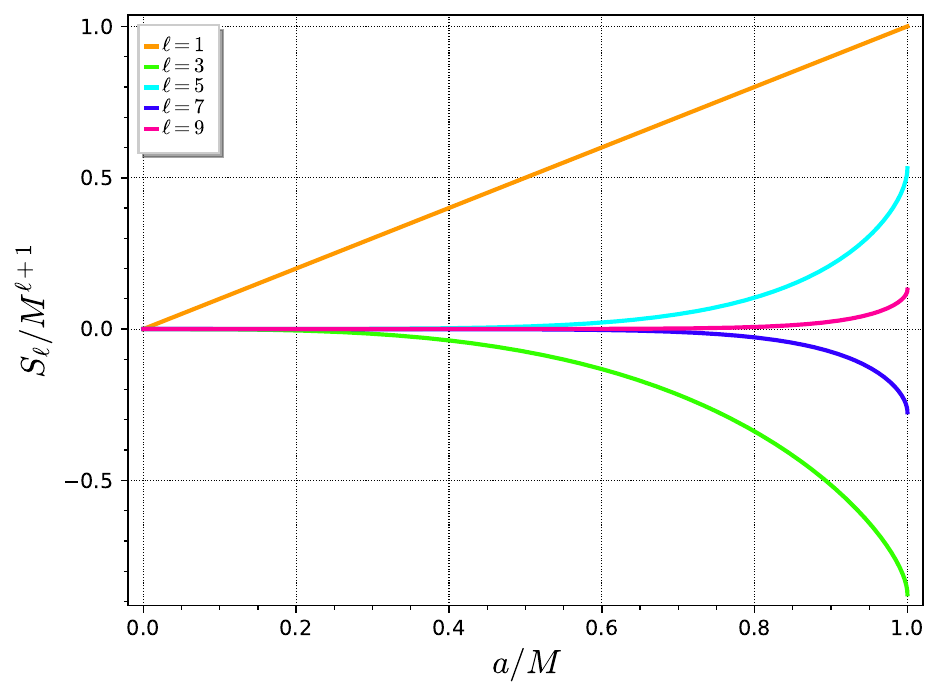}\quad
        \includegraphics[width=0.48\textwidth]{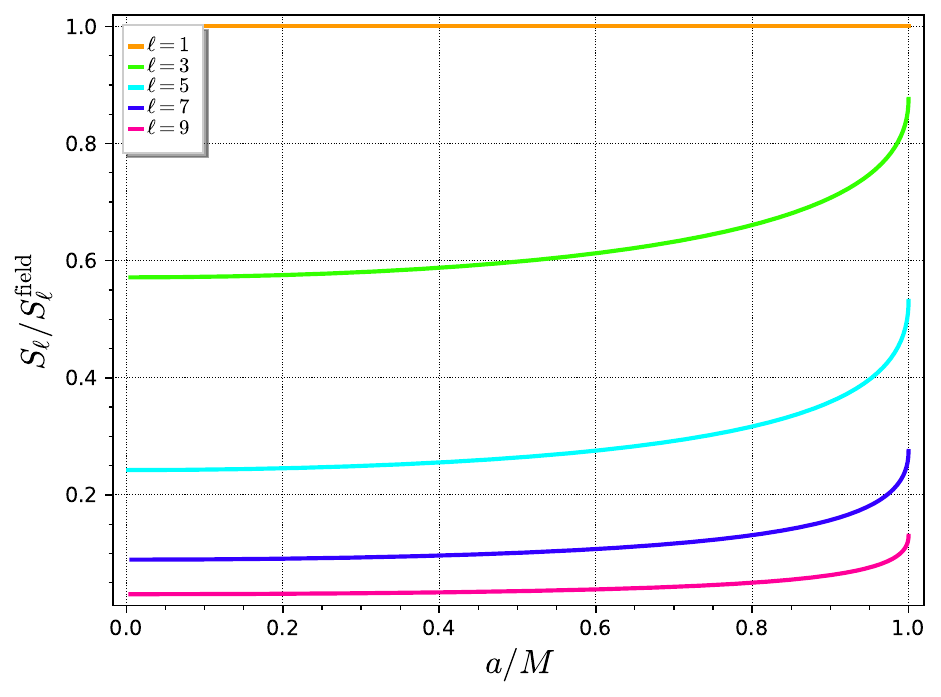}
        \caption{Current-type source multipole moments $S_\ell$ of the Kerr horizon as functions of the Kerr spin parameter $a$. The right panel shows $S_\ell$ rescaled by the current-type field multipole moments    given by Hansen's formula \eqref{Hansen}: $S_\ell^{\rm field} = (-)^{(\ell-1)/2} M a^\ell$ for $\ell$ odd.}
        \label{fig:Kerr_S_ell}
    \end{center}
\end{figure}

The source multipole moments $M_\ell$ and $S_\ell$ are plotted as functions of $a/M$ in Figs.~\ref{fig:Kerr_M_ell} and \ref{fig:Kerr_S_ell}, for $0\leqslant \ell\leqslant 9$.
The right panels of these figures display the comparison to the Hansen field multipole moments $M_\ell^{\rm field}$ and $S_\ell^{\rm field}$, as given by Eq.~\eqref{Hansen}.
Note that $M_0 = M_0^{\rm field} = M$ (the mass of Kerr spacetime) and $S_1 = S_1^{\rm field} = a M$ (the angular momentum of Kerr spacetime).
But starting from $\ell=2$, the relative discrepancy between the source and field multipoles gets quite large: for the mass quadrupole $M_2$, it is $20\%$ for $a$ small\footnote{This somehow tempers the claim made in Ref.~\cite{As.al.04} that the ``difference is insignificant.''} and up to $40\%$ for $a$ large (cf. Fig.~\ref{fig:Kerr_M_ell}),
while for the angular momentum octopole $S_3$, it is $40\%$ for $a$ small and up to $15\%$ for $a$ large (cf. Fig.~\ref{fig:Kerr_S_ell}).
Actually, the ratios $M_\ell/M_\ell^{\rm field}$ and $S_\ell/S_\ell^{\rm field}$ tend to zero for $\ell \to +\infty$. Indeed, from Eq.~\eqref{source_multipoles_Kerr}, the large-$\ell$ behaviors \eqref{G_l_large_l} and  $\ell!(\ell+2)!/(2\ell + 1)! \sim \sqrt{\pi} \, \ell^{3/2} / 2^{2\ell+1}$, we get
\beq
    \frac{M_\ell}{M_\ell^{\rm field}} \quad\mbox{or}\quad \frac{S_\ell}{S_\ell^{\rm field}}
    \sim \sqrt{\pi} \left(\frac{\ell}{2}\right) ^{3/2} \frac{(1 + \hat{a}^2)^{\lfloor\ell/2\rfloor + 3/4}}{(1 + \sqrt{1 + \hat{a}^2})^{\ell+1/2}} \quad\mbox{for}\quad \ell \to +\infty \, ,
\eeq
which tends to zero for $\ell \to +\infty$, whatever the value of $\hat{a}^2 \in [0,1]$.

\subsection{Unit round metric conformal to the physical one}\label{subsec:conformal}

Let us now move to the computation of the horizon multipole moments according to the generic definition discussed in Sec.~\ref{sec:generic_multipoles}. The first step is to determine the
unit round metric $\mathring{q}_{ab}$ conformal to the metric $q_{ab}$ induced by $g_{ab}$ on the cross-section $\calS$ [Eq.~\eqref{conformal_q}].
The physical metric $q_{ab}$ is given in terms of the Kerr coordinates $x^A = (\theta,\phi)$ by Eq.~\eqref{q_AB}.
On the other hand, any round metric $\mathring{q}_{ab}$ on $\calS$ is given in terms of adapted coordinates $x^{\mathring{A}} = (\vartheta,\varphi)$ by \eqref{canonical}.
Let us search for a metric $\mathring{q}_{ab}$ that  is compatible with the Kerr axisymmetry, i.e. that has the same rotational Killing vector $\partial_\phi$;
this yields $\varphi = \phi + \mathrm{const}$. Moreover, as we shall
see below, this choice restricts the degrees of freedom on $\mathring{q}_{ab}$ towards the vanishing area dipole metric advocated in Sec.~\ref{subsec:dipole}.
By introducing the coordinates $x^{\mathring{A}'} = (z,\phi)$ with $z \equiv \cos\vartheta$, we may recast Eq.~\eqref{canonical} (with $\varphi = \phi + \mathrm{const}$) as
\beq
    \mathring{q}_{\mathring{A}'\mathring{B}'} \, \ud x^{\mathring{A}'} \ud x^{\mathring{B}'} = \frac{\ud z^2}{1 - z^2} + (1 - z^2) \, \ud \phi^2  \, .
\eeq
By comparing with expression~\eqref{q_Kerr_zeta} of $q_{ab}$ in terms of $x^{A'} = (\zeta,\phi)$, we see that the
conformal relation $\mathring{q}_{ab} = \psi^2 q_{ab}$ [Eq.~\eqref{conformal_q}] is equivalent to
the following system of two equations:
\begin{subequations}\label{system}
    \begin{align}
        \frac{\ud z^2}{1 - z^2} &= \psi^2 R^2 \left( \frac{1}{1-\zeta^2} - \beta^2 \right) \ud \zeta^2 \, , \label{conf_ident1} \\
        1 - z^2 &= \psi^2 R^2 \left( \frac{1}{1-\zeta^2} - \beta^2 \right) ^{-1} . \label{conf_ident2}
    \end{align}
\end{subequations}
Extracting $\psi^2 R^2$ from Eq.~\eqref{conf_ident2}, substituting into Eq.~\eqref{conf_ident1}
and taking the square root while assuming that $z$ is an increasing function of $\zeta$, we get
\beq\label{trick}
  \frac{\ud z}{1 - z^2} =  \left( \frac{1}{1 - \zeta^2} - \beta^2 \right) \ud \zeta \, .
\eeq
\begin{figure}[t]
    \begin{center}
        \includegraphics[width=0.45\textwidth]{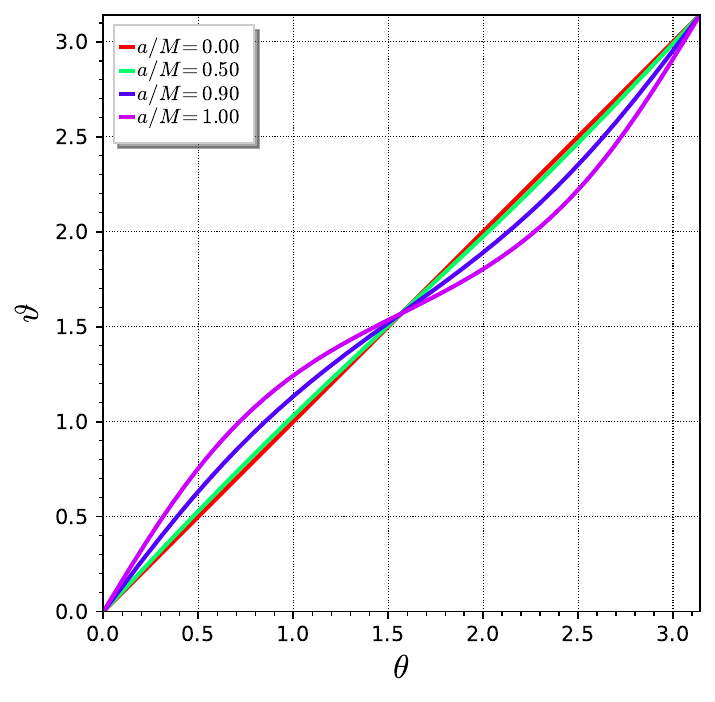}
        \caption{Colatitude coordinate $\vartheta$ adapted to the unit round metric of cross-sections of the Kerr horizon, as a function of the Kerr coordinate $\theta$, for various values of the Kerr spin parameter $a$.}
        \label{fig:vartheta}
    \end{center}
\end{figure}

The integration leads to $\mathrm{artanh}\, z = \mathrm{artanh}\, \zeta - \beta^2 \zeta $,
where the integration constant has been set to zero to make $z$ an odd function of $\zeta$.
Hence, we get the equivalent closed-form expressions
\beq \label{z-zeta}
  z(\zeta) = \tanh \left( \mathrm{artanh}\, \zeta - \beta^2 \zeta \right) =  \frac{\zeta - \tanh{(\beta^2 \zeta)}}{1 - \zeta \, \tanh{(\beta^2 \zeta)}} = \frac{1 + \zeta - (1 - \zeta) \mathrm{e}^{2\beta^2 \zeta}}{1 + \zeta + (1 - \zeta) \mathrm{e}^{2\beta^2 \zeta}} \, .
\eeq
Notice that \eqref{system} imply $z'(\zeta) = R^2 \psi^2$. The relation between $\vartheta = \arccos z$ and $\theta = \arccos \zeta$ resulting from \eqref{z-zeta} is depicted in Fig.~\ref{fig:vartheta}. It shows a linear relation increasingly modulated as $a \to M$; for $a=M$, the discrepancy $|\vartheta - \theta|$ is at most $0.28$, with a relative difference $|\vartheta - \theta|/\theta$
reaching $0.65$ for small values of $\theta$.

The expression for the conformal factor $\psi$ is deduced from Eq.~\eqref{conf_ident2}. Using $\zeta = \cos\theta$, $R^2 = r_+^2 + a^2$ and $\beta^2 = a^2/(r_+^2 + a^2)$ [Eq.~\eqref{def_beta}], we may cast it as
\beq\label{conformal}
    \psi = \frac{\sqrt{(r_+^2 + a^2 \cos^2\theta)(1-z^2)}}{(r_+^2 + a^2)\sin\theta} \, .
\eeq
For a given $z$, expression \eqref{conformal} for $\psi$ agrees with formula (A.8) in Ref.~\cite{As.al.22}. However, the function $z(\theta)$ that we found
[Eq.~\eqref{z-zeta} with $\zeta=\cos\theta$]
is different (and simpler) from that given in Ref.~\cite{As.al.22} (which is denoted $Z$ there). We checked that the conformal factor \eqref{conformal} is a solution of the partial differential equation \eqref{plout}, and that the scalar curvature of the metric $\psi^2 q_{ab}$ is equal to 2, as it should be (see below). Consequently we believe that there is a typo in the expression for $z(\theta)$ in Eq.~(A.8) of Ref.~\cite{As.al.22}.

We deduce from \eqref{z-zeta} that
$\sin{\theta} / \sqrt{1 - z^2} = \cosh{(\beta^2\cos{\theta})} - \cos{\theta} \, \sinh{(\beta^2\cos{\theta})}$, so that Eq.~\eqref{conformal}
can be turned into an explicit expression for $\psi$ in terms of $\theta$:
\beq\label{psi_Kerr}
    R \, \psi(\theta) = \frac{\sqrt{1 - \beta^2 \sin^2{\theta}}}{\cosh{(\beta^2\cos{\theta)}} - \cos{\theta} \, \sinh{(\beta^2 \cos{\theta})}} \, .
\eeq
By combining this result with Eq.~\eqref{q_AB}, we get the components of the unit round metric $\mathring{q}_{ab} = \psi^2 q_{ab}$ with
respect to the Kerr angular coordinates $x^A = (\theta,\phi)$:
\beq  \label{round_metric_Kerr}
    \mathring{q}_{AB} \, \ud x^A \ud x^B =
    \frac{(1 - \beta^2 \sin^2\theta)^2 \, \ud\theta^2 + \sin^2\theta \, \ud\phi^2}{\left[ \cosh{(\beta^2\cos{\theta})} - \cos{\theta} \, \sinh{(\beta^2\cos{\theta})} \right]^2} \, .
\eeq
One can check that the above metric has a constant scalar curvature equal to 2, as expected (see notebook~1 in App.~\ref{app:Sage}).
Notice that expression \eqref{round_metric_Kerr} is not the `canonical' one \eqref{canonical} for a unit round metric, except for $\beta=0$ (i.e. $a=0$). In other words, the Kerr angular coordinates $(\theta,\phi)$ are \textit{not} polar coordinates adapted to $\mathring{q}_{ab}$, contrary to $(\vartheta,\varphi)$.

It turns out the obtained round metric \eqref{round_metric_Kerr} is the `canonical' one among the family of unit round metrics conformally related to $q_{ab}$, i.e., it fulfills the criterion of vanishing area dipole moment introduced in Sec.~\ref{subsec:dipole}. Indeed, by definition, the modes \eqref{d modes} of the area dipole moment read
\beq
    d_{1,m} = \oint_{\calS} \mathring{Y}_{1,m} \, \ud S = R^2 \int_0^{2\pi} \ud \phi \int_{-1}^1 \ud \zeta \; Y_{1,m}(z(\zeta),\phi) \, .
\eeq
The nonaxisymmetric modes $m = \pm 1$ cancel out by averaging over $\phi$ of the periodic function $\exp{(\pm \ui \phi)}$. For the axisymmetric mode $m = 0$, we have instead $Y_{1,0}(z(\zeta),\phi) \propto z(\zeta)$. According to Eq.~\eqref{z-zeta}, the function $z(\zeta)$ is of odd parity [thanks to the choice of a zero integration constant while deriving Eq.~\eqref{z-zeta}], which readily implies $d_{1,0} = 0$.

According to the analysis in Sec.~\ref{subsec:round}, any other unit round metric $\mathring{q}'_{ab}$ conformally related to $q_{ab}$ must obey $\mathring{q}'_{ab} = \alpha^2 \mathring{q}_{ab}$, where $\alpha$ is given in terms of the three free parameters $\vec{\alpha} = (\alpha_1,\alpha_2,\alpha_3)$ by \eqref{alpha}. The corresponding conformal factor is then $\psi' \!=\! \alpha \psi$. While $(\mathring{q}_{ab},\psi)$ in \eqref{round_metric_Kerr} and \eqref{psi_Kerr} manifestly respect the continuous and discrete symmetries of the Kerr horizon cross-sections, namely the axisymmetry and symmetry with respect to the equatorial plane, a generic pair $(\mathring{q}'_{ab},\psi')$ does not, as $\alpha_1 \neq 0$ or $\alpha_2 \neq 0$ (resp. $\alpha_3 \neq 0$) explicitly breaks the continuous (resp. discrete) symmetry.

\begin{figure}[t]
    \begin{center}
        \includegraphics[width=0.55\textwidth]{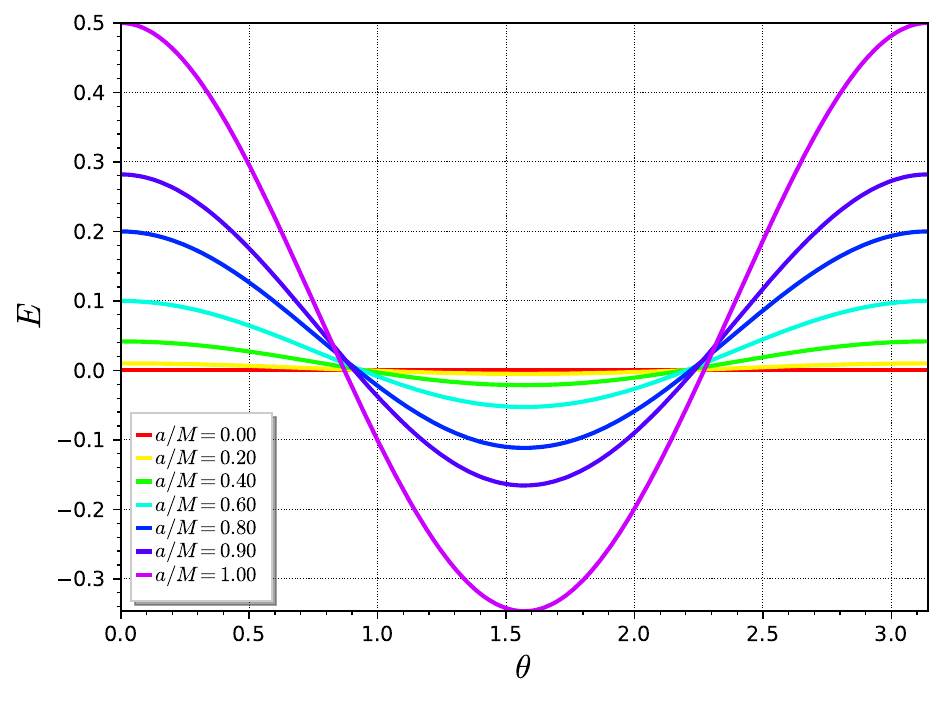}
        \caption{Electric potential $E = \ln(R\psi)$  of the Kerr horizon, resulting from Eq.~\eqref{psi_Kerr}, for various values of the Kerr spin parameter $a$.}
        \label{fig:E_theta}
    \end{center}
\end{figure}

The logarithm of $R \, \psi(\theta)$, as given by Eq.~\eqref{psi_Kerr}, i.e. the function $E$ defined by Eq.~\eqref{def_E_electric}, is plotted in Fig.~\ref{fig:E_theta}. Note that $\psi$ is an even function of the Kerr parameter $a$, and consequently does not depend on the direction of the black hole spin. It is also manifestly independent of $\phi$, thus reflecting the axisymmetry of the Kerr metric \eqref{eq:metric}, and is invariant under $\theta \to \pi - \theta$, corresponding to a reflection across the equatorial plane $\theta = \pi / 2$. For a Schwarzschild black hole, $\beta = 0$ and $R = 2M$ so $\psi = 1/(2M)$ is constant (spherical symmetry). For a slowly spinning black hole, we have
\beq\label{psi_small_a}
    R \, \psi(\theta) = 1 + \beta^2 P_2(\cos{\theta}) + O(\beta^4) \, ,
\eeq
where $P_2(\cos{\theta})$ is the Legendre polynomial of order $\ell = 2$, which captures an axisymmetric, quadrupolar deviation from spherical symmetry, with amplitude $\beta^2$. For an extremal Kerr black hole ($a=M$) we have $r_+ = M$, and thus $R = \sqrt{2} M$ and $\beta^2 = \tfrac{1}{2}$. Then $M \, \psi(\theta) = f(\theta)$, where $f$ is a strictly decreasing function varying in $[\sqrt{e},1/\sqrt{2}]$ over the interval $\theta \in [0,\pi/2]$.

\begin{figure}[t]
    \begin{center}
        \includegraphics[width=0.47\textwidth]{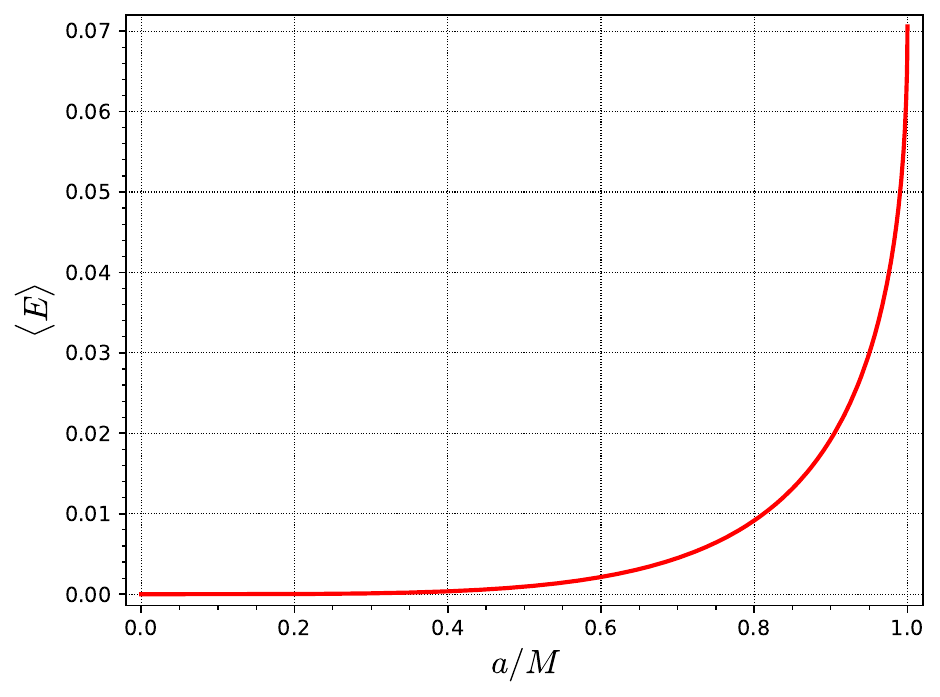}\quad
        \includegraphics[width=0.47\textwidth]{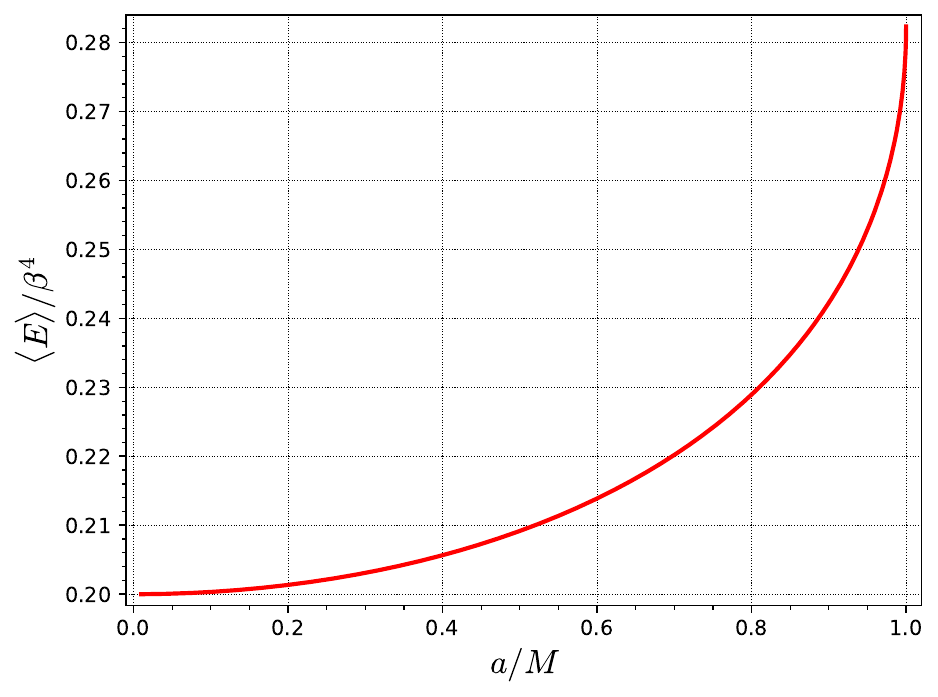}
        \caption{Mean value $\langle E \rangle$ of $E$ as a function of the Kerr spin parameter $a$; the right panel shows the ratio $\langle E \rangle / \beta^4$.}
        \label{fig:E_mean}
    \end{center}
\end{figure}

The average $\langle E \rangle$ of the electric potential $E$ with respect to the measure $\mathring{\varepsilon}_{ab}$ is given by
\beq
    \langle E \rangle \equiv \frac{1}{4\pi} \oint_\calS E \, \ud \mathring{S} =  \frac{1}{4\pi} \oint_\calS E \, \psi^2 \, \ud S =  \frac{1}{2} \int_{-1}^1 \ln{(R\psi)} \, (R\psi)^2 \, \ud (\cos{\theta}) = \frac{1}{5} \, \beta^4 + O(\beta^6) \, ,
\eeq
which vanishes `quickly' as $\beta \to 0$ (small-spin regime), as can be seen in Fig.~\ref{fig:E_mean}. Moreover, the right panel shows that the dependence of $\langle E \rangle / \beta^4$ over the Kerr spin parameter $a$ is weak. Notice that $\langle E \rangle$ is small, in the sense that $\langle E \rangle \simeq 0.07$ at most for $a=M$ while $\mathrm{max}\, E = 0.5$ for that value of $a$.

\subsection{Magnetic-type potential}\label{subsec:magnetic}

The \Hajicek{} 1-form $\Omega_a$ associated to the slicing of the Kerr horizon by the $v=\mathrm{const}$ cross-sections (cf. Sec.~\ref{s:rotation_form}) is
expressed in terms of the coordinates $x^{A'} = (\zeta,\phi)$ by
\beq \label{Hajicek_Kerr}
        \Omega_{A'} \, \ud x^{A'} = \frac{\hat{a}}{1 + \hat{a}^2 \zeta^2} \left[ \hat{a} \zeta \, \ud \zeta
            - \frac{3 + \hat{a}^2 + \hat{a}^2 ( 1 - \hat{a}^2) \zeta^2}{2(1 + \hat{a}^2\zeta^2)} (1 - \zeta^2) \, \ud \phi \right] ,
\eeq
where we have let the parameter $\hat{a} \equiv a / r_+ = \beta / \sqrt{1 - \beta^2}$ [Eqs.~\eqref{def_hat_a}--\eqref{hat_a_beta}] appear instead of $\beta$.
Formula \eqref{Hajicek_Kerr} can be found by setting $\theta = \arccos \zeta$ in Eqs.~(D33)--(D34) of Ref.~\cite{GoJa.06}. It
can also be derived directly from expression \eqref{Hajicek_n_nabla_l_q} of $\Omega_a$ in terms of the null vectors $\ell^a$ and $n^a$ generating at each point the 2-plane orthogonal to $\calS$, with $\ell^a \eqNEH (\partial_v)^a + \varOmega_{\calH} (\partial_\phi)^a$;
cf. notebook~2 in App.~\ref{app:Sage}.

The 1-form $\Omega_a$ is not divergence-free. However, by defining $\hat{\Omega}_a \equiv \Omega_a + D_a \ln f$, with $f \equiv (1 + \hat{a}^2 \zeta^2)^{-1/2}$, we get
a divergence-free 1-form: $D^a \hat{\Omega}_a = 0$ [Eq.~\eqref{Omega_divfree}]. Note that $\hat{\Omega}_a$ is the \Hajicek{} 1-form corresponding to the null normal
$\hat{\ell}^a = f {\ell}^a$.
Explicitly, we deduce from  Eq.~\eqref{Hajicek_Kerr} and the above expression of $f$ that
\beq\label{hat_Omega_Kerr}
    \hat{\Omega}_{A'} \, \ud x^{A'} = - \frac{\hat{a}(3 + \hat{a}^2 + \hat{a}^2 ( 1 - \hat{a}^2) \zeta^2)}{2(1 + \hat{a}^2\zeta^2)^2} (1 - \zeta^2) \, \ud \phi  \, .
\eeq
\begin{figure}[t]
    \begin{center}
        \includegraphics[width=0.55\textwidth]{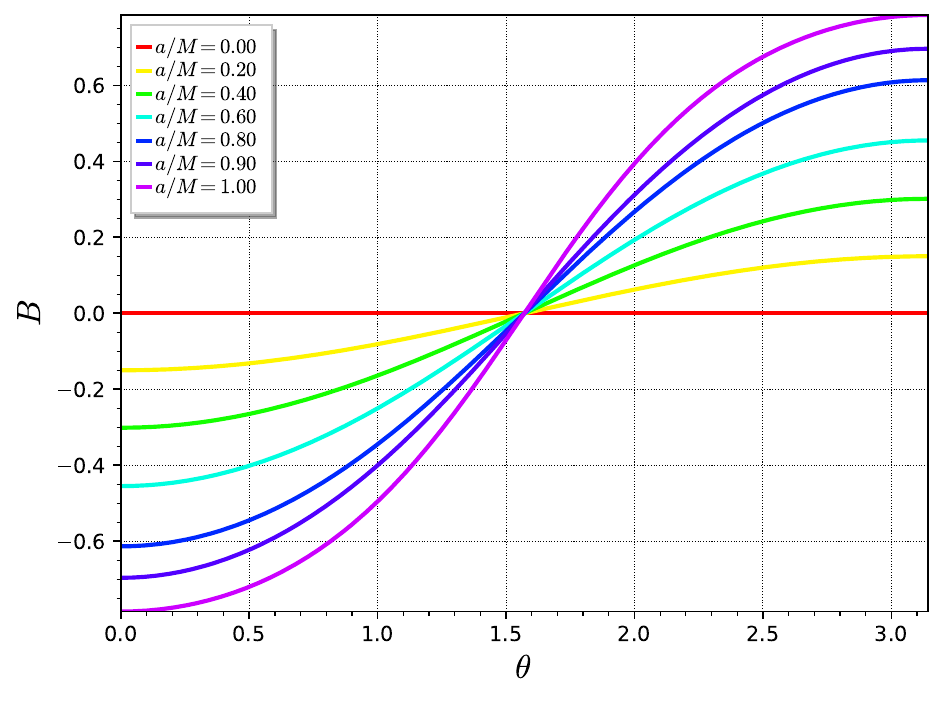}
        \caption{Magnetic potential $B$  of the Kerr horizon, as given by Eq.~\eqref{B_Kerr} with $\zeta=\cos\theta$, for various values of the Kerr spin parameter $a$.}
        \label{fig:B_theta}
    \end{center}
\end{figure}

By definition, the magnetic potential $B$ obeys Eq.~\eqref{bar_Omega_coexact}: $q^{bc} \varepsilon_{ac} D_b B = \hat{\Omega}_a$,
where $\varepsilon_{ab}$ is the area 2-form of $(\calS,q_{ab})$, as given by Eq.~\eqref{epsilon_S_Kerr_zeta}.
Thanks to the diagonal form of the metric components $q_{A'B'}$ [recall Eq.~\eqref{q_Kerr_zeta}], Eq.~\eqref{bar_Omega_coexact} is equivalent to the system
$q^{\phi\phi} \varepsilon_{\zeta\phi} \partial_\phi B = \hat{\Omega}_\zeta$ and
$q^{\zeta\zeta} \varepsilon_{\phi\zeta} \partial_\zeta B = \hat{\Omega}_\phi$, with $\varepsilon_{\zeta\phi} = - \varepsilon_{\phi\zeta} = -R^2$.
Given the components \eqref{q_Kerr_zeta} and \eqref{hat_Omega_Kerr} of respectively $q_{ab}$ and $\hat{\Omega}_a$, as well as the identity
$\beta^2 = \hat{a}^2 / (1+\hat{a}^2)$, we get $ \partial_\phi B = 0$ and
\[
    \left( \frac{1}{1-\zeta^2} - \frac{\hat{a}^2}{1+\hat{a}^2} \right) ^{-1} \frac{\partial B}{\partial\zeta} =  - \frac{\hat{a}(3 + \hat{a}^2 + \hat{a}^2 ( 1 - \hat{a}^2) \zeta^2)}{2(1 + \hat{a}^2\zeta^2)^2} \, (1 - \zeta^2) \, .
\]
After simplification, there comes
\beq
    \frac{\partial B}{\partial\zeta} = - \frac{\hat{a}}{1 + \hat{a}^2\zeta^2} - \frac{\hat{a}(1 - \hat{a}^2)}{2(1 + \hat{a}^2)} \, .
\eeq
In view of $ \partial_\phi B = 0$, the integration is immediate:
\beq\label{B_Kerr}
    B = - \arctan{(\hat{a} \zeta)} - \frac{1 - \hat{a}^2}{2(1 + \hat{a}^2)} \, \hat{a} \zeta \, .
\eeq
The integration constant (which has no physical significance since $B$ is a potential) has been set to zero. This makes $B$ an odd function of $\zeta$. Moreover, $B$ is an odd function of $\hat{a}$ that scales as $\sim \hat{a}$ for small spin values. In App.~\ref{app:magnetic} we provide an alternative proof of the key result \eqref{B_Kerr}. This equation yields $B$ as a function of $\theta = \arccos{\zeta}$, which we plot  in Fig.~\ref{fig:B_theta}.

\subsection{Horizon multipole moments}\label{subsec:moments}

\begin{figure}[t]
    \begin{center}
        \includegraphics[width=0.48\textwidth]{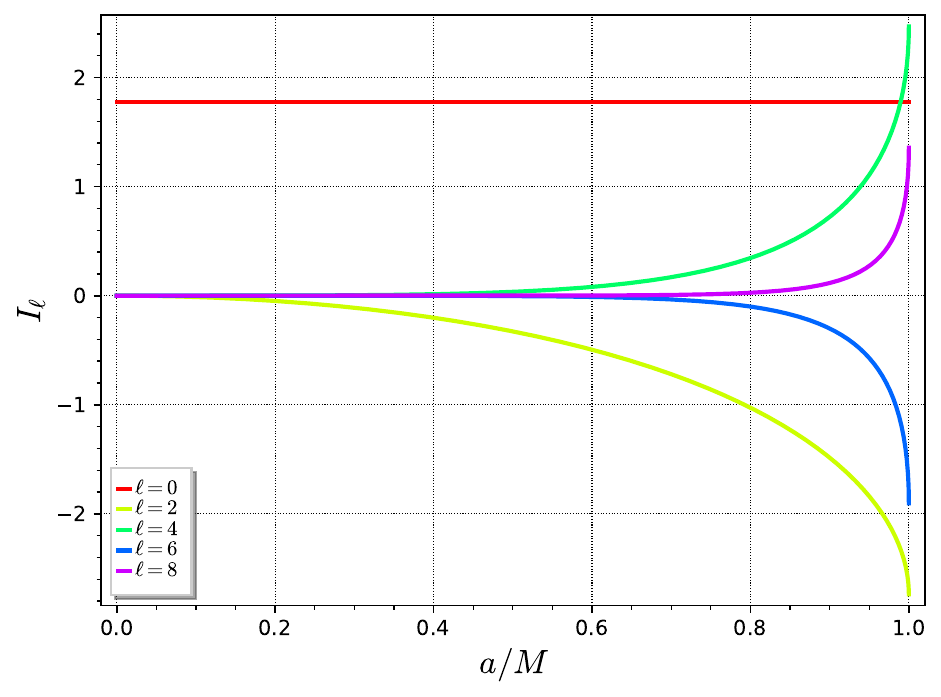}\quad
        \includegraphics[width=0.48\textwidth]{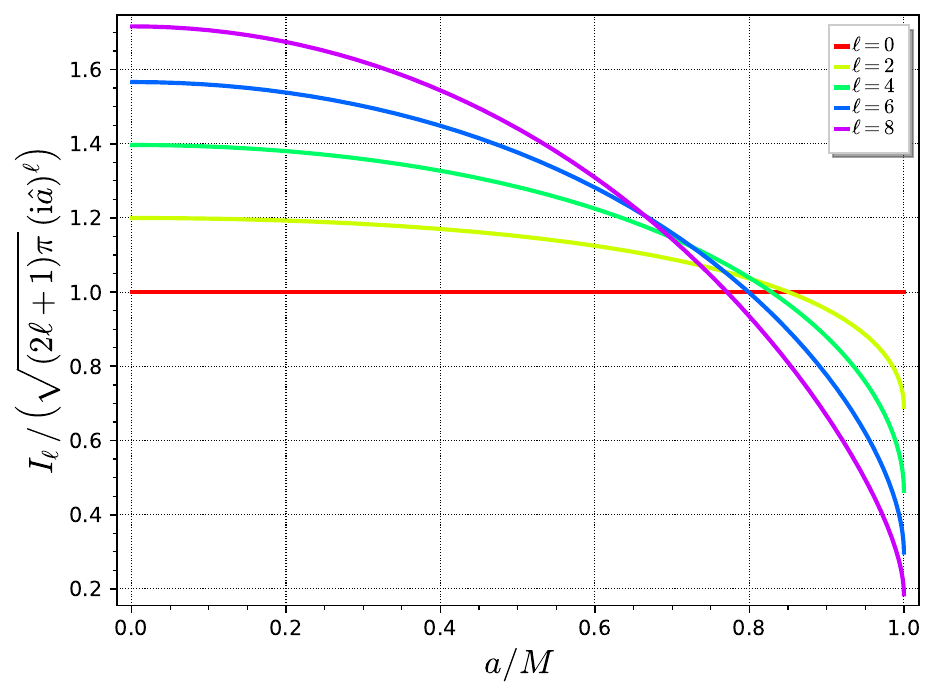}
        \caption{Shape multipole moments $I_{\ell}$ of the Kerr horizon, as functions of the Kerr spin parameter $a$; the right panel depicts $I_{\ell}$ rescaled by $\sqrt{(2\ell+1)\pi} \, (\ui \hat{a})^\ell$.}
        \label{fig:Kerr_I}
    \end{center}
\end{figure}

\begin{figure}[t]
    \begin{center}
        \includegraphics[width=0.48\textwidth]{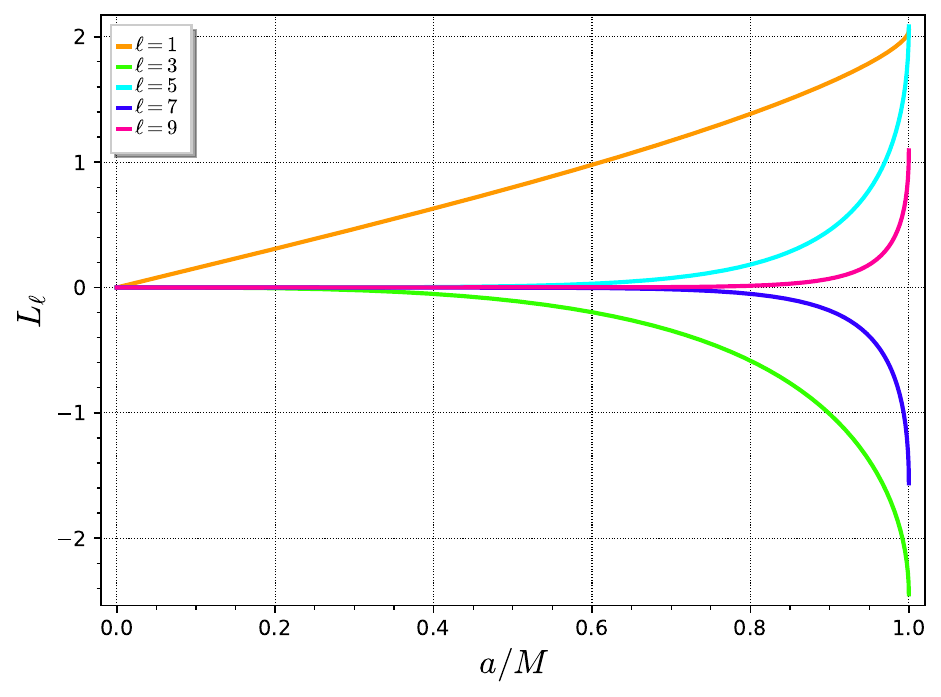}\quad
        \includegraphics[width=0.48\textwidth]{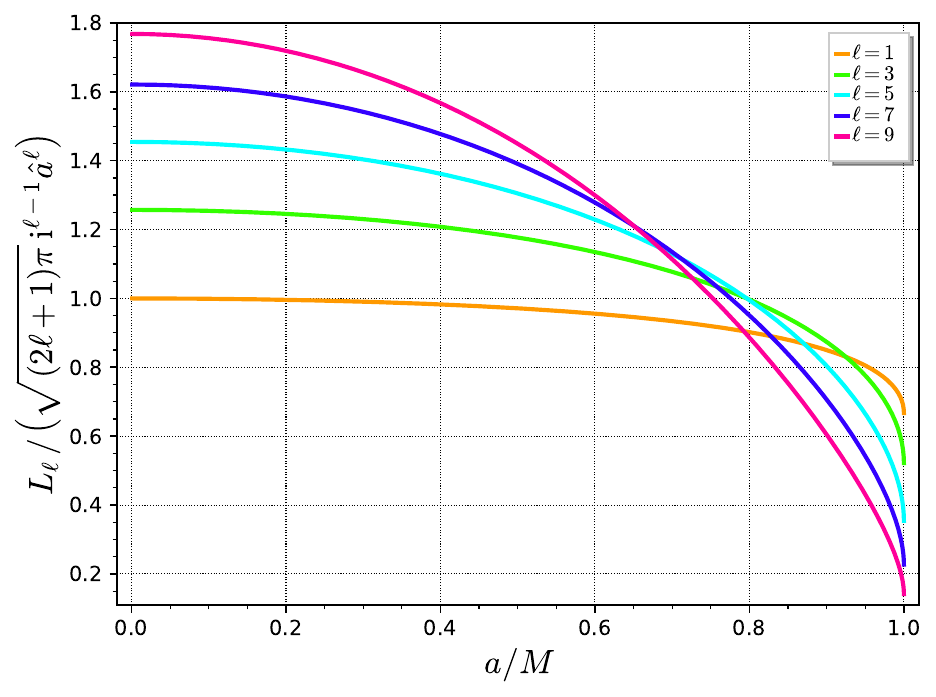}
        \caption{Current multipole moments $L_{\ell}$ of the Kerr horizon, as functions of the Kerr spin parameter $a$; the right panel depicts $L_{\ell}$ rescaled by $\sqrt{(2\ell+1)\pi} \; \ui^{\ell-1} \hat{a}^\ell$.}
        \label{fig:Kerr_L}
    \end{center}
\end{figure}

Substituting \eqref{Psi2_Kerr} and \eqref{def_mathringYlm} into the generic-NEH definition \eqref{I+iL} of the complex-valued multipole moments $K_{\ell m}$, we readily find
\beq\label{I-L_Kerr}
    K_{\ell,m} = - M  \oint_\calS \varrho_+^3 \, \mathring{Y}_{\ell,m}\, \ud S  =
    - M R^2 \int_0^{2\pi} \ud\phi \int_{-1}^{1} \ud \zeta \, \varrho_+^3(\zeta) \, Y_{\ell,m}(\vartheta(\zeta),\phi) \, ,
\eeq
where we have used Eq.~\eqref{epsilon_S_Kerr_zeta} to write $\ud S = - R^2 \, \ud\zeta \, \ud\phi$. Since $\varrho_+$, as given by Eq.~\eqref{Psi2_Kerr} at $r = r_+$, does not depend on the angle $\phi$, we deduce
\beq\label{I-L_Kerr2}
    K_{\ell,m} = (I_\ell + \ui L_\ell) \, \delta_{m,0} \, , \quad \text{with} \quad I_\ell + \ui L_\ell = \tfrac{1}{2} \, {(1+\hat{a}^2)}^2 \sqrt{(2\ell+1)\pi}  \int_{-1}^1 \ud\zeta \, \frac{P_\ell(z(\zeta))}{(1 - \ui \hat{a} \zeta)^3} \, ,
\eeq
where the function $z(\zeta)$ is given by Eq.~\eqref{z-zeta} and $\hat{a}  \equiv a / r_+$ [recall Eq.~\eqref{def_hat_a}].
We have used the identities $MR^2/r_+^3 = \tfrac{1}{2} \, (1 + \hat{a}^2)^2 $ and $Y_{\ell,0}(\vartheta(\zeta),\phi) = \sqrt{(2\ell + 1)/(4\pi)} \, P_\ell(z(\zeta))$.
The quantities $I_\ell$ and $L_\ell$ introduced in \eqref{I-L_Kerr2} are nothing but the multipole moments
$I_{\ell,0}$ and $L_{\ell,0}$ of the Kerr horizon [cf. Eq.~\eqref{I+iL}].
Notice that formula \eqref{I-L_Kerr2} is identical to formula \eqref{multipoles_axi} for the axisymmetry-based multipole moments, except for $\zeta$ replaced by
the function $z(\zeta)$ in the argument of the Legendre polynomial $P_\ell$.
This function is too complicated to allow us to express the integral in \eqref{I-L_Kerr2} in terms of known functions like the hypergeometric ones used in Sec.~\ref{subsec:axi_Kerr}
for $I_\ell^{\rm axi}$ and $L_\ell^{\rm axi}$, except in the small $\hat{a}$ limit [Eqs.~\eqref{I-L_Kerr3}--\eqref{alpha_ell} below].
Instead, we have written a SageMath code capable of numerically evaluating $I_\ell$ and $L_\ell$ to an arbitrary precision\footnote{Having a precision beyond the usual 16 digits (``double precision'')
is required to numerically evaluate quantities like $I_\ell/(\ui \hat{a})^\ell$ (right panel of Fig.~\ref{fig:Kerr_I}) for small $\hat{a}$ and $\ell$ larger than $\sim 5$.}
and have made it publicly available via the notebook~3 listed in App.~\ref{app:Sage}.
The values of $I_\ell$ and $L_\ell$ obtained with this code are plotted in terms of $a/M$ in Figs.~\ref{fig:Kerr_I} and \ref{fig:Kerr_L}, for $0 \leqslant \ell \leqslant 9$.


Many properties of the multipoles moments $K_{\ell,m}$ can be inferred from the integral formula \eqref{I-L_Kerr2}.
First of all, the $K_{\ell,m}$'s are nonzero only for $m=0$, with $I_\ell$ and $L_\ell$ real valued and functions of $\hat{a}$ only, or equivalently of $a/M$ only (see Figs.~\ref{fig:Kerr_I} and \ref{fig:Kerr_L}).
Secondly, the generic monopole properties \eqref{monopole_universal} are easily recovered from \eqref{I-L_Kerr2}: $I_0 = \sqrt{\pi}$ and $L_0 = 0$.
Thirdly, the reflection symmetry across the equatorial plane $\zeta=0$ and the parity properties of the Legendre polynomials $P_\ell(x)$ and the function $z(\zeta)$ given by Eq.~\eqref{z-zeta}
imply $I_\ell = 0$ when $\ell$ is odd and $L_\ell = 0$ when $\ell$ is even, similarly to the axisymmetry-based case [Eq.~\eqref{parity_axi}]:
\beq\label{parity}
   \forall n \in \mathbb{N},\quad  I_{2n+1} = 0 \quad \text{and} \quad L_{2n} = 0 \, .
\eeq
Moreover, the integral in \eqref{I-L_Kerr2} has a simple behavior under the change $a \to - a$ in the black hole spin direction. Indeed, using $z(-\zeta) = - z(\zeta)$ and $P_\ell(-x) = (-)^\ell P_\ell(x)$ we readily find
\beq
    I_\ell(-a) + \ui L_\ell(-a) = (-)^\ell \left[ I_\ell(a) + \ui L_\ell(a) \right] .
\eeq
Combined with Eqs.~\eqref{parity} we deduce that the shape multipoles $I_\ell$ only depends on the spin amplitude $|a|$, while the current multipoles $L_\ell$ change sign under $a \to -a$. More precisely, the horizon multipoles $I_\ell$ (resp. $L_\ell$) are even (resp. odd) in the Kerr spin parameter $a$, just like the Hansen (field) multipole moments \eqref{Hansen} of the Kerr solution \cite{Ha.74}. In the nonspinning limit $a \to 0$, we have $\hat{a} = 0$ and $\beta = 0$, so that according to Eq.~\eqref{z-zeta} $z = \zeta$. Then $I_\ell = L_\ell = 0$ for all $\ell \geqslant 1$. In other words, the only nonvanishing horizon multipole moment of a Schwarzschild black hole is its shape monopole $I_0 = \sqrt{\pi}$.

For small spin values, the behavior of the multipoles can be determined in closed form, and we find (see App.~\ref{app:asymptotics})
\beq\label{I-L_Kerr3}
    I_\ell + \ui L_\ell = \sqrt{(2\ell+1)\pi} \; (\ui \hat{a})^\ell \left[ \alpha_\ell + O(\hat{a}^2) \right] ,
\eeq
where
\beq\label{alpha_ell}
    \alpha_\ell = \frac{2^\ell \, \gamma_{\lceil \ell/2 \rceil}}{\binom{2\ell+1}{\ell}} \times
    \begin{cases}
        1 & \text{($\ell$ even)} \\
        \tfrac{1}{2} & \text{($\ell$ odd)}
    \end{cases}
    \quad \text{with} \quad \gamma_n \equiv \sum_{k=0}^n \binom{2n}{k} = \frac{2^{2n} + \binom{2n}{n}}{2} \, .
\eeq
The first 15 numerical values of the sequence $(\alpha_\ell)_{\ell \in \mathbb{N}}$ are listed in Table~\ref{table:sequence}. In the limit $\ell \to +\infty$ we have $\alpha_\ell \sim \tfrac{1}{2} \, \sqrt{\pi \ell}$, so that $I_\ell + \ui L_\ell \sim  (\pi/\sqrt{2}) \, \ell \, (\ui \hat{a})^\ell$. The scaling property \eqref{I-L_Kerr3} of the multipoles $I_\ell$ and $L_\ell$ as functions of $\hat{a}$ is confirmed graphically by the right panels of Figs.~\ref{fig:Kerr_I} and \ref{fig:Kerr_L}, with the value of $\alpha_\ell$ from Table~\ref{table:sequence} being read at the intersection of the curves with the axis $a=0$.

\begin{table}[t]
    \begin{center}
        \begin{tabular}{lcccc}
            \hline
            $\ell$ && $\alpha_\ell$ (exact) && $\alpha_\ell$ (approximate) \\
            \hline
            0  && $1$                        && $1.00000000\dots$ \\
            1  && $1$                        && $1.00000000\dots$ \\
            2  && $\frac{6}{5}$              && $1.20000000\dots$ \\
            3  && $\frac{44}{35}$            && $1.25714285\dots$\\
            4  && $\frac{88}{63}$            && $1.39682539\dots$ \\
            5  && $\frac{16}{11}$            && $1.45454545\dots$ \\
            6  && $\frac{224}{143}$          && $1.56643356\dots$ \\
            7  && $\frac{10432}{6435}$       && $1.62113442\dots$ \\
            8  && $\frac{20864}{12155}$      && $1.71649536\dots$ \\
            9  && $\frac{7424}{4199}$        && $1.76804001\dots$ \\
            10 && $\frac{163328}{88179}$     && $1.85223239\dots$ \\
            11 && $\frac{1285120}{676039}$   && $1.90095541\dots$ \\
            12 && $\frac{514048}{260015}$    && $1.97699363\dots$ \\
            13 && $\frac{10145792}{5014575}$ && $2.02326059\dots$ \\
            14 && $\frac{20291584}{9694845}$ && $2.09302820\dots$ \\
            \hline
        \end{tabular}
        \caption{The first 15 numerical values of the sequence $(\alpha_\ell)_{\ell \in \mathbb{N}}$ of numbers \eqref{alpha_ell} that appears in the small-spin limit \eqref{I-L_Kerr3} of the Kerr horizon multipole moments $I_\ell$ and $L_\ell$.}
        \label{table:sequence}
    \end{center}
\end{table}

\begin{figure}[t]
    \begin{center}
        \includegraphics[width=0.48\textwidth]{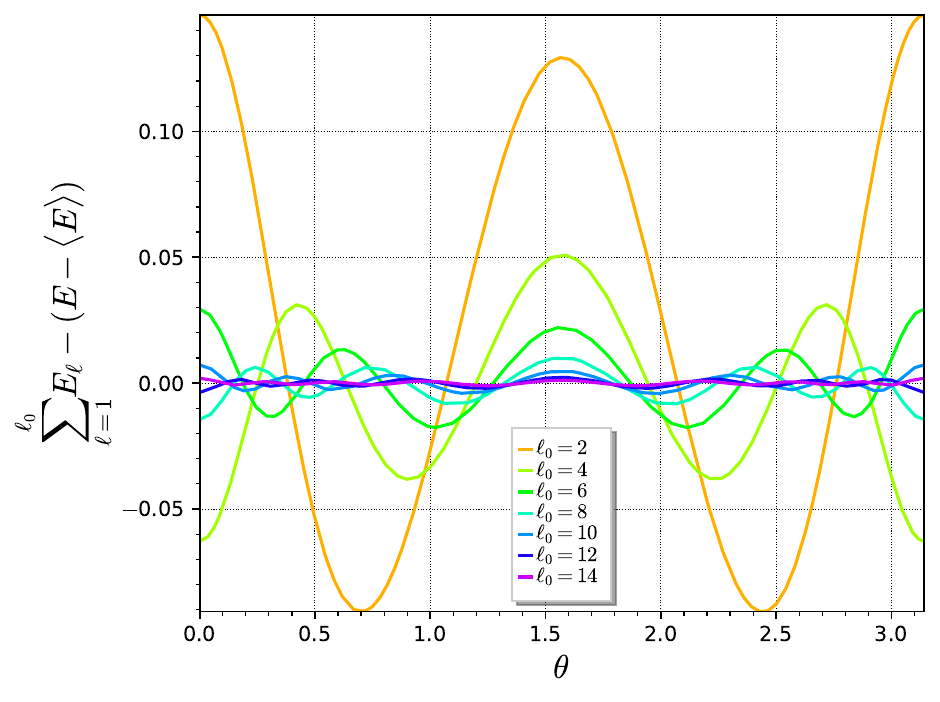}\quad
        \includegraphics[width=0.48\textwidth]{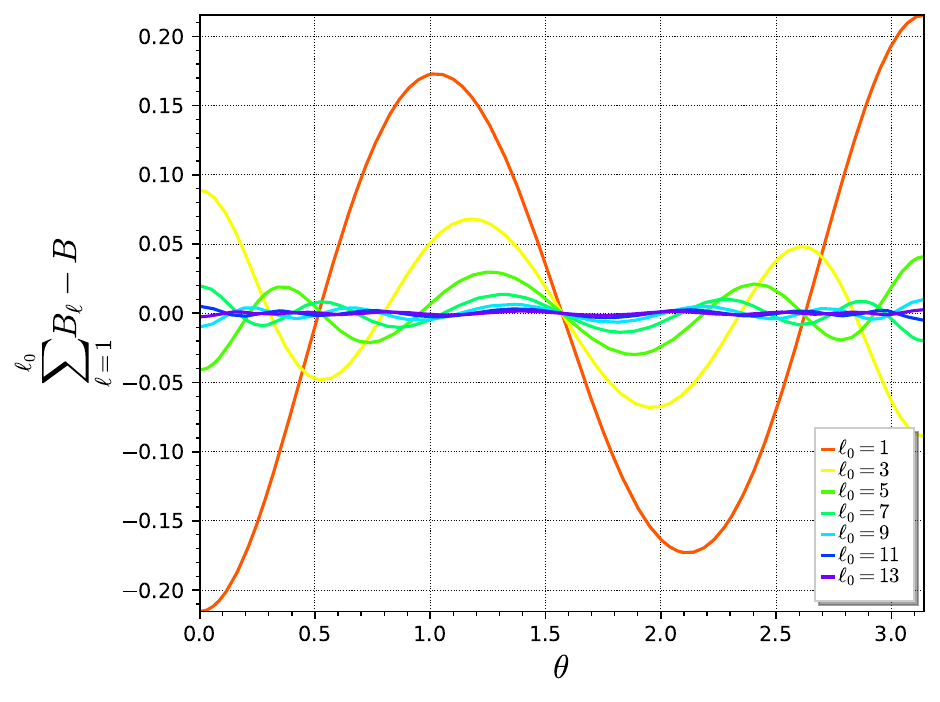}
        \caption{Convergence of the multipole expansions \eqref{ell-sums} for the extremal Kerr horizon ($\hat{a}=1$). Here, $\sum_{\ell=1}^{\ell_0} E_\ell$ and $\sum_{\ell=1}^{\ell_0} B_\ell$ stand for the sums appearing in the formulas \eqref{ell-sums}, but truncated at a given order $\ell_0$.}
        \label{fig:Kerr_convergence_E_B}
    \end{center}
\end{figure}

Finally, given the multipoles $I_\ell$ and $L_\ell$, the mode decomposition \eqref{E+iB} yields the electric and magnetic potentials $E$ and $B$ of a Kerr black hole horizon as series expansions over Legendre polynomials, according to
\begin{subequations}\label{ell-sums}
    \begin{align}
        E - \langle E \rangle &= - \frac{1}{\sqrt{\pi}} \sum_{\ell = 1}^{+\infty} \frac{\sqrt{2\ell+1}}{\ell(\ell+1)} \, I_\ell(\hat{a}) \, P_\ell(\cos{\vartheta}) \, , \\
        B &= - \frac{1}{\sqrt{\pi}} \sum_{\ell = 1}^{+\infty} \frac{\sqrt{2\ell+1}}{\ell(\ell+1)} \, L_\ell(\hat{a}) \, P_\ell(\cos{\vartheta}) \, .
    \end{align}
\end{subequations}
Of course $E$ and $B$ are axisymmetric and have a reflection (anti)symmetry with respect to the equatorial plane, in agreement with the continuous and discrete spatial symmetries of the Kerr metric. Recall that the electric potential $E = \ln{(R \, \psi)}$ and the magnetic potential $B$ are known in closed form from Eqs.~\eqref{psi_Kerr} and \eqref{B_Kerr}, according to which
\begin{subequations}\label{E_B_Kerr}
    \begin{align}
        E &= \tfrac{1}{2} \, \ln{(1-\beta^2 \sin^2{\theta})} - \ln{[\cosh{(\beta^2 \cos{\theta})} - \cos{\theta} \, \sinh{(\beta^2 \cos{\theta})}]} \, , \label{E_Kerr} \\
        B &= - \arctan{(\hat{a} \cos{\theta})} - \frac{1-\hat{a}^2}{2 \left(1+\hat{a}^2\right)} \, \hat{a} \cos{\theta} \, .
    \end{align}
\end{subequations}
Note that in the above formulas, $\theta$ is the Kerr colatitude coordinate, while in \eqref{ell-sums}, $E$ and $B$ are expressed in terms of $\vartheta$---the colatitude adapted to the unit round metric $\mathring{q}_{ab}$, the two angles being related by \eqref{z-zeta}, where $\zeta=\cos\theta$ and $z=\cos\vartheta$ (see also Fig.~\ref{fig:vartheta}).
The functions $E$ and $B$ are depicted in Figs.~\ref{fig:E_theta} and \ref{fig:B_theta} respectively, while Fig.~\ref{fig:Kerr_convergence_E_B} shows the convergence of the  $\ell$-sums \eqref{ell-sums} towards the closed-form formulas \eqref{E_B_Kerr} as $\ell \to +\infty$.

\subsection{Comparison of the two families of horizon multipoles} \label{subsec:comparison_Kerr}

\begin{figure}[t]
    \begin{center}
        \includegraphics[width=0.48\textwidth]{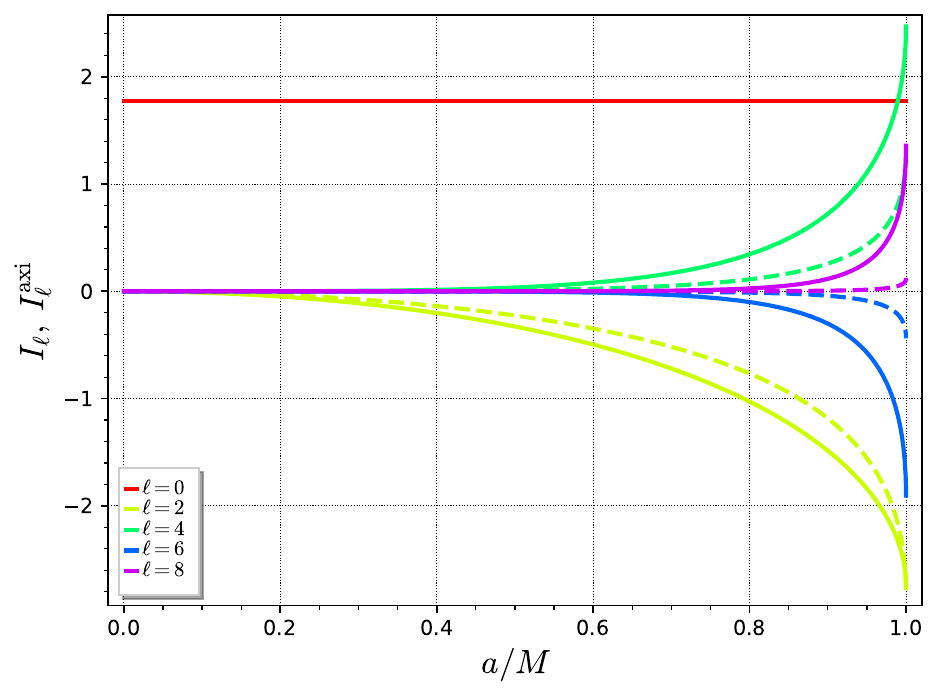}\quad
        \includegraphics[width=0.48\textwidth]{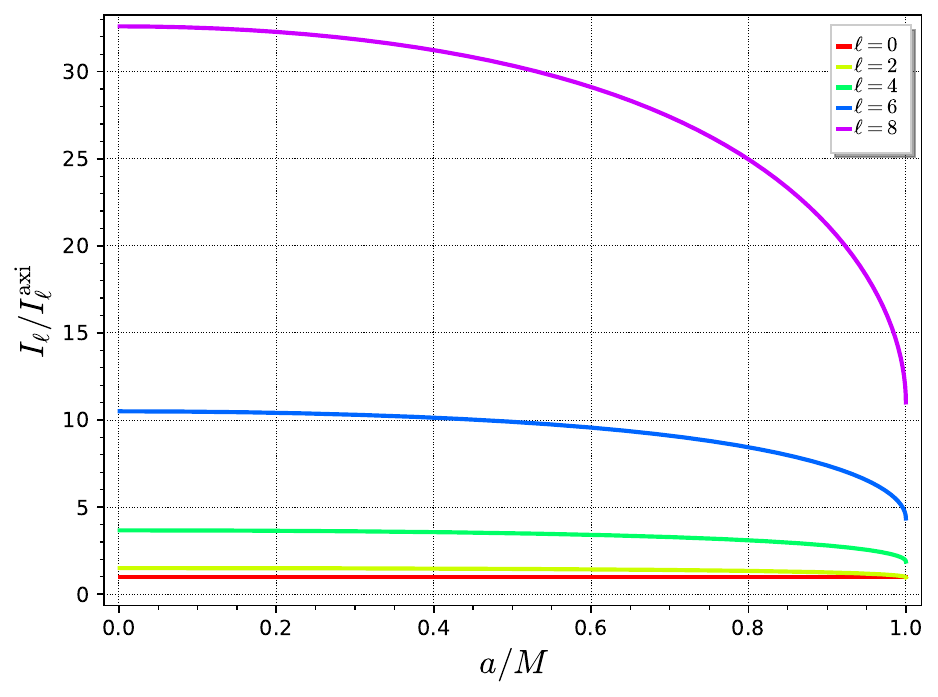}
        \caption{Comparison, for the Kerr horizon, between the shape multipole moments $I_\ell$ resulting from the generic-NEH definition (solid curves in the left panel) and the axisymmetry-based moments $I_{\ell}^{\rm axi}$ (dashed curves, same as solid ones in Fig.~\ref{fig:Kerr_I_L_axi}). The right panel depicts the
        ratio $I_\ell/I_{\ell}^{\rm axi}$.}
        \label{fig:Kerr_I_comp_axi}
    \end{center}
\end{figure}

\begin{figure}[t]
    \begin{center}
        \includegraphics[width=0.48\textwidth]{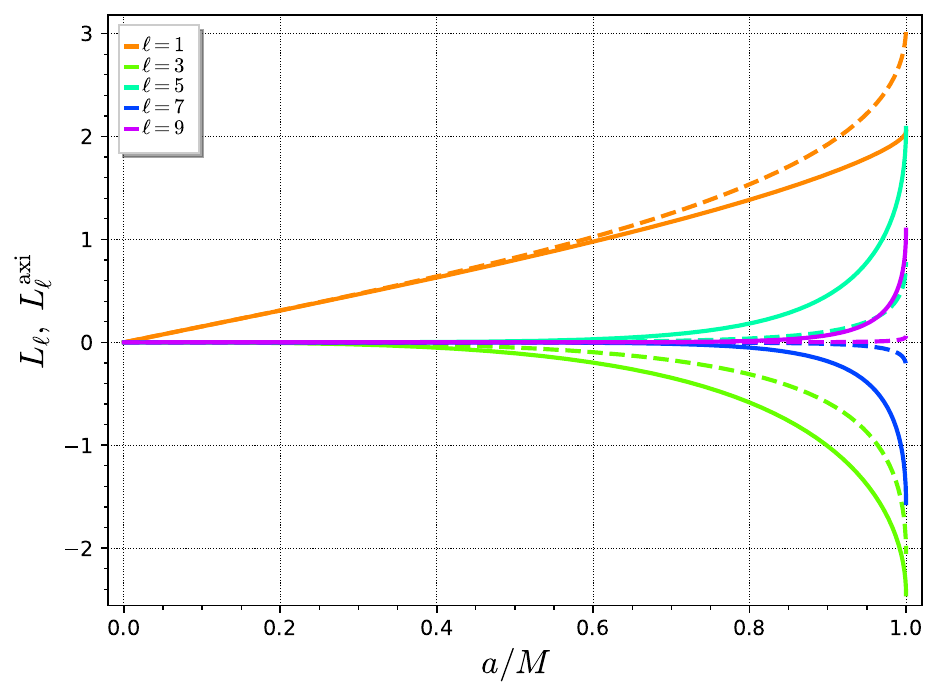}\quad
        \includegraphics[width=0.48\textwidth]{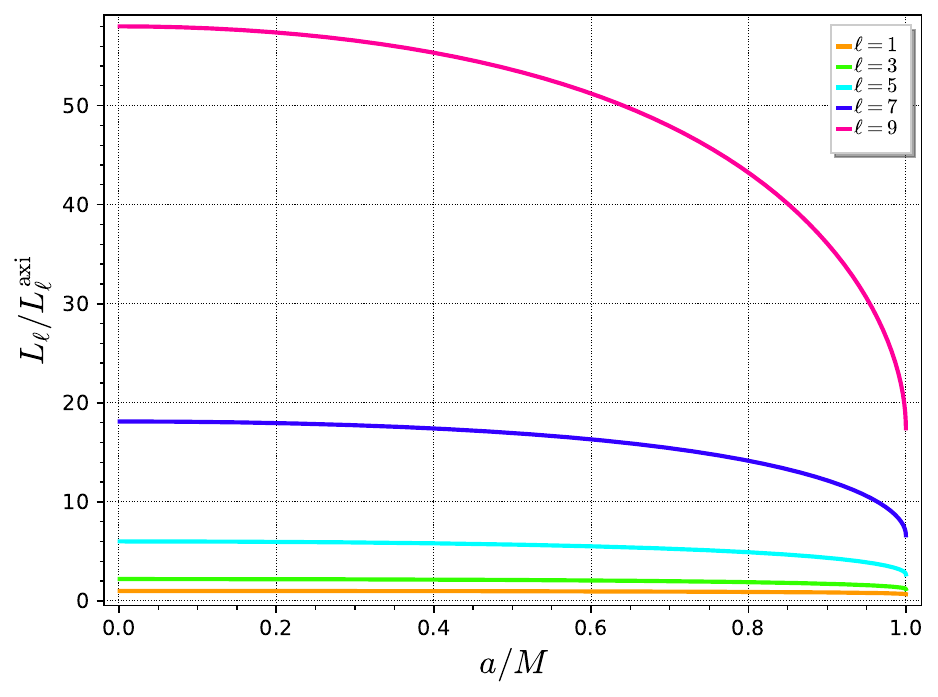}
        \caption{Same as Fig.~\ref{fig:Kerr_I_comp_axi} but for the current multipole moments $L_\ell$ and $L_\ell^{\rm axi}$.}
        \label{fig:Kerr_L_comp_axi}
    \end{center}
\end{figure}

The multipole moments $(I_\ell, L_\ell)$ are graphically compared to the axisymmetric-based multipole moments $(I_\ell^{\rm axi}, L_\ell^{\rm axi})$
computed in Sec.~\ref{subsec:axi_Kerr} in Figs.~\ref{fig:Kerr_I_comp_axi} and \ref{fig:Kerr_L_comp_axi}.
In the small spin regime, we read from Eqs.~\eqref{I-L_Kerr3-axi}, \eqref{I-L_Kerr3} and \eqref{alpha_ell} that both families behave like a real-valued coefficient times $(\ui \hat{a})^\ell$,
but with a generically different coefficient, as one can see from the ratios
\beq\label{compare_axi_low_a}
    \frac{I_\ell}{I_\ell^\text{axi}}\bigg\vert_{\ell~\text{even}} \ \mbox{or}\quad \frac{L_\ell}{L_\ell^\text{axi}}\bigg\vert_{\ell~\text{odd}} = \frac{2^\ell + \binom{\ell}{\lceil \ell/2 \rceil}}{\ell+2} + O(\hat{a}^2) \, .
\eeq
Since $z(\zeta) \sim \zeta$ for $a\to 0$ and the only difference between
formulas \eqref{multipoles_axi} and \eqref{I-L_Kerr2} is $P_\ell(\zeta)$ replaced by $P_\ell(z(\zeta))$,
one might have thought naively that in the limit of small spins, $I_\ell$ and $L_\ell$ coincide with $I^{\rm axi}_\ell$ and $L^{\rm axi}_\ell$. However formula~\eqref{compare_axi_low_a} shows that this is not the case, except for $\ell=0$ and $\ell=1$. The reason is that, in the limit $a \to 0$, the part $z = \zeta$ in the expansion of $z(\zeta)$ in powers of $a$ contributes only to the multipoles $\ell=0$ and $\ell=1$, i.e. its integral in \eqref{I-L_Kerr2} vanishes for higher values of $\ell$, so that the integrals depend on nonzero powers of $a$ in the expansion of $z(\zeta)$, which do not appear in $I^{\rm axi}_\ell$ and $L^{\rm axi}_\ell$.

Let us take the example of the shape quadrupole, $I_2$. From Eq.~\eqref{z-zeta} with $\beta = \hat{a} / \sqrt{1+\hat{a}^2}$ [Eq.~\eqref{hat_a_beta}], we have
\beq
     z(\zeta) = \zeta - \hat{a}^2 \zeta (1 - \zeta^2) + O(\hat{a}^4) \, .
\eeq
Accordingly,
\beq \label{integrand_I_2}
 \text{Re}\,  \frac{P_2(z(\zeta))}{(1 - \ui \hat{a} \zeta)^3} = \tfrac{1}{2} \, (3 \zeta^2 - 1) - 6 \hat{a}^2 \zeta^4 + O(\hat{a}^4) \, ,
\eeq
while
\beq \label{integrand_I_2_axi}
 \text{Re}\,  \frac{P_2(\zeta)}{(1 - \ui \hat{a} \zeta)^3} = \tfrac{1}{2} \, (3 \zeta^2 - 1) + 3 \hat{a}^2 \zeta^2 (1 - 3\zeta^2) + O(\hat{a}^4) \, .
\eeq
The integral of $\tfrac{1}{2} \, (3\zeta^2 - 1) = P_2(\zeta)$ yields $0$ so that the value of $I_2$ is given at leading order by the term $\propto \hat{a}^2$ in Eq.~\eqref{integrand_I_2}, while the value of $I_2^{\rm axi}$ is given by the term $\propto \hat{a}^2$ in Eq.~\eqref{integrand_I_2_axi}. These two terms are distinct quartic polynomials in $\zeta$ and hence, once integrated over $[-1,1]$, they lead to distinct values
for $I_2$ and $I_2^{\rm axi}$:
\beq
    I_2 = - 6\sqrt{\frac{\pi}{5}} \, \hat{a}^2 + O(\hat{a}^4) \qquad \text{and} \qquad
    I_2^{\rm axi} = - 4 \sqrt{\frac{\pi}{5}} \, \hat{a}^2 + O(\hat{a}^4) \, .
\eeq
It follows that $I_2 / I_2^{\rm axi} \to  3/2$ in the limit $\hat{a} \to 0$, in agreement with formula \eqref{compare_axi_low_a} for $\ell=2$.


In the small spin limit, Eq.~\eqref{compare_axi_low_a} implies that the generic-NEH multipole moments are increasingly larger than the axisymmetry-based ones, as $\ell$ increases:
\beq \label{diverging_multipole_ratio}
    \lim_{\ell\to +\infty} \frac{I_\ell}{I_\ell^\text{axi}} = \lim_{\ell\to +\infty} \frac{L_\ell}{L_\ell^\text{axi}} = +\infty \quad \mbox{for} \quad \hat{a} \to 0 \, .
\eeq
The numerical results displayed in Figs.~\ref{fig:Kerr_I_comp_axi} and \ref{fig:Kerr_L_comp_axi} suggest that this behavior remains true for all values of $a/M$.

\section{Conclusions and prospects} \label{sec:concl}

We have presented the first computation [see Figs.~\ref{fig:Kerr_I}--\ref{fig:Kerr_L} and Eqs.~\eqref{I-L_Kerr3}--\eqref{alpha_ell}] of the horizon multipole moments of a Kerr black hole, following the generic definition for non-expanding horizons proposed by
Ashtekar, Khera, Kolanowski and Lewandowski in 2022 \cite{As.al.22}.
In particular, we have obtained a rather simple expression [Eq.~\eqref{round_metric_Kerr}],
in terms of the Kerr angular coordinates $(\theta,\phi)$, of the unique axisymmetric unit round metric $\mathring{q}_{ab}$ conformally related to the physical metric
$q_{ab}$ of the horizon cross-sections.
Besides, we have extended to all spherical harmonic degrees $\ell\in\mathbb{N}$ the computation [Eqs.~\eqref{truc}--\eqref{G_ell_general_form}] of the Kerr horizon multipole moments following the
definition for axisymmetric isolated horizons proposed by Ashtekar, Engle, Pawlowski and Van Den Broeck in 2004 \cite{As.al.04}, whereas previous results in the literature were limited to $\ell\leqslant 8$ \cite{As.al.04,Gu.al.18}. We have compared the two families of horizon multipoles resulting from those definitions and shown that they differ, except for $\ell = 0$ and,  in the small spin limit, also for $\ell=1$. We have also compared these horizon multipoles to the (Hansen) field multipole moments. The results of these comparisons are detailed in Sec.~\ref{sec:Kerr} and summarized in Sec.~\ref{sec:main_results}.

At first glance, it may be surprising that the two families of horizon multipole moments do not coincide for the Kerr black hole. This merely reflects the lack of a unique (canonical) construction of multipole moments, even in the particular case of axisymmetric horizons. It shall be kept in mind that the same situation occurs for the \emph{field} multipole moments: for a non-spherically symmetric static body, the Geroch multipole moments \cite{Ge2.70} are different from the Hansen ones \cite{Ha.74}. What matters is
that the whole intrinsic and extrinsic geometry of the horizon can be reconstructed from the knowledge of any of the two multipole sets \cite{As.al.04,As.al.22},
similarly to the asymptotic spacetime geometry being fully determined by the knowledge of the field multipole moments \cite{BeSi.81}.

Astrophysical black holes are neither perfectly isolated, nor stationary. For instance, a supermassive black hole in a binary system of compact objects is subject to the tidal field of its companion. Just like for a Newtonian self-gravitating body, the tidal deformability of an astrophysical black hole can be characterized by means of two families of so-called \textit{tidal Love numbers} (TLNs). In particular, the \textit{field} TLNs characterize the linear response of the black hole at the level of the gravitational field itself. For stationary tidal perturbations, they can for instance be defined and computed from the linear perturbation in the (Geroch-Hansen) field multipole moments at spatial infinity \cite{Le.al.21,LeCa.21}.

By contrast, the \textit{surficial} TLNs characterize in a local manner the linear response of the black hole at the level of the event horizon, akin to the surface of a Newtonian self-gravitating body. The surficial TLNs of nonspinning (Schwarzschild) black holes have been defined and computed by several authors
\cite{DaLe.09,LaPo.14,Ri.al.24}. For spinning (Kerr) black holes, a geometrically-motivated definition of surficial TLNs is still lacking, except in the small spin limit \cite{Ri.al.24}.
The horizon multipole moments of the second family explored here, namely those for generic non-expanding horizons, pave the way to the definition and computation of surficial TLNs for Kerr black holes, since the definition of these multipoles \cite{As.al.22} is valid beyond axisymmetry, contrary to the definition proposed previously in Ref.~\cite{As.al.04}. The surficial TLNs should provide an invariant characterization of the shape and angular momentum structure of the horizon of a spinning black hole subject to a weak, but otherwise arbitrary tidal perturbation.  This will be the topic of a forthcoming paper.

\section*{Acknowledgments}

ALT acknowledges the hospitality of the Brazilian Center for Research in Physics (CBPF) and the ICTP-SAIFR (FAPESP grant 2021/14335-0), where part of this work was carried out. EG acknowledges funding by l’Agence Nationale de la Recherche, projects StronG ANR-22-CE31-0015-01 and Einstein PPF ANR-23-CE40-0010-02.

\appendix

\section{Integral for the axisymmetry-based multipoles of the Kerr horizon} \label{app:integ_Kerr_axi}

The integral appearing in the right-hand side of formula~\eqref{multipoles_axi} for the axisymmetry-based multipole moments of the Kerr horizon is\footnote{We denote it $J_{\ell,0}$ because this is the
case $n=0$ of a more general integral $J_{\ell,n}$ introduced in \eqref{pilou} below.}
\beq\label{def_J_ell_0}
    J_{\ell,0} \equiv \int_{-1}^1 \ud x \, \frac{P_\ell(x)}{(1 - \ui \hat{a} x)^3} \, .
\eeq
Explicit values of $J_{\ell,0}$ in terms of $\hat{a}$ can be found in App.~A of Ref.~\cite{Gu.al.18} for $2 \leqslant \ell \leqslant 8$. In what follows, we derive a closed-form expression valid for any value of $\ell$, first in terms of the hypergeometric function ${}_2F_1$, and then in terms of polynomials and the arctangent function.

\subsection{Expression in terms of the hypergeometric function}

Our starting point is the Rodrigues formula for the Legendre polynomials, namely
\beq\label{Rodrigues}
  P_\ell(x) = \frac{1}{2^\ell \ell!} \, \frac{\ud^\ell}{\ud x^\ell} \left[ (x^2-1)^\ell \right] .
\eeq
Substituting this formula into Eq.~\eqref{def_J_ell_0} and integrating by parts $\ell$ times, while noticing that the boundary contributions at $x = \pm 1$ vanish thanks to the factor $(x^2-1)^\ell$ in Eq.~\eqref{Rodrigues}, we obtain
\beq
    J_{\ell,0} =
    \frac{(-)^{\ell}}{2^{\ell}\ell!}\int_{-1}^1 \ud x\, (x^2-1)^{\ell} \, \frac{\ud^\ell}{\ud x^\ell} \left[(1-\ui\hat{a}x)^{-3} \right] .
\eeq
Given that $\ud^\ell/\ud x^\ell  \left[(1-\ui\hat{a}x)^{-3} \right] = \tfrac{1}{2} \, (\ell + 2)! \, (\ui \hat{a})^\ell \, (1-\ui\hat{a}x)^{-(\ell+3)}$, there comes
\beq
     J_{\ell,0} = \frac{(\ell+2)(\ell+1)}{2^{\ell+1}} \, (\ui \hat{a})^\ell \int_{-1}^1 \ud x \, \frac{(1 - x^2)^\ell}{(1 -\ui\hat{a}x)^{\ell + 3}} \, .
\eeq
The change of variable $t \equiv \tfrac{1}{2} (1+x)$ leads us to an integral between $0$ and $1$:
\beq \label{J_ell0_integ_t}
    J_{\ell,0} =  2^\ell (\ell+2)(\ell+1) \, \frac{(\ui \hat{a})^\ell }{(1 + \ui \hat{a})^{\ell + 3}}
    \int_0^1 \ud t \, \frac{t^\ell (1 - t)^\ell}{\left( 1 - \frac{2\ui\hat{a}}{1 + \ui\hat{a}} \, t \right)^{\ell+3}} \, ,
\eeq
and we may use the Euler formula (see DLMF-15.6.1 \cite{NIST:DLMF} with $a=\ell+3$, $b=\ell+1$ and $c=2(\ell+1)$) to express this integral in terms of the hypergeometric function ${}_2F_1$:
\beq
     \int_0^1 \ud t \, \frac{t^\ell (1 - t)^\ell}{\left( 1 - \frac{2\ui\hat{a}}{1 + \ui\hat{a}} \, t \right)^{\ell+3}} =
     \frac{(\ell!)^2}{(2\ell + 1)!} \;  {}_2F_1 \left(\ell+3, \ell+1, 2(\ell+1); \frac{2\ui\hat{a}}{1 + \ui\hat{a}} \right) .
\eeq
Then, since $(\ell+2)(\ell+1)\ell! = (\ell+2)!$ and $(2\ell+1)! = 2^\ell \ell! (2\ell+1)!!$, Eq.~\eqref{J_ell0_integ_t} becomes
\beq
     J_{\ell,0} =  \frac{(\ell+2)!}{(2\ell+1)!!} \,  \frac{(\ui \hat{a})^\ell }{(1 + \ui \hat{a})^{\ell + 3}}
      \;  {}_2F_1 \left(\ell+3, \ell+1, 2(\ell+1); \frac{2\ui\hat{a}}{1 + \ui\hat{a}} \right) .
\eeq
Next, let us introduce the variable $z\equiv 2\ui\hat{a}/(1 + \ui\hat{a})$, which appears as the argument of ${}_2F_1$. It fulfills
$\ui\hat{a} = z/(2 - z)$ and $1 + \ui\hat{a} = (1 - z/2)^{-1}$, so that
\beq
     J_{\ell,0} =  \frac{(\ell+2)!}{(2\ell+1)!!} \,  (\ui \hat{a})^\ell \left(1 - \frac{z}{2} \right) ^{\ell+3}
        {}_2F_1 \left(\ell+3, \ell+1, 2(\ell+1); z \right) .
\eeq
Now, by virtue of a standard quadratic transformation formula of the hypergeometric function (formula DLMF-15.8.13 \cite{NIST:DLMF} with $a=\ell+3$ and $b = \ell+1$), we have
the identity
\beq
    \left(1 - \frac{z}{2} \right) ^{\ell+3}   {}_2F_1 \left(\ell+3, \ell+1, 2(\ell+1); z \right) =
    {}_2F_1\left(\frac{\ell+3}{2},\frac{\ell+4}{2},\ell+\frac{3}{2}; \frac{z^2}{(2 - z)^2} \right) .
\eeq
Given that $z^2/(2 - z)^2 = (\ui \hat{a})^2 = - \hat{a}^2$, we get the final expression of the integral $J_{\ell,0}$ in terms of the hypergeometric function ${}_2F_1$:
\beq \label{J_l0_var}
    J_{\ell,0} =  \frac{(\ell+2)!}{(2\ell+1)!!} \,  (\ui \hat{a})^\ell \, {}_2F_1\left(\frac{\ell+3}{2},\frac{\ell+4}{2},\ell+\frac{3}{2}; -\hat{a}^2\right) .
\eeq

In expression \eqref{multipoles_axi} for the multipole moments, $J_{\ell,0}$ is multiplied by the factor $(1 + \hat{a}^2)^2$. Thanks to the Euler transformation
formula
${}_2F_1(a,b,c;z) = (1-z)^{c-a-b} \, {}_2F_1(c-a,c-b,c;z)$ (cf. DLMF-15.8.1 \cite{NIST:DLMF}) with $a=\ell/2$, $b=(\ell-1)/2$, $c=\ell+3/2$ and $z=-\hat{a}^2$, the product happens to
be the hypergeometric function with different arguments:
\beq \label{G_ell_F_ell}
 G_\ell(\hat{a}^2) \equiv {}_2F_1\left(\frac{\ell}{2}, \frac{\ell-1}{2},\ell+\frac{3}{2};-\hat{a}^2\right)  = (1 + \hat{a}^2)^2 \; {}_2F_1\left(\frac{\ell+3}{2}, \frac{\ell+4}{2},\ell+\frac{3}{2};-\hat{a}^2\right) .
\eeq
In conjunction with Eq.~\eqref{J_l0_var}, this yields \eqref{truc}. The function $G_\ell(\hat{a}^2)$ is depicted in Fig.~\ref{fig:G_ell}.
The low $\ell$ values are
\begin{subequations}
\label{G_l_low_l}
\begin{align}
    & G_0(\hat{a}^2) =  G_1(\hat{a}^2) = 1 \, ,  \\
    & G_2(\hat{a}^2) = - \frac{5}{8\hat{a}^4} \left[ 5 \hat{a}^2 + 3 - 3 (1 + \hat{a}^2)^2 \, \frac{\arctan \hat{a}}{\hat{a}} \right] , \\
    & G_3(\hat{a}^2) = \frac{7}{8\hat{a}^6} \left[ 8 \hat{a}^4 + 25 \hat{a}^2 + 15 - 15 (1 + \hat{a}^2)^2 \, \frac{\arctan \hat{a}}{\hat{a}}  \right] , \\
    & G_4(\hat{a}^2) = - \frac{21}{32\hat{a}^8} \left[ 81 \hat{a}^4 + 190 \hat{a}^2 + 105 - 15 (1 + \hat{a}^2)^2 (7 + \hat{a}^2) \, \frac{\arctan \hat{a}}{\hat{a}} \right] , \\
    & G_5(\hat{a}^2) = \frac{33}{32\hat{a}^{10}} \left[ 32 \hat{a}^6 + 343 \hat{a}^4 + 630 \hat{a}^2 + 315 - 105 (1 + \hat{a}^2)^2 (3 + \hat{a}^2) \, \frac{\arctan \hat{a}}{\hat{a}} \right] .
\end{align}
\end{subequations}
Other expressions of $G_\ell(\hat{a}^2)$, up to $\ell=12$, are listed in the notebook~3 of App.~\ref{app:Sage}.
For large values of $\ell$, the following asymptotic behavior holds:
\beq \label{G_l_large_l}
    G_\ell(\hat{a}^2) \sim (1 + \hat{a}^2)^{3/4} \left( \frac{2}{1 + \sqrt{1 + \hat{a}^2}} \right)^{\ell + 1/2}
    \quad\mbox{for}\quad \ell \to +\infty \, .
\eeq
To prove \eqref{G_l_large_l}, let us use Euler's integral representation of the hypergeometric function ${}_2F_1$ (formula DLMF-15.6.1 \cite{NIST:DLMF}) to rewrite Eq.~\eqref{G_ell_F_ell} as
\beq \label{G_l_for_Laplace}
    G_\ell(\hat{a}^2) = \underbrace{\frac{\Gamma(\ell+3/2)}{\Gamma((\ell-1)/2) \Gamma(\ell/2+2)}}_{C_\ell} \;
        \underbrace{\int_0^1 \ud t \, h(t) \, \mathrm{e}^{\frac{\ell}{2} f(t)}}_{E_\ell(\hat{a}^2)} \, ,
\eeq
with
\beq
    \quad h(t) \equiv \frac{1 - t}{t^{3/2}} \qquad\mbox{and}\qquad
        f(t) \equiv \ln \left( \frac{t(1-t)}{1 + \hat{a}^2 t} \right) .
\eeq
Since $f(t)$ has a unique maximum over $]0,1[$, at $t=t_0 \equiv (1 + \sqrt{1 + \hat{a}^2})^{-1}$, we may use Laplace's method of approximating the integrand in $E_\ell(\hat{a}^2)$ by a Gaussian function
centered at $t_0$, of standard deviation $\sigma = \sqrt{2/(\ell |f''(t_0)|)}$ and of height $h(t_0)  \, \mathrm{e}^{\frac{\ell}{2} f(t_0)}$,
thereby getting an equivalent to the integral $E_\ell(\hat{a}^2)$ for large $\ell$:
\beq
    E_\ell(\hat{a}^2) \sim \sqrt{\frac{4\pi}{\ell |f''(t_0)|}} \, h(t_0) \, \mathrm{e}^{\frac{\ell}{2} f(t_0)}
        \quad\mbox{for}\quad \ell \to +\infty \, .
\eeq
Given that $f(t_0) = - 2 \ln(1 + \sqrt{1 + \hat{a}^2})$, $f''(t_0) = - 2 (1 + \hat{a}^2)^{-1/2} ( 1 + \sqrt{1 + \hat{a}^2})^2$ and $h(t_0) =(1 + \hat{a}^2)^{1/2} ( 1 + \sqrt{1 + \hat{a}^2})^{1/2}$, there comes
\beq \label{I_l_Laplace}
    E_\ell(\hat{a}^2) \sim \sqrt{\frac{2\pi}{\ell}} \, (1 + \hat{a}^2)^{3/4} \left( 1 + \sqrt{1 + \hat{a}^2} \right) ^{-(\ell+1/2)}
            \quad\mbox{for}\quad \ell \to +\infty \, .
\eeq
On the other hand, thanks to Stirling's formula, it is easy to see that the prefactor $C_\ell$ in \eqref{G_l_for_Laplace} behaves as
\beq \label{C_l_Stirling}
    C_\ell \sim 2^\ell \sqrt{\frac{\ell}{\pi}} \quad\mbox{for}\quad \ell \to +\infty \, .
\eeq
Combining \eqref{I_l_Laplace} and \eqref{C_l_Stirling} leads to \eqref{G_l_large_l}.

\subsection{Hypergeometric function in terms of arctangent}

Here we express the hypergeometric function $F_{\ell}(\hat{a}^2)\equiv {}_2F_1\left(\frac{\ell+3}{2},\frac{\ell+4}{2},\ell+\frac{3}{2};-\hat{a}^2\right)$ appearing in \eqref{J_l0_var} in terms of $\arctan\hat{a}$ and  rational functions of $\hat{a}$.

Let us first show the derivation for the case where $\ell$ is even. That is, we set $\ell=2n$, with $n\in \mathbb{N}$,  so that
$F_{\ell}(\hat{a}^2)={}_2F_1\left(n+\frac{3}{2},n+2,2n+\frac{3}{2}; -\hat{a}^2\right)$.
We first apply DLMF-15.5.2~\cite{NIST:DLMF} (with `$n$' there taking on the value of the `$n+1$' here), which yields
\beq
F_{\ell}(\hat{a}^2)=\frac{\left(n+1/2\right)_{n+1}}{\left(1/2\right)_{n+1}(n+1)!} \, \frac{\ud^{n+1}}{\ud z^{n+1}}  \, {}_2F_1\left(\frac{1}{2},1,n+\frac{1}{2}; z\right),
\eeq
where $z\equiv -\hat{a}^2<0$ and $(\alpha)_n\equiv \Gamma(\alpha+n)/\Gamma(\alpha)$ (with $\alpha \not\in \mathbb{Z}_{\leqslant 0}$) is Pochhammer’s symbol.
We next apply DLMF-15.5.6~\cite{NIST:DLMF} (with `$n$' there taking on the value of the `$n-1$' here), yielding
\beq
F_{\ell}(\hat{a}^2)= A_\ell \,
\frac{\ud^{n+1}}{\ud z^{n+1}} \left[ (1-z)^{n-1}\frac{\ud^{n-1}}{\ud z^{n-1}} \, {}_2F_1\left(\frac{1}{2},1,\frac{3}{2}; z\right) \right],
    \quad A_\ell \equiv \frac{2^\ell (2\ell + 1)!!}{(\ell + 2)!\, (\ell-2)!} \, .
\eeq
From DLMF-15.4.3~\cite{NIST:DLMF}, we know that ${}_2F_1\left(\frac{1}{2},1,\frac{3}{2}; z\right)=\arctan(\hat{a})/\hat{a}$, which is how our sought-after $\arctan\hat{a}$ makes its appearance.
We now apply the general Leibniz rule to obtain
\beq
F_{\ell}(\hat{a}^2)= A_\ell
\sum_{m=0}^{n+1}
\binom{n+1}{m} \sum_{k=0}^{n-1}
\binom{n-1}{k}
\frac{\ud^{n+1-m}}{\ud z^{n+1-m}}\left((1-z)^{n-1}
\frac{\ud^{n-1-k}\hat{a}^{-1}}{\ud z^{n-1-k}}\right)\frac{\ud^{k+m}\arctan\hat{a}}{\ud z^{k+m}} \, .
\eeq
It is easy to recognize that the only term with $\arctan\hat{a}$ in this double sum is the one with $k=m=0$. Thus,
\beq\label{eq:2F1-arctan,l even}
F_{\ell}(\hat{a}^2)=
P^{(\text{e})}_{\ell}(\hat{a})+Q^{(\text{e})}_{\ell}(\hat{a})\arctan\hat{a}  \qquad  (\ell~ \text{even}) \, ,
\eeq
where
\begin{subequations}
 \label{P_Q_l_even}
\begin{align}
P^{(\text{e})}_{\ell}(\hat{a}) & \equiv  A_\ell \!\!
\sum_{\substack{0 \le m \le n+1 \\ 0 \le k \le n-1 \\ (m,k) \neq (0,0)}}
\!\! \binom{n+1}{m}
\binom{n-1}{k}
\frac{\ud^{n+1-m}}{\ud z^{n+1-m}}\left((1-z)^{n-1}
\frac{\ud^{n-1-k}\hat{a}^{-1}}{\ud z^{n-1-k}}\right)\frac{\ud^{k+m}\arctan\hat{a}}{\ud z^{k+m}} \, ,
\\
Q^{(\text{e})}_{\ell}(\hat{a}) & \equiv A_\ell \,
\frac{\ud^{n+1}}{\ud z^{n+1}} \left((1-z)^{n-1}
\frac{\ud^{n-1}\hat{a}^{-1}}{\ud z^{n-1}}\right) .
\end{align}
\end{subequations}
Note that in these expressions, $\hat{a}$ is to be considered as a function of $z$, according to $\hat{a} = \sqrt{-z}$.
Is it easy to show that $Q^{(\text{e})}_{\ell}(\hat{a})$ is equal to $\hat{a}^{-2\ell-1}$ times a polynomial of degree $\frac{\ell}{2}-1$ in $\hat{a}^2$. Thus $Q^{(\text{e})}_{\ell}(\hat{a})$ is odd in $\hat{a}$ (as expected, since it multiplies the odd function $\arctan\hat{a}$ in \eqref{eq:2F1-arctan,l even} and $F_{\ell}(\hat{a}^2)$ is even).
Similarly, one can see\footnote{In fact, we can show that $P^{(\text{e})}_{\ell}(\hat{a})$ is equal to $\hat{a}^{-2\ell}\left(1+\hat{a}^2\right)^{-\ell/2-1}$ times a polynomial of degree $\ell-1$ in $\hat{a}^2$. However, on this instance, by giving various specific values to $\ell$ (cf. notebook~4 in App.~\ref{app:Sage}) we can then see that this latter polynomial can be factored out as $\left(1+\hat{a}^2\right)^{\ell/2-1}$ times a polynomial of degree $\ell/2$ in $\hat{a}^2$, so that it is as we state in the main text, namely, that $P^{(\text{e})}_{\ell}(\hat{a})$ is equal to $\hat{a}^{-2\ell}\left(1+\hat{a}^2\right)^{-2}$ times a polynomial of degree $\ell/2$ in $\hat{a}^2$.} that $P^{(\text{e})}_{\ell}(\hat{a})$ is equal to $\hat{a}^{-2\ell}\left(1+\hat{a}^2\right)^{-2}$ times a polynomial of degree $\ell/2$ in $\hat{a}^2$.
Thus $P^{(\text{e})}_{\ell}(\hat{a})$ is even in $\hat{a}$ (as expected, since  $F_{\ell}(\hat{a}^2)$  in \eqref{eq:2F1-arctan,l even} is even).

Let us now turn to the case where $\ell$ is odd, i.e. $\ell=2n+1$ with $n\in \mathbb{N}$. We
first write $F_{\ell}\left(\hat{a}^2\right)={}_2F_{1}(n+2,n+5/2, 2n+5/2,-\hat{a}^2)={}_2F_{1}(n+5/2, n+2,2n+5/2,-\hat{a}^2)$  and then apply
DLMF-15.5.7~\cite{NIST:DLMF}  (with `$n$' there set to 1, $a=n+3/2$, $b=n+2$ and $c=2n+3/2$) to readily get
\begin{equation}
\label{F_l_odd}
F_{\ell}\left(\hat{a}^2\right)=
B_\ell \left[\left(n+\frac{3}{2}\right)
F_{2n}(\hat{a}^2)
-(1-z)\frac{\ud
F_{2n}(\hat{a}^2)
}{\ud z}\right], \quad B_\ell \equiv \frac{2(2\ell + 1)}{(\ell+2)(\ell - 2)} \, .
\end{equation}
Using now Eq.~\eqref{eq:2F1-arctan,l even}, it readily follows that
\beq\label{eq:2F1-arctan,l odd}
F_{\ell}(\hat{a}^2)=
P^{(\text{o})}_{\ell}(\hat{a})+Q^{(\text{o})}_{\ell}(\hat{a})\arctan\hat{a}
\qquad  (\ell~\text{odd}) \, ,
\eeq
where
\begin{subequations}
 \label{P_Q_l_odd}
\begin{align}
P^{(\text{o})}_{\ell}(\hat{a})&\equiv B_\ell \left[\left(n+\frac{3}{2}\right)P^{(\text{e})}_{2n}(\hat{a})-(1+\hat{a}^2)\frac{\ud P^{(\text{e})}_{2n}(\hat{a})}{\ud z}+\frac{Q^{(\text{e})}_{2n}(\hat{a})}{2\hat{a}}\right],
\\ \label{eq:Qlo}
Q^{(\text{o})}_{\ell}(\hat{a})&\equiv
B_\ell \left[\left(n+\frac{3}{2}\right)Q^{(\text{e})}_{2n}(\hat{a})-(1+\hat{a}^2)\frac{\ud Q^{(\text{e})}_{2n}(\hat{a})}{\ud z}\right].
\end{align}
\end{subequations}
We next use the facts, shown above, that  $Q^{(\text{e})}_{\ell}(\hat{a})$ is equal to $\hat{a}^{-2\ell-1}$ times a polynomial---call it $q^{(\text{e})}_{\ell}(\hat{a})$---of degree $\frac{\ell}{2}-1$ in $\hat{a}^2$ and that $P^{(\text{e})}_{\ell}(\hat{a})$ is equal to $\hat{a}^{-2\ell}\left(1+\hat{a}^2\right)^{-2}$ times a polynomial of degree $\ell/2$ in $\hat{a}^2$,
to derive the following properties for $P^{(\text{o})}_{\ell}(\hat{a})$ and $Q^{(\text{o})}_{\ell}(\hat{a})$.
The term  $P^{(\text{o})}_{\ell}(\hat{a})$ is readily seen to be equal to
$\hat{a}^{-4n-2}\left(1+\hat{a}^2\right)^{-2}$ times a polynomial of degree $n+1$ in $\hat{a}^2$. In its turn, the coefficient $Q^{(\text{o})}_{\ell}(\hat{a})$ could be thought to be equal, naively and in principle,  to
$\hat{a}^{-4n-3}$ times a polynomial---call it $q^{(\text{o})}_{\ell}(\hat{a})$---of degree $n$ in $\hat{a}^2$. However,
a more detailed inspection shows that the coefficient of the highest-order term in the
polynomial $q^{(\text{o})}_{\ell}(\hat{a})$  of would-be degree  $n$ actually vanishes, so that it really is a polynomial of degree  $n-1$ in $\hat{a}^2$. Indeed, denoting by $c_{\text{h}}$ the coefficient of the highest-order term in $q^{(\text{e})}_{\ell}(\hat{a})$ (that is,  $q^{(\text{e})}_{\ell}(\hat{a})$ is equal to $c_{\text{h}}\hat{a}^{n-1}$ plus lower order terms), the
contribution to the would-be term $\hat{a}^{2n}$ in $q^{(\text{o})}_{\ell}(\hat{a})$  from the first and second  terms  in \eqref{eq:Qlo} are equal to, respectively, $Ac_{\text{h}}(n+3/2)\hat{a}^{2n}$
and $-Ac_{\text{h}}(n+3/2)\hat{a}^{2n}$, so that they actually cancel and  $q^{(\text{o})}_{\ell}(\hat{a})$  is a polynomial of degree $n-1$ in $\hat{a}^2$.
Thus, $Q^{(\text{o})}_{\ell}(\hat{a})$ is actually  equal to
$\hat{a}^{-4n-3}$ times a polynomial of degree $n-1$ in $\hat{a}^2$.

By combining the above results, one obtains formula~\eqref{G_ell_general_form} for $G_\ell(\hat{a}^2) = (1 + \hat{a}^2)^2 \, F_\ell(\hat{a}^2)$, which is valid whatever the parity of $\ell$.


\section{Electric and magnetic potentials of the Kerr horizon}\label{app:magnetic}

In this appendix we give an alternative derivation of the electric and magnetic potentials $E$ and $B$ of a Kerr black hole horizon (Secs.~\ref{subsec:conformal} and \ref{subsec:magnetic}), by solving for the linear partial differential equation \eqref{psi2_epsilon} obeyed by the complex-valued linear combination $F \equiv E + \ui B$, namely
\beq\label{PDE}
    \mathring{D}^2 F = - 2 \psi^{-2} \Psi_2 - 1 \, ,
\eeq
where $\mathring{D}^2$ stands for the Laplace operator associated with the `canonical' unit round metric $\mathring{q}_{ab}$. From Eqs.~\eqref{Psi2_Kerr} and \eqref{psi_Kerr} we note that the conformal factor $\psi$ and the Weyl curvature scalar $\Psi_2$ only depend on the Kerr polar coordinate $\theta$. We thus look for a solution of the form $F(\theta)$. Actually, the polar coordinate adapted to $\mathring{q}_{ab}$ is $\vartheta$ [Eq.~\eqref{canonical}], and not $\theta$. Hence
\beq
    \mathring{D}^2 F = \frac{1}{\sin\vartheta} \frac{\ud}{\ud\vartheta} \left( \sin\vartheta \, \frac{\ud F}{\ud\vartheta} \right) = \frac{\ud}{\ud z} \left( (1-z^2) \, \frac{\ud F}{\ud z} \right) ,
\eeq
where $z = \cos{\vartheta}$ is related to $\zeta = \cos{\theta}$ via Eq.~\eqref{z-zeta}. Then, using expression~\eqref{Psi2_Kerr} for $\Psi_2$, the relation $R^{-2} \psi^{-2} = \ud\zeta/\ud z$, which follows
from the system~\eqref{system}, and $R^2 = r_+^2+a^2$, Eq.~\eqref{PDE} becomes
\beq
    \frac{\ud}{\ud z} \left( (1-z^2) \, \frac{\ud F}{\ud z} \right) = \frac{C}{(1 - \ui \hat{a} \zeta)^3} \, \frac{\ud \zeta}{\ud z} - 1 \, ,
\eeq
where $C \equiv 2MR^2/r_+^3 = (1+\hat{a}^2)^2$ is a dimensionless constant. Remarkably, this differential equation can be integrated to yield
\beq
    (1-z^2) \, F'(z) = C \int^\zeta \frac{\ud \zeta'}{(1 - \ui \hat{a} \zeta')^3} - \int^z \ud z' = \frac{C}{2\ui\hat{a}} \, \frac{1}{(1 - \ui \hat{a} \zeta)^2} - z + z_0 \, ,
\eeq
where $z_0$ is a complex-valued constant of integration. Now by using Eq.~\eqref{trick} we readily get the following relationship between the differentials $\ud F$, $\ud \zeta$ and $\ud z$:
\beq
    \ud F = \frac{C}{2\ui\hat{a}} \, \frac{1}{(1 - \ui \hat{a} \zeta)^2} \left( \frac{1}{1-\zeta^2} - \beta^2 \right) \ud \zeta - \frac{z - z_0}{1-z^2} \; \ud z \, .
\eeq
Performing decompositions of the integrants in the right-hand side into simple elements and integrating, while using the identities $\beta^2=\hat{a}^2/(1+\hat{a}^2)$ and $C = \left(1+\hat{a}^2\right)^2$, then gives the general solution for $F$ in terms of $\zeta$ and $z(\zeta)$, in closed form:
\begin{align}\label{F}
    F &= \ln{(1-\ui \hat{a}\zeta)} - \frac{(1+\ui\hat{a})^2}{4\ui\hat{a}} \, \ln{|1-\zeta|} + \frac{(1-\ui\hat{a})^2}{4\ui\hat{a}} \, \ln{|1+\zeta|} \nonumber \\ &\quad + \frac{1}{2} \, \ln{(1-z^2)} + \frac{z_0}{2} \, \ln{\left| \frac{1-z}{1+z} \right|} + F_0 \, ,
\end{align}
where $F_0$ is a second complex-valued constant of integration. This expression contains terms that diverge as $\zeta \to \pm 1$. In those limits, we have the asymptotic behaviors [recall Eq.~\eqref{z-zeta}]
\beq
    z(\zeta) =
        \begin{cases}
            1 + e^{2\beta^2} (\zeta - 1) + O[(\zeta-1)^2]  & \text{if}~\zeta \to +1 \\
            - 1 + e^{2\beta^2} (\zeta + 1) + O[(\zeta+1)^2] & \text{if}~\zeta \to -1
    \end{cases} \, .
\eeq
By requiring that the coefficients of the diverging terms vanish in those two limits, i.e. by imposing global regularity of the general solution \eqref{F}, we obtain
\beq
    z_0 = \frac{1 - \hat{a}^2}{2\ui\hat{a}} \in \ui \mathbb{R} \, .
\eeq
Finally, by substituting for this value of $z_0$ into the expression \eqref{F}, while using the identities $(1-z)/(1-\zeta) \times (1+\zeta)/(1+z) = e^{2\beta^2 \zeta}$ and $\beta^2 = \hat{a}^2/(1+\hat{a}^2)$, we find the remarkably compact expression\footnote{Note that the second term in Eq.~\eqref{Fbis} can also be written explicitly in terms of $\zeta$, according to $$\frac{1}{2} \, \ln{\left(\frac{1-z^2}{1-\zeta^2}\right)} = - \ln{\left[ (1-\zeta) \, e^{\beta^2\zeta} + (1+\zeta) \, e^{-\beta^2\zeta} \right]} + \ln{2} \, .$$}
\begin{align}\label{Fbis}
    F &= \ln{(1-\ui \hat{a}\zeta)} + \frac{1}{2} \, \ln{\left(\frac{1-z^2}{1-\zeta^2}\right)} - \frac{\ui\hat{a}\zeta}{2} \, \frac{1-\hat{a}^2}{1+\hat{a}^2} + F_0 \, .
\end{align}
The logarithms therein suggest to exponentiate Eq.~\eqref{Fbis}. According to the definition \eqref{def_E_electric}, the electric potential $E$ is itself the logarithm of the rescaled conformal factor $R\,\psi > 0$. We thus have $e^F = (R\,\psi) \, e^{\ui B}$, which implies (with $\zeta = \cos{\theta}$)
\begin{subequations}
    \begin{align}
        R\,\psi = \left| e^F \right| &= \left| e^{F_0} \right| \frac{\sqrt{(1 + \hat{a}^2 \cos^2{\theta})(1-z^2(\theta))}}{\sin{\theta}} \, , \\
        B = \arg{\left(e^F\right)} &= - \arctan{(\hat{a} \cos{\theta})} - \frac{1-\hat{a}^2}{2(1+\hat{a}^2)} \, \hat{a} \cos{\theta} + B_0 \, ,
    \end{align}
\end{subequations}
where $B_0 \equiv \text{Im}\,F_0$ is a physically irrelevant constant since $B$ is a (pseudo-scalar) potential. These expressions are regular throughout $\theta \in (0,\pi)$, and are in perfect agreement with the results \eqref{conformal} [with $|e^{F_0}| = (1+\hat{a}^2)^{-1/2}$] and \eqref{B_Kerr} established in Secs.~\ref{subsec:conformal} and \ref{subsec:magnetic}.

\section{Asymptotic behavior in the small-spin regime}\label{app:asymptotics}

In this appendix we establish that the horizon multipoles \eqref{I-L_Kerr2} of a Kerr black hole have the asymptotic behavior \eqref{I-L_Kerr3}--\eqref{alpha_ell} in the small-spin regime. Let $J_\ell$ denote the definite integral appearing in Eq.~\eqref{I-L_Kerr2}, i.e.,
\beq\label{J}
    J_\ell(\hat{a}) \equiv \int_{-1}^1 \ud x \, \frac{P_\ell(z(x;\beta))}{(1 - \ui \hat{a} x)^3} \, ,
\eeq
with $\hat{a} = a/r_+$ a given parameter in the range $[0,1)$ and $\beta = \hat{a}/\sqrt{1+\hat{a}^2}$. Recall that $P_\ell(z)$ is the Legendre polynomial of order $\ell$ and the function $z(x;\beta)$ is given by formula \eqref{z-zeta}.\footnote{In this appendix we make explicit the dependence of that function on the deformation parameter $\beta$ and we denote $\zeta$ by $x$.} We thus wish to prove that, in the regime $\hat{a} \ll 1$ of small spin values,
\beq\label{J2}
    J_\ell \sim 2\alpha_\ell \, (\ui \hat{a})^\ell \, .
\eeq
We shall first establish the scaling with spin, before determining in closed form the numerical prefactor $\alpha_\ell$.

\subsection{Scaling with spin}

Define $\epsilon(x;\beta) \equiv z(x;\beta) - x = O(\beta^2)$ and perform a Taylor series expansion of $P_\ell(z(x;\beta))$ about $z = x$. Then the integral \eqref{J} becomes
\beq\label{pilou}
    J_\ell = \sum_{n=0}^{+\infty} J_{\ell,n} \quad \text{with} \quad J_{\ell,n} = \frac{1}{n!} \! \int_{-1}^1 \ud x \, \frac{\epsilon^n(x;\beta) \, P_\ell^{(n)}(x)}{(1 - \ui \hat{a} x)^3} \, .
\eeq
Next, using the fact that $\tanh'(x) = 1 - \tanh^2(x)$, we may perform a Taylor series expansion of $\epsilon(x;\beta)$ itself near $\beta = 0$, at fixed $x$, according to
\beq\label{epsilon_exp}
     \epsilon(x;\beta) = \sum_{k=1}^{+\infty} \frac{(-\beta^2 x)^k}{k!} \, \tanh^{(k)}\!{(\mathrm{artanh}\,x)} = \beta^2 \, (x^2-1) \, x \sum_{k=0}^{+\infty} \frac{\beta^{2k} \, (-x)^k}{(k+1)!} \, Q_k(x) \, ,
\eeq
with $Q_k(x)$ a polynomial of degree $k$ that obeys the recurrence formula $Q_{k+1} = (1-x^2) Q'_k - 2x Q_k$. Its coefficients involve the (even) Bernoulli numbers, and can be computed explicitly if necessary. Substituting expression \eqref{epsilon_exp} into \eqref{pilou} yields
\beq\label{J_ln1}
    J_{\ell,n} = \frac{\beta^{2n}}{n!} \! \int_{-1}^1 \ud x \, f_n(x) \, P_\ell^{(n)}(x) \, ,
\eeq
with
\beq\label{f_n}
    f_n(x) = (x^2-1)^n \, x^n \left[ \sum_{k=0}^{+\infty} \frac{\beta^{2k} \, (-x)^k}{(k+1)!} \, Q_k(x) \right]^n \left[ \sum_{p=0}^{+\infty} \frac{(p+1)(p+2)}{2} \, (\ui \hat{a} x)^p \right] .
\eeq
For later use, it is convenient to rewrite \eqref{f_n} in the following canonical form, which makes explicit both the minimal power of $x$ and the associated scaling with $\hat{a}$ and $\beta$:
\beq\label{f_n_bis}
    f_n(x) = (x^2-1)^n \sum_{m=0}^{+\infty} (\ui \hat{a})^m A_{n,m}(\beta^2) \, x^{n+m} \, ,
\eeq
where the coefficients $A_{n,m}$ are analytic in $\beta^2$.

We may now integrate by parts $n$ times the integral \eqref{J_ln1}, while noticing that the boundary contributions at $x = \pm 1$ vanish thanks to the factor $(x^2-1)^n$ in \eqref{f_n_bis}. This gives
\beq\label{J_ln2}
    J_{\ell,n} = (-)^n \, \frac{\beta^{2n}}{n!} \! \int_{-1}^1 \ud x \, f^{(n)}_n(x) \, P_\ell(x) \, .
\eeq
We have thus reduced the problem of evaluating the integral \eqref{J} to that of determining the projections over the Legendre polynomials of the (derivatives of the) functions \eqref{f_n_bis}.

To extract the leading-order behavior of each $J_{\ell,n}$ as $\hat{a} \to 0$ (or equivalently as $\beta \to 0$), we now make use of the Rodrigues formula \eqref{Rodrigues} for the Legendre polynomials. By substituting for this formula into Eq.~\eqref{J_ln2} and integrating by parts $\ell$ times, we notice that the boundary contributions at $x = \pm 1$ vanish, once again, thanks to the factor $(x^2-1)^\ell$ in Eq.~\eqref{Rodrigues}. We thus obtain
\beq\label{J_ln3}
    J_{\ell,n} = (-)^{\ell+n} \, \frac{\beta^{2n}}{2^\ell\ell!n!} \int_{-1}^1 \ud x \, (x^2-1)^{\ell} f^{(\ell+n)}_n(x) \, .
\eeq

To evaluate the scaling of that integral as a function of $(\ell,n)$ in the regime where $\hat{a} \to 0$, we must control the derivative of order $\ell+n$ of the function \eqref{f_n_bis}. Using the Leibniz rule, we readily find
\beq\label{f_n^(l+n)}
    f^{(\ell+n)}_n(x) = \sum_{m=0}^{+\infty} (\ui \hat{a})^m A_{n,m}(\beta^2) \sum_{k=0}^{\ell+n} \binom{\ell+n}{k} \, \frac{\ud^k}{\ud x^k} \left[ (x^2-1)^n \right] \, \frac{\ud^{n+\ell-k}}{\ud x^{n+\ell-k}} \left[ x^{m+n} \right] .
\eeq
The derivative of order $k$ of the polynomial $(x^2-1)^n$ vanishes if $k > 2n$, and the derivative of order $n+\ell-k$ of the monomial $x^{m+n}$ vanishes if $\ell-k > m$. These conditions imply that the terms with $m < \ell - 2n$ do not contribute to \eqref{f_n^(l+n)}, so that whenever $2n \leqslant \ell$, Eq.~\eqref{J_ln3} with $\beta \sim \hat{a}$ implies the asymptotic behavior (coming from $m = \ell - 2n \geqslant 0$)
\beq\label{done}
    J_{\ell,n} \sim (-)^n \beta^{2n} \, c_{\ell,n} \, (\ui \hat{a})^{\ell-2n} \sim c_{\ell,n} \, (\ui \hat{a})^\ell \, ,
\eeq
for some real-valued coefficient $c_{\ell,n}$. If $2n > \ell$, however, then the series in Eq.~\eqref{f_n^(l+n)} does not truncate to a lower bound for $m$, but the factor of $\beta^{2n} \sim \hat{a}^{2n}$ in \eqref{J_ln3} implies a subdominant contribution with respect to the leading-order scaling behavior \eqref{done} of the integrals $J_{\ell,n}$ with $2n \leqslant \ell$. We thus conclude that the sum \eqref{pilou} scales as $(\ui \hat{a})^\ell$, as claimed.

Notice that in the specific case where $n = 0$, the above analysis can be made more precise, and we readily find
\beq \label{J_l0}
    J_{\ell,0} = \frac{(-)^\ell}{2^\ell\ell!} \sum_{p=0}^{+\infty} \frac{(p+2)(p+1)}{2} \, (\ui \hat{a})^p \int_{-1}^1 \ud x \, (x^2-1)^{\ell} \, \frac{\ud^\ell x^p}{\ud x^\ell} \sim \frac{(\ell+2)!}{(2\ell+1)!!} \; (\ui \hat{a})^\ell \, .
\eeq
This coincides with the asymptotic behavior of the closed-form expression \eqref{J_l0_var} as $\hat{a} \to 0$.

\subsection{Numerical prefactor}

Having established that $J_\ell \sim (\ui \hat{a})^\ell$ in the regime where $\hat{a} \ll 1$, we would like to determine in closed form the numerical prefactor, i.e. the coefficient $\alpha_\ell$ in Eq.~\eqref{J2}. Given the intricate expression for the function $f_n(x)$ appearing in \eqref{f_n}, obtaining a closed-form formula for $\alpha_\ell$ from the analysis above is challenging. Instead, we shall proceed by inductive reasoning as follows.

For a given $\ell \in \mathbb{N}$, we perform a Taylor-series expansion of the integrand in \eqref{J} in powers of $\hat{a} \ll 1$ (e.g. by means of computer algebra; cf. notebook~3 in App.~\ref{app:Sage}), and proceed to compute the integral in closed form. Doing so e.g. for all $\ell \in \{0,\dots,14\}$, we find that \eqref{J2} holds, with the numerical values for $\alpha_\ell$ listed in Table~\ref{table:sequence} above. Next, we look for patterns in that sequence of rational numbers and try to infer a general formula which would be valid for all $\ell \in \mathbb{N}$.

To do so, we first note that by multiplying $(\alpha_\ell)_{\ell \in \{0,\dots,14\}}$ by the binomials $\binom{2\ell+1}{\ell}$, we obtain a sequence of \textit{integers} $\beta_\ell \equiv \binom{2\ell+1}{\ell} \, \alpha_\ell$, namely
\begin{align}\label{betas}
    {(\beta_\ell)}_{\ell \in \{0,\dots,14\}} &= (1,3,12,44,176,672,2688,10432,41728,163328, \nonumber \\ &\qquad\! 653312,2570240,10280960,40583168,162332672) \, .
\end{align}
Second, we observe a simple pattern relating even and odd members of the sequence \eqref{betas}, namely $\beta_{2n} = 4\beta_{2n-1}$ for all $n \in \{1,\dots,7\}$. Moreover, all even members of $(\beta_\ell)$ are divisible by $2^\ell$. We can thus write $\beta_{2n} = 2^{2n} \, \gamma_n$, and therefore $\beta_{2n-1} = 2^{2(n-1)} \, \gamma_n$, where
\beq\label{gammas}
    {(\gamma_n)}_{n \in \{0,\dots,7\}} = (1,3,11,42,163,638,2510,9908) \, .
\eeq
Third, we notice that this specific sequence of numbers matches perfectly the first 8 elements of a particular sum of binomials, namely \cite{OEIS:A032443}
\beq
    \gamma_n = \sum_{k=0}^n \binom{2 n}{k} = \frac{2^{2n} + \binom{2n}{n}}{2} \, .
\eeq
This leads us to infer the general formula \eqref{alpha_ell} for $\alpha_\ell$. We then use this formula to \textit{predict} the next few numerical values, e.g. for $\ell \in \{15,16,17\}$:
\beq\label{prediction}
    \alpha_{15} = \frac{642301952}{300540195} \, , \qquad \alpha_{16} = \frac{1284603904}{583401555} \, , \qquad \alpha_{17} = \frac{3080192}{1372525} \, .
\eeq
As a check,
one can repeat the Taylor-series expansion detailed above to determine $J_{15}$, $J_{16}$ and $J_{17}$ to leading order in $\hat{a}$; see notebook~3 in App.~\ref{app:Sage}. The pattern \eqref{J2} is found to hold, with the corresponding numerical prefactors given by the values \eqref{prediction}, as predicted. Beware that the inductive reasoning that we followed here does not provide a rigorous (i.e. hypothetico-deductive) mathematical proof, but rather gives convincing arguments that the closed-form formula \eqref{alpha_ell} is indeed correct for any $\ell \in \mathbb{N}$.

\section{SageMath notebooks} \label{app:Sage}

Some exact and numerical computations in this article have been performed by means of the free Python-based mathematical software system SageMath \cite{SageMath}.
The relevant Jupyter notebooks are publicly available, from the Zenodo repository \cite{Zenodo} or directly from the links below. These notebooks have also been used to generate Figs.~\ref{fig:G_ell}--\ref{fig:Kerr_L_comp_axi}.
Notebooks 1 and 2 are using SageMath differential geometry tools developed through the SageManifolds project \cite{GoMa.18,SageManifolds}, while notebook~3 is using SageMath's interface to FLINT \cite{FLINT} for arbitrary precision computations of complex-valued integrals with error bounds.

\begin{enumerate}
    \item Unit round metric of a cross-section of the Kerr horizon conformally related to the physical 2-metric:\\
    \url{https://nbviewer.org/url/zenodo.org/records/18511585/files/Kerr_cross_section.ipynb}
    \item \Hajicek{} 1-form and magnetic potential $B$ on the Kerr horizon:\\
    \url{https://nbviewer.org/url/zenodo.org/records/18511585/files/Kerr_Hajicek_form_B.ipynb}
    \item Evaluation of the multipole moments of the Kerr horizon:\\
    \url{https://nbviewer.org/url/zenodo.org/records/18511585/files/Kerr_multipoles.ipynb}
    \item Expression of the hypergeometric function $F_\ell(\hat{a}^2)$ (App.~\ref{app:integ_Kerr_axi}) in terms of arctangent:\\
    \url{https://nbviewer.org/url/zenodo.org/records/18511585/files/hypergeom_arctan.ipynb}
\end{enumerate}

\bibliography{main}

\end{document}